\documentclass[12pt,prd,amsmath,amssymb,nofootinbib,tightenlines,floatfix,preprintnumbers]{revtex4-1}

\usepackage{epsfig,latexsym,cancel}
\usepackage{color}
\usepackage{graphicx}

\newcommand{\be}{\begin{equation}}
\newcommand{\ee}{\end{equation}}
\newcommand{\bea}{\begin{eqnarray}}
\newcommand{\eea}{\end{eqnarray}}
\newcommand{\ba}{\begin{array}}
\newcommand{\ea}{\end{array}}

\newcommand{\Lint}{{\mathcal{L}_{\textrm{int}}}}

\long\def\symbolfootnote[#1]#2{\begingroup%
\def\thefootnote{\fnsymbol{footnote}}\footnote[#1]{#2}\endgroup} 

\newcommand{\eq}[1]{Eq.~\eqref{#1}}
\newcommand{\eqs}[2]{Eqs.~\eqref{#1} and \eqref{#2}}

\renewcommand{\sec}[1]{Sec.~\ref{#1}}

\newcommand{\vect}[1]{\mathbf{#1}}
\newcommand{\abs}[1]{\left\lvert #1\right\rvert}

\newcommand{\pd}[2]{\frac{\partial #1}{\partial #2}}

\newcommand{\eosc}{\epsilon_{\text{osc}}}
\newcommand{\ew}{\epsilon_{\text{wall}}}
\newcommand{\eint}{\epsilon_{\text{coll}}}
\newcommand{\tw}{\tau_{\text{w}}}

\DeclareMathOperator{\Tr}{Tr}

\DeclareMathOperator{\PV}{PV}
\newcommand{\Id}{\begin{pmatrix} 1 & 0 \\ 0 & 1 \end{pmatrix}}

\begin{document}

\preprint{LAUR-09-08118}
\preprint{NPAC-09-16}
\preprint{UCB-PTH-09/37}

\title{Flavored Quantum Boltzmann Equations}

\author{Vincenzo Cirigliano}
\affiliation{Theoretical Division, Los Alamos National Laboratory, Los Alamos, NM, 87545, USA}
\author{Christopher Lee} 
\affiliation{Center for Theoretical Physics, University of California, \\ and Theoretical Physics Group, Lawrence Berkeley National Laboratory,  Berkeley, CA, 94720, USA}
\author{Michael J. Ramsey-Musolf} 
\affiliation{Department of Physics, University of Wisconsin--Madison, 1150 University Ave., Madison, WI, 53706, USA \\ and \\
Kellogg Radiation Laboratory, California Institute of  
Technology, Pasadena, CA, 91125, USA}
\author{Sean Tulin} 
\affiliation{Theory Group, TRIUMF, 4004 Wesbrook Mall, Vancouver, BC, V6T 2A3, Canada}

\date{February 15, 2010}

\begin{abstract}
\vspace{1cm}

We derive from first principles, using non-equilibrium field theory, the quantum Boltzmann equations that describe the dynamics of flavor oscillations, collisions, and a time-dependent mass matrix in the early universe.  
Working to leading non-trivial order in ratios of relevant time scales, 
we study in detail a toy model for weak scale baryogenesis: two scalar species that mix through a  slowly varying  
time-dependent and $CP$-violating mass matrix, and interact with a thermal bath.  
This model clearly illustrates how the $CP$ asymmetry arises through
coherent flavor oscillations in a non-trivial background.  We solve the Boltzmann equations numerically for the density matrices, investigating the impact of collisions in various regimes.

\end{abstract}

\maketitle

\newpage

\section{Introduction}
\label{sec:intro}

The origin of the baryon asymmetry of the Universe (BAU) is one of the great unsolved puzzles in 
particle and nuclear physics and cosmology.
The BAU has been measured through studies of (i) Big Bang Nucleosynthesis (BBN), the epoch of light element formation at time $t_{\mathrm{BBN}} \sim 1$ minute, and (ii) the cosmic microwave background (CMB), a relic of hydrogen recombination at $t_{\mathrm{CMB}} \sim 10^5$ years.  Characterized by $n_B/s$, the ratio of baryon number density to entropy density, the BAU has the value
\be
n_B/s = \left\{ 
\begin{array}{ccc}
(6.7 \; - \; 9.2) \times 10^{-11}  & \; & \textrm{BBN} \; \text{\cite{Yao:2006px}} \\
(8.36  \; - \; 9.32) \times 10^{-11}  & \; & \textrm{CMB} \; \text{\cite{Yao:2006px, Dunkley:2008ie}}  \end{array} \right.
\ee
at 95\% C.L.  It is a triumph for cosmology that both measurements agree, despite the fact that $t_{\mathrm{BBN}} \ll t_{\mathrm{CMB}}$.  At the same time, it is a challenge for particle and nuclear physics, as the Standard Model (SM) cannot account for the observed BAU~\cite{ckn-review,Kajantie:1995kf}. 
Successful baryogenesis requires the concurrence of three necessary conditions~\cite{Sakharov:1967dj}: 
violation of baryon number ($B$); violation of $C$ and $CP$ symmetries, where $C$ is charge-conjugation and $P$ is parity; 
and a departure from thermal equilibrium (or a violation of $CPT$ symmetry, where $T$ is time-reversal).  
Therefore, on general grounds  a quantitative understanding of baryogenesis mechanisms   
requires setting up transport equations  for quantum systems in out-of-equilibrium conditions. 
More specifically,  the genuinely quantum phenomena of particle mixing and  flavor oscillations  play an important 
role in a number of baryogenesis mechanisms, from weak scale baryogenesis~\cite{krs85} 
to leptogenesis~\cite{Fukugita:1986hr},  where one needs to follow the evolution of $CP$ asymmetries  
in lepton flavor space~\cite{Abada:2006fw,Nardi:2006fx}.
Mostly motivated by applications to supersymmetric electroweak baryogenesis (EWB), in this paper 
we study the formulation of quantum kinetic  equations for mixing species from first principle 
non-equilibrium quantum field theory.
We focus explicitly on mixing scalars in a time-varying mass background, although the main concepts and techniques 
will apply to fermionic systems as well. 

In electroweak baryogenesis (EWB), the baryon asymmetry is generated during the electroweak phase transition, 
the era at temperature $T\sim 100$ GeV  when the Higgs field acquires a vacuum expectation value (vev). 
The generation of  a net baryon number occurs as follows~\cite{ckn-review}:
If the phase transition is first order, bubbles of broken electroweak symmetry 
($ \langle \Phi \rangle \neq 0$) nucleate and expand in a background of unbroken symmetry, 
providing the necessary departure from equilibrium.  
$CP$-violating interactions between the Higgs field(s) and other particle species 
may lead to the production of  $CP$-asymmetries in certain particle densities,  
within the  expanding  domain wall separating the two phases (bubble wall).  
These $CP$-asymmetries diffuse ahead of the expanding bubble,
and through inelastic scattering get partially converted into 
a $CP$-asymmetry of left-handed SM doublets ($n_L \neq 0$),    which is then 
eventually converted into a baryon asymmetry by weak sphaleron processes.
Finally, the baryon asymmetry is captured into the broken phase by the expanding bubble, 
where it persists without washout only if the 
phase transition is strongly first order ($\langle \Phi \rangle/T \geq 1$ so that electroweak
 sphalerons are shut off).
Within the SM,  it turns out that for values of the Higgs mass experimentally allowed 
the phase transition is not first order~\cite{Kajantie:1995kf}, 
hence EWB is not viable. 

In  SM extensions, the viability of this scenario depends on two main considerations. 
First, is the electroweak phase transition strongly first order?  
Second, is there enough $CP$-violation to generate the observed BAU?  

An answer to the first question follows from a study of the Higgs finite temperature effective potential~\cite{Quiros:1999jp}. 
The requirement of a strong first-order phase transition in general 
implies strong restrictions on the spectrum. 
For example, in the Minimal Supersymmetric Standard Model (MSSM), this requirement implies a relatively light Higgs boson ($m_h < 127$ GeV) and stop (supersymmetric partner of the top quark), 
with $m_{\tilde{t}} <  125$ GeV~\cite{Cline:1998hy,Quiros:1999jp,Carena:2008vj}.  Other models lead to various predictions within an extended Higgs sector~\cite{Profumo:2007wc, Menon:2004wv, Huber:2006wf, Huber:2000mg}.  The LHC should definitely be able to probe these scenarios.

On the other hand, addressing the second issue remains an open problem that requires 
a quantitative understanding of particle transport at the phase boundary in presence of $CP$ 
violation.  In general, one has to  write down a network of  kinetic equations for the relevant species accounting 
for all relevant processes  ($CP$ violation,  elastic and inelastic scattering,  $B$ violation),  
and solve them under the boundary condition of equilibrium far away from the domain wall. 
Current state-of-the-art treatments utilize the Closed Time Path (CTP) formalism~\cite{Schwinger:1960qe,Keldysh:1964ud,
Mahanthappa:1962ex,Bakshi:1962dv}
of finite-temperature quantum field theory~\cite{Carena:1997gx,Riotto:1998zb,Carena:2000id,Carena:2002ss,Cirigliano:2006wh, Prokopec:2003pj, Prokopec:2004ic, konstandin}.
Yet, within this formalism, different groups rely upon different simplifying approximations to compute the BAU.  
Within  the MSSM, the most well-studied scenario for EWB, 
given the same underlying parameters, the predicted BAU  spans 
nearly two orders of magnitude~\cite{Carena:2002ss, konstandin, Cirigliano:2006wh}. 
Our ability to test the viability of supersymmetric EWB through a combined analysis of  
collider and Electric Dipole Moment experiments~\cite{edmpheno} 
depends critically on the resolution of this theoretical confusion.  

Within the MSSM, the expanding bubble wall has a thickness $L_w \sim 20/T$~\cite{Moreno:1998bq},  much larger than the 
typical  thermal  de Broglie  wavelength,  the  intrinsic length  $L_{\rm  int} \sim O(1/T)$. 
This circumstance enables one to use the so-called gradient expansion in $L_{\rm int}/L_w $ in studying the quantum transport dynamics. 
To leading (first) order in  this gradient expansion, the $CP$ asymmetries arise through 
particle mixing~\cite{konstandin}.  At second order in the gradient expansion, one finds 
a number of additional contributions, including the ``semiclassical force"  term  discussed 
extensively in Refs.~\cite{Cline:1997vk, Cline:2000nw, Kainulainen:2001cn,Kainulainen:2002th,Zhou:2008yd}.
Here we confine ourselves to leading order in the gradient expansion, in which 
the dominant non-equilibrium effect  is that the  bubble wall induces  
space-time dependent particle mixing phenomena for all species coupling to the Higgs.
Therefore,  a remarkably simple picture emerges: 
physically,  the sourcing of  $CP$ asymmetries  is   essentially reduced to the 
phenomenon of  coherent flavor oscillation in a non-trivial background.
On the technical side,  mixing particles are to be described 
by density matrices rather than particle  number distributions, in order  to properly take 
into account the coherent  $CP$-violating oscillations and the collisions 
that tend to break such coherence.
The diagonal entries  of the density matrix   keep track of the populations of the individual ``flavor states'', 
while the off-diagonal terms  keep track of the coherence between the two mixing flavor states.
We emphasize that it is only by evolving the full density matrices that one can account for the relevant physics 
and avoid technical issues such as the basis-dependence (flavor versus mass basis) of the final results. 
Although the relevance of flavor oscillations has been pointed out in Refs.~\cite{konstandin}, 
none of the existing treatments in the literature includes correctly 
the physics of both $CP$-violating flavor oscillations and collisions. 

The present work is the first step towards such a complete treatment of the dynamics of mixing particles.  
We study a toy model of mixing scalars $\Phi_{L,R}$  that incorporates a number of 
salient aspects of the full EWB problem, namely: 
(i)  time-dependent, flavor non-diagonal   and $CP$-violating mass matrix, 
which leads (through flavor oscillations)  to the generation of a $CP$-violating  asymmetry from a $CP$-symmetric 
equilibrium initial state; 
(ii)   flavor-dependent collisions in the plasma, described  by interactions of $\Phi_{L,R}$ with a thermal background 
of  scalar bosons ($A$) in equilibrium.
Compared to the realistic EWB problem, the main simplification in the toy model is the use of a time-dependent mass matrix, 
rather than one that varies in both space and time. 
In the latter case, diffusion currents arise and tend to 
enhance the BAU, as they transport the $CP$ asymmetries into the unbroken phase 
where  electroweak sphalerons are active~\cite{Cohen:1994ss}. 
Diffusion currents are not fully incorporated in the existing oscillation-based treatment of Ref.~\cite{konstandin}. 
The all-important generalization to a space-dependent mass matrix will be the subject of a separate publication, while
here we focus on the time-dependent case.

We note that we are not the first to study the problem of flavor mixing  in the presence of a time-varying mass matrix. Ê
In different physical contexts, earlier  works~\cite{Nilles:2001fg,Gumrukcuoglu:2008fk, Garbrecht:2002pd,Garbrecht:2003mn,Garbrecht:2007gf}  have approached the problem by utilizing the method of
Bogoliubov transformations to derive equations of motion for particle number operators in vacuum. Ê
In the present study, we follow a more kinetic theory-oriented  approach, as we seek to determine 
the impact of plasma interactions on the evolution of the system (damping of flavor oscillations, equilibration). 
We thus view our results as complementary to those of 
Refs.~\cite{Nilles:2001fg,Gumrukcuoglu:2008fk, Garbrecht:2002pd,Garbrecht:2003mn,Garbrecht:2007gf}. 

Our discussion is organized in the following way.
In Sec.~\ref{sec:prelim}, we describe our toy model: a two-flavor scalar system with a time-dependent mass matrix and interactions with a thermal bath, modelled by a third scalar $A$, assumed to be in equilibrium.  Additionally, we provide an expansion scheme in the ratios of time scales in the problem.  The relevant time scales are the collisional mean free time $\tau_{\rm coll}$, the flavor oscillation time scale $\tau_{\rm osc}$, and the wall time scale\footnote{We borrow the word ``wall'' from the EWB problem, even though it doesn't strictly apply to our toy model scenario.  In this work, ``wall'' denotes the time region over which the time-dependent mass matrix is varying; the ``wall time scale'' is the time scale over which this variation occurs.} $\tw$; we assume that all three time scales are much larger than the inverse frequencies of $\Phi_{L,R}$, generically denoted $\tau_{\rm int}$.  For a slowly varying wall ($\tw \gg \tau_{\rm int}$), this is the physically interesting regime.

In Sec.~\ref{sec:analytic}, we derive the quantum Boltzmann equation for the two-flavor density matrices in our toy model, utilizing the Closed Time Path formalism.  Here, our expansion scheme plays a key role, allowing for a straightforward generalization of previous one-flavor treatments.  Our results provide a significant improvement over previous treatments of flavored Boltzmann equations; we derive the two-flavor Boltzmann equations from first principles, including collisions, flavor oscillations, and a time-dependent mass matrix in one unified framework.

In Sec.~\ref{sec:numerical}, we solve the quantum Boltzmann equations numerically for our toy model.  Previous EWB treatments have relied upon various unproven ans\"atze for the form of the density matrices; in our work, we obtain directly the two-flavor density matrices as functions of momentum and time.  
First, we illustrate of the nature of the (leading-order) $CP$-violating source in EWB, showing explicitly how a $CP$-asymmetry can arise through coherent flavor oscillations in a time-varying background.  We show that the resulting $CP$-asymmetry is maximized for $\tau_{\rm osc} \sim \tw$, reminiscent of the ``resonant" EWB scenario studied in Refs.~\cite{Carena:1997gx,Riotto:1998zb,Carena:2000id,Carena:2002ss,Cirigliano:2006wh}.  Next, we explore the impact of collisions with the thermal bath of $A$ bosons.  Generally speaking, we find that collisions lead to decoherence of flavor oscillations and relaxation of the density matrices to their equilibrium forms.  In the case of flavor-sensitive interactions, all $CP$-asymmetries induced by the wall are ultimately damped away at late times; for flavor-blind interactions, $CP$-asymmetry persists.  Lastly, we study how the two-flavor dynamics depends on the underlying parameters of the mass matrix.  In particular, we find that the EWB resonance found in previous treatments~\cite{Carena:1997gx,Riotto:1998zb,Cirigliano:2006wh} does persist somewhat in our more exact formalism.

We provide our conclusions in Sec.~\ref{sec:conclude}.  In addition, we provide three appendices.  Appendix A provides a brief review of the Closed Time Path formalism.  Appendix B describes how to derive the usual one-flavor Boltzmann equation using the CTP approach.  Appendix C is an addendum to Sec.~\ref{sec:analytic}, providing additional technical details related to the derivation of the two-flavor Boltzmann equations.

\section{Preliminaries}

\label{sec:prelim}

\subsection{Toy Model}

We consider a two-flavor scalar system, with fields $(\Phi_{L},\Phi_R) \equiv \Phi$, described by the Lagrangian
\be
\mathcal{L}(x) = \partial_\mu \Phi^\dagger \, \partial^\mu \Phi - \Phi^\dagger \, M^2 \, \Phi + \Lint \;, \label{eq:model}
\ee
with spacetime coordinate $x = (x^0\equiv t, \, \mathbf x)$.  We couple each flavor scalar field to a real scalar $A$ 
via the interaction
\be
\Lint = \, - \, \frac{1}{2} \: A^2 \, \Phi^\dagger \, y \, \Phi  \label{eq:lint} \; .
\ee
The $A$ field is assumed to be in thermal equilibrium and we will not consider 
its evolution equations. The role of $A$ in this toy model is simply to provide 
a thermal bath of scatterers for the $\Phi$ fields. 
We take the coupling constants
\be
y=\left( \ba{cc} y_L & 0 \\ 0  & y_R \ea \right)
\ee
to be flavor diagonal; indeed, this defines the ``flavor-basis'' fields $\Phi$.  In addition, we assume that the mass matrix $M^2(t)$ is 
flavor non-diagonal and is a function of time:
\be
M^2(t) = \left( \ba{cc} m_L^2 & v(t) \, e^{-i \, a(t)} \\ v(t) \, e^{i \, a(t)} & m_R^2 \ea \right) \; .  \label{eq:wow}
\ee
The off-diagonal elements, assumed to be a function of a time-dependent background field (e.g., the Higgs vev), are parametrized in terms of the magnitude $v(t)$ and phase $a(t)$.  Motivated by EWB~\cite{Moreno:1998bq}, we assume the following forms
\bea
v(t) &=& \frac{v_0}{2} \left( \, 1+ \tanh(t/\tw) \, \right) \\
a(t) &=& \frac{a_0}{2} \left( \, 1+ \tanh(t/\tw) \, \right) \  , 
\label{eq:tauw}
\eea
that introduce the external ``wall time scale'' $\tw$.  A necessary condition for $CP$-violation is $\dot a \ne  0$ (see 
below).  Our formalism can be adapted to other functional forms as well, as long as the relevant time 
scale is sufficiently long.

The important ingredients of this model are as follows:
\begin{itemize}
\item time-dependent, $CP$-violating mass matrix, which leads to the generation of a $CP$-violating flavor asymmetry from a  $CP$-symmetric initial state; 
\item flavor oscillations, due to the flavor non-diagonal mass matrix; 
\item flavor-dependent collisions in the plasma, described here by interactions of $\Phi_{L,R}$ with a thermal background of $A$ bosons.
\end{itemize}

It is convenient to transform Eq.~\eqref{eq:model} into the ``local mass basis.''  We
diagonalize the mass matrix with the time-dependent transformation $U(t)$, such that
\be
U^\dagger(t) \, M^2(t) \, U(t) = \left( \ba{cc} m^2_1(t) & 0 \\ 0 & m_2^2(t) \ea \right) \equiv m^2(t) \; ,  \quad 
 U(t) = \begin{pmatrix} \cos\theta(t) & -\sin\theta(t)\, e^{-i\sigma(t)} \\ \sin\theta(t)\, e^{i\sigma(t)} & \cos\theta(t) \end{pmatrix}  \, , 
\label{eq:diagonalization}
\ee
with
\bea
m^2_{1,2}(t) &=& \frac{1}{2} \, (m_L^2 + m_R^2) \   \pm \  \frac{1}{2} \,  \sqrt{(m_L^2 - m_R^2)^2  + 4 \, v^2(t) } \\
\tan 2\theta (t) & =&  \frac{2 \, v(t)}{m_L^2 - m_R^2}   ,      \qquad  \qquad  \sigma(t)  =  a(t)   \ . 
\eea
This diagonalization defines the mass basis fields $\phi \equiv (\phi_1, \, \phi_2) \equiv U^\dagger \Phi$.
The Lagrangian, in the mass basis, is
\be
\mathcal{L}(x) = \partial_\mu \phi^{\dagger} \, \partial^\mu \phi - \phi^{\dagger} \, m^2  \, \phi - \phi^\dagger \, \Sigma \, \dot\phi + \dot\phi^\dagger \, \Sigma \, \phi - \phi^\dagger \, \Sigma^2 \, \phi + \Lint \;, \label{eq:massbasis}
\ee
where $\Sigma(t) \equiv U^\dagger \dot U$ and the dot denotes $\partial_t$.  
   
One can study the condition for $CP$ invariance of this model.  The action is invariant under the transformation $\phi_i \xrightarrow{CP} \eta_i \, \phi_i^\dagger$, where $|\eta_i|^2 = 1$, if
\begin{equation}
\eta_i^*  \, \eta_j  \, \Sigma^{ij} \; = \; - \, \Sigma^{ji} \;  . 
\end{equation}
This condition implies that $\Sigma^{11} = \Sigma^{22} = 0$.  From the explicit expression  
\begin{equation}
\Sigma   = 
\begin{pmatrix}
0 & -e^{-i\sigma} \\
e^{i\sigma} & 0
\end{pmatrix}
\dot\theta 
+
\begin{pmatrix}
i\sin^2\theta & \frac{i}{2}\sin 2\theta e^{-i\sigma} \\
\frac{i}{2}\sin 2\theta e^{i\sigma} & -i\sin^2\theta
\end{pmatrix}
\dot  \sigma \ , 
\end{equation}
we see that $CP$ invariance requires $\dot \sigma = 0$, or equivalently, $\dot a = 0$. 
Once this condition is satisfied, the one for $i \neq j$ is easily satisfied by an appropriate 
choice of $\eta_{1,2} (\sigma)$  (now $\sigma$ is a constant phase). 

Neglecting the interactions, the free field equations are
\begin{subequations}
\label{eq:field}
\bea
\left[ \frac{}{} \partial_x^2 + m^2 + 2\, \Sigma \, \partial_{x^0} + \Sigma^2 + \dot\Sigma \, \right] \, \phi(x) &= 0 \; , \\
\phi^\dag(y)\left[\smash{\overset{\leftarrow}{\partial}}_y^2 + m^2 - 2\overset{\leftarrow}{\partial}_{y^0}\Sigma + \Sigma^2 - \dot\Sigma \right] &=0 \;.
\eea
\end{subequations}
The operators in square brackets are  Klein-Gordon operators, suitably modified due to the time-dependent mass matrix.  
In this basis, the interaction becomes
\be
\Lint = \, - \, \frac{1}{2} \: A^2 \, \phi^\dagger \, Y \, \phi \;,
\ee
with $Y(t) \equiv U^\dagger \, y \, U$.

\subsection{Time scales}

\label{sec:counting}

The quantum mechanical evolution of a system out of equilibrium  reduces to kinetic theory, 
described by Boltzmann-like  equations, in the limit that there exists a hierarchy between microscopic and macroscopic time scales.  
In our derivation of the Boltzmann equations below, 
we utilize  a perturbative expansion in the ratios of these time scales.

The microscopic scale is given by the ``intrinsic time'' corresponding to the inverse frequencies of $\phi_{1,2}$:
\be
\tau_{\textrm{int}} \sim \omega_{1,2}^{-1} = \left(k^2 + m_{1,2}^2\right)^{-1/2} \; .
\ee
In the thermal plasma of the early universe (with temperature $T$), we take $\tau_{\textrm{int}} \sim T^{-1}$.  (We assume
for   the zero temperature masses 
$m_{1,2} \sim T$, as typically holds for a subset of the superpartners in supersymmetric electroweak baryogenesis; 
for $m_{1,2} \gg T$, these particles are not active in the plasma.)  

Typical macroscopic scales in the problem are   
associated with (i) the time-varying mass (with time scale $\tw$),  
and (ii) collisional processes that lead to equilibration (with time scale $\tau_{\textrm{coll}}$).  
In the limit   $\tw$, $\tau_{\textrm{coll}} \gg \tau_{\textrm{int}}$, the usual gradient 
expansion is applicable.   In the two-flavor case, there arises an additional scale: the oscillation time scale $\tau_{\textrm{osc}} = 2 \pi  \Delta\omega^{-1}$, where $\Delta\omega = \abs{\omega_1 - \omega_2}$.
 
In our analysis, we work in the regime in which the following ratios are small parameters:
\be
\epsilon_{\textrm{wall}} \equiv \frac{\tau_{\textrm{int}}}{\tw} \, , \qquad \epsilon_{\textrm{coll}} \equiv \frac{\tau_{\textrm{int}}}{\tau_{\textrm{coll}}} \, , \qquad \epsilon_{\textrm{osc}} \equiv \frac{\tau_{\textrm{int}}}{\tau_{\textrm{osc}}} \; .
\label{eq:epsdef} 
\ee
We derive the quantum Boltzmann equations to first order in these parameters, generically denoted $\mathcal{O}(\epsilon)$.
This regime corresponds the following physical picture:
\begin{itemize}
\item The time-dependent background field is slowly varying, such that $\epsilon_{\textrm{wall}} \ll 1$.
\item The two scalars are nearly degenerate, such that $\epsilon_{\textrm{osc}} \ll 1$.  
\item The mean free time is sufficiently long, such that $\epsilon_{\textrm{coll}} \ll 1$.  This assumption is realized if the coupling constant $y$ is perturbative, as shown in Sec.~\ref{sec:analytic}.
\end{itemize}
Within a more realistic model, our assumption $\epsilon_{\textrm{osc}} \ll 1$ is not necessarily satisfied, depending on the spectrum.  However, for a slowly varying wall, we find that the $CP$-asymmetry is maximized for $\tau_{\textrm{osc}} \sim \tw$ (shown in Sec.~\ref{sec:numerical}); therefore, the regime $\epsilon_{\textrm{osc}} \ll 1$ is the physically interesting case in which to study these $CP$-violating flavor oscillation effects.  We speculate that when $\epsilon_{\textrm{osc}} \sim 1$, flavor decoherence occurs rapidly compared to the longer time scales $\tw, \, \tau_{\textrm{coll}}$; therefore, one can utilize the 
one-flavor Boltzmann equation for each of the mass eigenstates, thereby neglecting the off-diagonal elements in the mass-basis density matrix.  However, a complete treatment that is valid in both the small and large  $\epsilon_{\textrm{osc}}$
regimes and that would test this expectation, remains an open problem. 
 
In principle, in the multi-flavor case, one has to worry about an additional time scale beyond 
those discussed above: $\tau_{\rm coh}$,  the 
time  up to which  the quantum mechanical coherence among different mass eigenstates persists (and therefore 
up to which flavor oscillations can play a role). One can estimate 
$\tau_{\rm coh}$ by considering the time up to which 
wavepackets  of different mass eigenstates produced in the same process  
are still overlapping. 
The key physical input here is an estimate of the wavepacket 
length~\cite{Kayser:1981ye}, which in a thermal bath 
can be taken as the mean free path of the relevant particles~\footnote{We thank Boris Kayser and Petr Vogel for 
discussions on this point.}. Using this input we find that 
$\tau_{\rm coh}  \sim  \tau_{\rm osc}/\epsilon_{\rm coll} \sim \tau_{\rm coll}/\epsilon_{\rm osc}$, 
which is  very long compared to all other time-scales in the problem. 
So in the regime under study,   kinematical decoherence effects are negligible.

\section{Quantum Boltzmann Equations from First Principles}
\label{sec:analytic}

We derive the quantum Boltzmann equations using finite-temperature, non-equilibrium quantum field theory, provided by the Closed Time Path (CTP) formalism.  In Ref.~\cite{Calzetta:1986cq}, Calzetta and Hu derived in this way the Boltzmann equation for a single real scalar field with collisions~\footnote{The authors of Ref.~\cite{Calzetta:1986cq} used a derivation based on the 2 Particle Irreducible effective action, equivalent
to the approach described below.}.  In this section, we generalize their results in two ways.
First, we consider flavor; i.e., we study the case of two complex scalar fields, including the effects of mixing between flavor and mass eigenstates, quantum mechanical flavor oscillations, and flavor-dependent interactions with the background plasma.  The physics is similar to that  of neutrino oscillations in a dense medium~\cite{Sigl:1992fn,Stodolsky:1986dx}.  Second, we consider a time-dependent $CP$ violating mass matrix for the two-flavor scalar system.  As discussed in Sec.~\ref{sec:intro}, this is a toy model for the dynamics of baryogenesis during the electroweak phase transition, wherein one can observe some of the essential physics without making an unproven ansatz about the form of the distribution functions.

In this section, we derive the Boltzmann equations pedagogically.  A summary of our logic is as follows:
\begin{itemize}

\item Derive equations of motion for the CTP Green's functions for the scalar field $\phi$ accurate to all orders in $\epsilon$.  These equations are (i) the constraint equation, which gives the quantum spectrum of microscopic excitations, and (ii) the kinetic equation, which gives the macroscopic evolution of the distribution of states in the thermal bath. 

\item In order to derive the quantum Boltzmann equations at (leading) $\mathcal{O}(\epsilon)$, we truncate the kinetic equation at $\mathcal{O}(\epsilon)$ and the constraint equation at $\mathcal{O}(\epsilon^0)$.  At these orders, the constraint equation identifies excitations in the thermal bath by their tree-level dispersion relations, and the kinetic equation describes how their distribution functions evolve  due to the leading nontrivial effects of collisions, the variation of the ``wall'', and flavor oscillations. The effects of $\mathcal{O}(\epsilon)$ shifts in the dispersion relations  add subleading corrections to the evolution of distribution functions on top of the most important effects we include here.
 (We deal with the $\mathcal{O}(\epsilon)$ constraint equation in Appendix~\ref{appx:constraint}.)

\item By integrating the kinetic equation over energy, we obtain the quantum Boltzmann equations for the quasi-particle and antiparticle two-flavor density matrices.  These Boltzmann equations, given at $\mathcal{O}(\epsilon)$, describe the evolution of this two-flavor system in the presence of a time-dependent and complex mass matrix, accounting for flavor mixing, oscillations, and collisions with other particles in the thermal bath.

\end{itemize}

For the reader unfamiliar with the CTP formalism, we provide a brief review in Appendix~\ref{appx:CTP}.  In addition, in Appendix~\ref{appx:1flavor} we review the derivation of the one-flavor Boltzmann equations, following Ref.~\cite{Calzetta:1986cq}, to illustrate the implementation of the above logic in a simpler case.  In this section, we begin immediately with the Green's functions and equations of motion in this formalism.

\subsection{Closed Time Path Formalism}

In zero temperature equilibrium quantum field theory, only the time-ordered propagator is relevant for perturbation theory.  At finite temperature and away from equilibrium, in the CTP formalism, one requires four Green's functions, corresponding to all possible orderings of fields along the closed time path 
$(-\infty,+\infty) \cup (+\infty,-\infty)$~\cite{Schwinger:1960qe,Keldysh:1964ud,
Mahanthappa:1962ex,Bakshi:1962dv}.  
In addition, to study the two-flavor system introduced above, we define a matrix of CTP Green's functions in flavor- or mass-basis field space.
In the mass basis, we will use indices $i,j$ to label the components of the Green's functions, which we define by
\be
\widetilde{G}_{ij}(x,y) = \left( \ba{cc} G^t_{ij}(x,y) & -\, G^<_{ij}(x,y) \\[4pt] G^>_{ij}(x,y) & - \, G^{\bar{t}}_{ij}(x,y) \ea \right) \; ,
\ee
where
\begin{subequations}
\label{eq:props}
\bea
G^>_{ij}(x,y) &\equiv& \left\langle \phi_i(x) \, \phi_j^\dagger(y) \right\rangle  \\
G^<_{ij}(x,y) &\equiv& \left\langle \phi_j^\dagger(y) \, \phi_i(x) \right\rangle  \\
G^t_{ij}(x,y) &\equiv& \theta(x^0-y^0) \, G^>_{ij}(x,y) +  \theta(y^0-x^0) \, G^<_{ij}(x,y) \\
G^{\bar t}_{ij}(x,y) &\equiv& \theta(x^0-y^0) \, G^<_{ij}(x,y) +  \theta(y^0-x^0) \, G^>_{ij}(x,y) \; .
\eea
\end{subequations}
These Green's functions satisfy the Schwinger-Dyson equations
\begin{subequations}
\label{eq:schwing2}
\bea
\widetilde{G}(x,y) &=& \widetilde{G}^{(0)}(x,y) + \int d^4 z \; \int d^4 w \; \widetilde{G}^{(0)}(x,z) \, \widetilde{\Pi}(z,w) \, \widetilde{G}(w,y) \\
 &=& \widetilde{G}^{(0)}(x,y) + \int d^4 z \; \int d^4 w \; \widetilde{G}(x,z) \, \widetilde{\Pi}(z,w) \, \widetilde{G}^{(0)}(w,y)  \; ,
\eea
\end{subequations}
where $\widetilde{G}^{(0)}$ is the free Green's function and $\widetilde{\Pi}$ is the self-energy.   Furthermore, Eq.~\eqref{eq:schwing2} is a matrix equation in field space; we have suppressed the field indices.

Our task now is to recast the Schwinger-Dyson equations into the language of kinetic theory.  (Here, for completeness, we work to all orders in $\epsilon$; in the next sections, when we truncate at $\mathcal{O}(\epsilon)$, our results simplify considerably.)  First, by virtue of the free field equations \eqref{eq:field}, the free Green's functions satisfy 
\begin{subequations}
\label{eq:freegreens} 
\bea
\left( \frac{}{} \partial_x^2 + m^2(x^0) + 2 \, \Sigma(x^0) \, \partial_{x^0}  + \Sigma(x)^2 + \partial_{x^0} \Sigma(x) \, \right) \, \widetilde{G}^{(0)}(x,y) &=& - \, i \, \delta^4(x-y) \, \widetilde{I} \\
\widetilde{G}^{(0)}(x,y) \, \left( \frac{}{} \smash{\overset{\leftarrow}{\partial}}_y^2 + m^2(y^0)  -  2 \, \smash{\overset{\leftarrow}{\partial}}_{y^0} \, \Sigma(y^0) + \Sigma(y)^2 - \partial_{y^0} \Sigma(y) \,  \right) &=& - \, i \, \delta^4(x-y) \, \widetilde{I} \; ,
\eea 
\end{subequations}
where $\widetilde{I}$ is the identity matrix in both field and CTP component space.  Using Eq.~\eqref{eq:freegreens}, we act on the Schwinger-Dyson equations \eq{eq:schwing2} with the Klein-Gordon operator in \eq{eq:field}:
\begin{subequations}
\label{eq:fullgreens}
\begin{align}
\left( \frac{}{} \partial_x^2 + m^2(x^0) + 2 \, \Sigma(x^0) \, \partial_{x^0}  + \Sigma(x)^2 + \partial_{x^0} \Sigma(x) \, \right)\,  \widetilde G(x,y) & \notag \\
= -\; i \, \delta^4(x-y) \, \widetilde I -  i &\int\! d^4 z\, \widetilde\Pi(x,z) \, \widetilde G(z,y) \\
\widetilde G(x,y) \, \left( \frac{}{} \smash{\overset{\leftarrow}{\partial}}_y^2 + m^2(y^0)  -  2 \, \smash{\overset{\leftarrow}{\partial}}_{y^0} \, \Sigma(y^0) + \Sigma(y)^2 - \partial_{y^0} \Sigma(y) \,  \right) & \notag \\
=  - \; i\, \delta^4(x-y) \, \widetilde I - i &\int\! d^4 z\, \widetilde G(x,z) \, \widetilde \Pi(z,y)  \; .
\end{align}
\end{subequations}

The Green's functions \eq{eq:props} contain information both about the spectrum of excitations of the fields $\phi$ at the microscopic level, and the distribution of states in the thermal bath at the macroscopic level. We can decouple the micro- and macroscopic dynamics by, first, defining the average and relative coordinates
\be
X  =  (X^0 \equiv t, \, \mathbf X) \equiv \frac{x+y}{2}\; , \qquad r \equiv x - y \; , 
\ee
and, then,  Fourier transforming with respect to the relative coordinate $r$. This  procedure gives  the Wigner transform: e.g.,
\be
\widetilde{G}(k;X) = \int\! d^4 r \, e^{i \, k \cdot r} \, \widetilde{G}(x,y) \; . \label{eq:wignermania}
\ee
At zero temperature, the position-space Green's functions only depend on $r$.  At finite temperature and away from equilibrium, the average coordinate $X$, describing the macroscopic evolution of the system, also plays a role.  
In this study, we assume isotropy and homogeneity, so that quantities depend on $t$, 
but are independent of $\mathbf X$.

By taking the Wigner transform of the sum and difference, respectively, of Eqs.~\eqref{eq:fullgreens}, we obtain the constraint equation for $G^\gtrless(k;t)$:
\begin{align}
\label{constraint}
\left( \, 2 k^2 - \frac{\partial_{t}^2}{2} \, \right) \,  G^\gtrless(k;t) \; = \; e^{-i\Diamond} \, \biggl( \, \bigl\{ \, m^2(t)- 2i \, k^0 \, \Sigma(t) +\Sigma(t)^2, \: G^\gtrless \, \bigr\} +  \bigl[ \, \dot\Sigma(t), \, G^\gtrless \, \big] \; \biggr. \; \;  & \\
\qquad  + \, i \, \bigl\{\Pi^h,G^\gtrless\bigr\} +  i \, \bigl\{\Pi^\gtrless,G^h\bigr\} +  \frac{i}{2} \,\bigl[\Pi^>,G^<\bigr] +  \frac{i}{2} \, \bigl[G^>,\Pi^<\bigr] \, \biggr) &\; , \notag 
\end{align} 
and the kinetic equation:
\begin{align}
\label{kinetic}
2 k_0 \, \partial_t \, G^\gtrless(k;t) \; = \; e^{-i\Diamond} \biggl( \, - i \, \bigl[ \, m^2(t) - 2i \, k^0 \, \Sigma(t) + \Sigma(t)^2, \: G^\gtrless \, \bigr] - i \, \bigl\{ \, \dot\Sigma(t), \, G^\gtrless \, \bigr\} \; \biggr. \qquad & \\
+ \biggl. \, \bigl[\Pi^h,G^\gtrless\bigr] + \bigl[\Pi^\gtrless,G^h\bigr] + \frac{1}{2}\, \bigl\{\Pi^>,G^<\bigr\} - \frac{1}{2}\, \bigl\{\Pi^<,G^>\}\biggr) & \;, \notag
\end{align}
where  $[\cdot,\cdot]$ and $\{\cdot,\cdot\}$ denote the commutator and anticommutator. The diamond operator $\Diamond$ acts on pairs of Wigner transforms according to the definition
\begin{equation}
\label{diamond}
\Diamond\Bigl(A(k;X)B(k;X) \Bigr) \; = \; \frac{1}{2}\, \left(\pd{A}{X^\mu}\pd{B}{k_\mu} - \pd{A}{k_\mu}\pd{B}{X^\mu}\right)\,.
\end{equation}
This differential operator arises due to taking the Wigner transform of the spacetime convolutions appearing in \eq{eq:fullgreens}. In \eqs{constraint}{kinetic} we have introduced the combinations of CTP Green's functions and self-energies
\begin{equation}
G^h = \frac{1}{2}\,(G^t - G^{\bar t})\,,\quad \Pi^h = \frac{1}{2}\, (\Pi^t - \Pi^{\bar t})\,,
\end{equation} 
where $G^h$ contains information only about the microscopic spectrum, and $\Pi^h$ gives the shift in the dispersion relation for excitations in this spectrum due to interactions. See Appendix~\ref{appx:constraint} for further details. 
So far, the equations of motion \eqs{constraint}{kinetic} are valid to all orders in $\epsilon_{\textrm{wall}}$, 
$\epsilon_{\textrm{coll}}$, 
$\epsilon_{\textrm{osc}}$.  

\subsection{Lowest-order solutions of constraint and kinetic equations}

In solving for $G^{\gtrless}$, we utilize a  power counting in $\epsilon$.  In this section, we work to zeroth order in $\epsilon$.  First, we summarize the rules for how to determine the order of any term in \eqs{constraint}{kinetic}. 
Recalling  that  $\tau_{\textrm{int}} \sim 1/k_0 \sim 1/\omega_{1,2} (k)$ and the definitions of $\epsilon$'s 
in Eq.~(\ref{eq:epsdef}), one has that: 
\begin{itemize}
\item Each derivative $\partial_t$ acting on $U(t)$ or $m_{1,2}^2(t)$ carries one power of 
$1/\tau_{\rm  w}$.  As a consequence,  
$k_0 \Sigma$ scales as  $k_0^2  \times \mathcal{O}(\ew)$, $\dot\Sigma$ and $\Sigma^2$  as 
$k_0^2 \times \mathcal{O}(\ew^2)$, etc.
\item Each factor of the self-energy $\Pi$ scales as  $k_0^2 \times \mathcal{O}(\eint)$ (recall that 
$1/\tau_{\rm coll} \sim \Pi/k_0$).  When we evaluate the collision term explicitly below, it will be clear that this expansion is equivalent to an expansion in the coupling constant $y$.
\item Each factor of $\Delta m^2 \equiv (m^2_1 \! - \! m^2_2)$ scales as   
$(\omega_1 + \omega_2)^2 \times   \mathcal{O}(\eosc)$.
\item Each derivative acting on $G^\gtrless(k;t)$ carries a power of $\epsilon$, up to factors of $k_0$. 
\end{itemize}
The last rule follows from the kinetic equation \eqref{kinetic}, truncated at $\mathcal{O}(\epsilon^0)$.  In particular, note that the commutator term
\be
- i \left[ \, m^2(t) , \, G^\gtrless(k,t) \, \right] = - i \, \Delta m^2 \begin{pmatrix} 0 & G^\gtrless_{12}(k,t) \\ -  G^\gtrless_{21}(k,t) & 0 \end{pmatrix} \; , \label{eq:commterm}
\ee
is $\mathcal{O}(\eosc)$.  Using the first three rules, the $\mathcal{O}(\epsilon^0)$ kinetic equation is
\be
\label{kinetic0}
2k_0\, \partial_t \, G^{\gtrless}(k;t) \;  = \;  0 \; ;
\ee
therefore, we see that $\partial_t G^\gtrless$ is $\mathcal{O}(\epsilon)$.  Similar arguments at higher order imply that each derivative acting on $G^\gtrless$ gets an additional power of $\epsilon$.

We now turn our attention to the constraint equation.  At $\mathcal{O}(\epsilon^0)$, Eq.~\eqref{constraint} becomes
\be
\label{constraint0}
\left( \, k^2 - \bar{m}^2 \, \right) \, G_{ij}^{\gtrless}(k;t) \; = \;  0 \; .
\ee
At this order we can replace $m_{1,2}$ by the average mass $\bar{m} \equiv (m_1 + m_2)/2$, since deviations from this replacement are $\mathcal{O}(\eosc)$.  In other words, the Green's function $G^{\gtrless}(k;t)$ vanishes unless the appropriate dispersion relation
\be
k^0 \, = \, \pm \, \bar\omega_k \, \equiv \, \pm \, \sqrt{ |\mathbf k|^2 + \bar{m}^2(t) } \; 
\ee
is satisfied.\footnote{There is a further subtlety if one works in the limit $\ew, \,  \eint \ll 1$ limit, but for finite $\eosc$. One finds {\it four} different modes for the off-diagonal Green's functions: $k^0 = \pm (\omega_1+\omega_2)/2$ and $k^0 = \pm \Delta\omega/2$.  The latter modes become tachyonic with large $|\mathbf k|$.  However, we find that they decouple from the $\mathcal{O}(\epsilon)$ Boltzmann equations in the limit $\eosc \ll 1$.  A general analysis for $\eosc \sim \mathcal{O}(1)$ will have to account for these modes.}

A general form of the solution to the constraint equation is then
\begin{equation}
\label{treesolution}
\begin{split}
G_{ij}^{>(0)}(k;t)  &= 2\pi\delta(k^2 - \bar{m}^2) \, \left[\, \theta(k^0)(\delta_{ij}+ f_{ij}(\mathbf k, t)) + \theta(-k^0) \bar f_{ij}(-\mathbf k,t) \, \right] \\
G_{ij}^{<(0)}(k;t)  &= 2\pi\delta(k^2 - \bar{m}^2) \, \left[\,\theta(k^0) f_{ij}(\mathbf k,t) + \theta(-k^0)(\delta_{ij} + \bar f_{ij}(-\mathbf k,t))\, \right] \,,
\end{split}
\end{equation}
The constraint equation has determined the spectrum of excitations present in the thermal bath.  We identity $f_{ij}, \, \bar f_{ij}$ as the particle and antiparticle density matrices; here, in the free field case, they are given by the expectation values $\langle a^\dag_j \, a_i\rangle$ and $\langle b^\dag_i \, b_j\rangle$ of free particle and antiparticle mode operators.  The evolution of the density matrices is determined by the kinetic equation; at $\mathcal{O}(\epsilon^0)$, \eq{kinetic0} tells us that they remain static.

\subsection{Kinetic equation at $\mathcal{O}(\epsilon)$}

We now return to the equations of motion \eqs{constraint}{kinetic} for $G^\gtrless$
and solve them at $\mathcal{O}(\epsilon)$. We present the solution of the constraint equation at $\mathcal{O}(\epsilon)$ in Appendix~\ref{appx:constraint}. Here, we are interested in finding the leading nontrivial evolution of distribution functions $f_{ij},\bar f_{ij}$, which occurs at $\mathcal{O}(\epsilon)$. For this purpose we will need to know the spectrum, or solution of the constraint equation, only at $\mathcal{O}(\epsilon^0)$. The solutions to the equations of motion \eqs{constraint}{kinetic} then take the same general form as in \eq{treesolution}. However, the solution of the kinetic equation \eq{kinetic} at $\mathcal{O}(\epsilon)$ will now give the nontrivial evolution in time of the distributions $f_{ij},\bar f_{ij}$ due to  interactions, to the time-varying ``wall'', and  to flavor oscillations.

Let us now truncate the kinetic equation \eq{kinetic} to $\mathcal{O}(\epsilon)$. We note immediately that we can drop the terms containing $\Sigma(t)^2$ and $\dot\Sigma(t)$, which are $\mathcal{O}(\ew^2)$. Then, we may use the information from the previous order that $\partial_t$ acting on Green's functions introduces a suppression by at least one power of $\epsilon$. Therefore we may drop all $\Diamond$ operators in those terms already containing an explicit factor of $\mathcal{O}(\epsilon)$, namely,  $\Sigma$ or $\Pi$.  After these simplifications, the terms surviving in the kinetic equation \eq{kinetic} are
\begin{equation}
\begin{split}
2k_0 \, \partial_t \, G^\gtrless(k;t) + i\bigl[m^2(t)&,G^\gtrless(k;t)\bigr] + \frac{1}{2}\,\bigl\{\dot m^2(t),\partial_{k^0}G^\gtrless\bigr\} + 2\, k^0\bigl[\Sigma(t),G^{\gtrless}(k;t)\bigr] \\
&= \; \bigl[\Pi^h,G^\gtrless\bigr] + \bigl[\Pi^\gtrless,G^h\bigr] + \frac{1}{2}\Bigl(\bigl\{\Pi^>,G^<\bigr\} - \bigl\{\Pi^<,G^>\bigr\}\Bigr)\,.
\end{split}
\end{equation}
There is, however, one remaining term of $\mathcal{O}(\epsilon^2)$ that can be dropped. In the second term on the right-hand side, $G^h$ acts as a source for $G^\gtrless$. Since it multiplies a factor of $\Pi$, which is $\mathcal{O}(\epsilon)$, we can evaluate $G^h$ at $\mathcal{O}(\epsilon^0)$. Since $G^h = G^t - G^{\bar t}$, at tree-level we have
\begin{equation}
G^h_{ij} = i\PV\left(\frac{1}{k^2 - m_i^2}\right)\delta_{ij}\,,
\end{equation}
which is diagonal. Thus $[\Pi^\gtrless,G^h]$ is proportional to $\Delta m^2$ and so is suppressed additionally by $\eosc$. Therefore the term is overall $\mathcal{O}(\eint\eosc)$ and can be dropped from the $\mathcal{O}(\epsilon)$ kinetic equation.

After suitably rearranging the terms in the kinetic equation, we obtain the master equation 
on which we rely in the remainder of the analysis:
\begin{equation}
\label{kineticfinal}
\begin{split}
2k_0\partial_t G^\gtrless(k;t) + i\left[m^2(t) + i\Pi_h(k;t) - 2ik^0\Sigma(t) \; , \, G^\gtrless(k;t) \right] + \frac{1}{2}\bigl\{\dot m^2(t),\partial_{k^0}G^\gtrless(k;t)\bigr\} = \mathcal{C} \,,
\end{split}
\end{equation}
where the collision term $\mathcal{C}$ is defined
\begin{equation}
\label{collision}
\mathcal{C}(k;t) =  \frac{1}{2}\Bigl(\bigl\{\Pi^>,G^<\bigr\} - \bigl\{\Pi^<,G^>\bigr\}\Bigr)(k;t)\,.
\end{equation}
On the left-hand side of \eq{kineticfinal}, the term $i\Pi_h$ shifts the tree-level mass to the interaction-corrected value 
(this is true also in the spectral function, cf. \eq{GRdiag1}). As we will see, the $\Sigma$ term induces flavor mixing during the time the mass is actively varying, described by Eqs.~(\ref{eq:wow}-\ref{eq:tauw}). The last term on the left-hand side will disappear in the analysis below since it is a total derivative in $k_0$, over which we will integrate. The collision term $\mathcal{C}$ on the right-hand side causes equilibration of the distributions $f,\bar f$ through interactions amongst particles, antiparticles and $A$ bosons in the thermal bath.

The kinetic equation \eq{kineticfinal} is the master equation governing the quantum field theoretic evolution of the distribution of states in the thermal bath, expanded consistently to linear order in the perturbations of order $\ew,\eint,\eosc$. 
In the next section, we derive the quantum Boltzmann equations for the evolution of the density matrices $f,\bar f$ themselves directly from the quantum field theoretic kinetic equation \eq{kineticfinal}.

\subsection{From Kinetic Equation to Kinetic Theory}

In this section, we show how the kinetic equation \eqref{kineticfinal} has the structure of a quantum Boltzmann equation. Using  \eq{treesolution}, we identify the following positive and negative frequency integrals (which project out the particle and antiparticle modes) as the particle and antiparticle density matrices:
\begin{subequations}
\label{eq:densitymat}
\bea
f_{ij}(\mathbf k, t) &\equiv& \int_0^\infty \frac{dk^0}{2\pi} \: 2 \, k^0 \: G_{ij}^<(k,t)  \\
\bar{f}_{ij}(-\mathbf k, t) &\equiv& \int^0_{-\infty} \frac{dk^0}{2\pi} \: (- \, 2 \, k^0) \: G^>_{ij}(k,t) \; \; .
\eea 
\end{subequations}
We obtain the Boltzmann equations by taking the positive and negative frequency integrals of \eq{kineticfinal}.
Taking the positive frequency integral of the $ <$ component of the kinetic equation, we have
\be
\label{oscterm}
\int^\infty_0 \frac{dk^0}{2\pi} \:  2 k^0 \, \partial_t \, G^<(k,t) = \int^\infty_0 \frac{dk^0}{2\pi} \, \left( \, - \, i \left[ \, m^2(t) + i \Pi^h, \, G^< \, \right] - 2  k^0 \, \left[ \, \Sigma, \, G^< \, \right] \, + \, \mathcal{C}\frac{}{} \right) \; .
\ee
Since $G^<(k,t)$ vanishes at the boundaries, the derivative term $\partial G^</\partial k^0$  in \eq{kineticfinal} integrates to zero.  Using Eqs.~\eqref{eq:densitymat}, we can express this equation in terms of $f$ and $\bar{f}$.  In particular, there is a trick for evaluating the commutator term:
\be
\int^\infty_0 \frac{dk^0}{2\pi} \;  \left[ \, m^2(t) , \, G^< \, \right] = \int^\infty_0 \frac{dk^0}{2\pi} \;  \left( \frac{2k^0}{2\bar\omega_k} \right) \,  \left[ \, m^2(t) , \, G^< \, \right] = \left[ \, \omega_{k}(t) , \, f(\mathbf k, t) \, \right] \; . \label{realoscterm}
\ee
In the first step, we have inserted a factor of unity (since the spectral function, restricted to positive frequencies, implies $k^0 = \bar\omega$); in the second step, we have used Eqs.~(\ref{eq:commterm}, \ref{eq:densitymat}) and defined
\be
\omega_k(t) \equiv \left( \ba{cc} \omega_{1k}(t) & 0 \\ 0 & \omega_{2k}(t) \ea \right) \; .
\ee
Since the term \eq{realoscterm} is already $\mathcal{O}(\epsilon_{\textrm{osc}})$, it suffices to use the zeroth-order constraint equation.
All in all, the Boltzmann equation for the particle density matrix is, to $\mathcal{O}(\epsilon)$,
\be
\frac{\partial f(\mathbf k,t) }{\partial t} = \; - \, i \, \left[ \, \omega_k(t) - i \, \Sigma(t), \, f(\mathbf k, t) \, \right] + \int^\infty_0 \frac{dk^0}{2\pi} \, \left( \, \mathcal{C} + \left[\Pi^h, G^< \right] \, \right) \; . \label{eq:boltzpart}
\ee
Similarly, by taking the negative frequency integral of the $>$ component of the kinetic equation, we obtain the antiparticle Boltzmann equation
\be
\frac{\partial \bar f(\mathbf k,t) }{\partial t} = \;  i \, \left[ \, \omega_k(t)  
+ i \, \Sigma(t), \, \bar f(\mathbf k, t) \, \right] - \int_{-\infty}^0 \frac{dk^0}{2\pi} \, \left( \, \mathcal{C} + \left[\Pi^h, G^> \right] \, \right)  \; .
\ee
In the next section, we evaluate in detail the remaining interaction-induced terms under the integrals.

The simplest physical application of these Boltzmann equations is the case of vacuum flavor oscillations.  If we neglect both interactions and the $\Sigma$ term (induced by the time-dependent mixing matrix), the particle Boltzmann equation simplifies to
\be
\frac{\partial f}{\partial t} = - \, i \, \left[ \, \omega_k, \, f \right] \; ,  \label{eq:freeboltz}
\ee
which is the familiar density matrix formulation of the Schrodinger equation.  Given an initial condition $f(\mathbf k, 0)$ at $t=0$, the solution to Eq.~\eqref{eq:freeboltz} is
\be
f(\mathbf k,t) = \left( \ba{cc} f_{11}(\mathbf k,0) & f_{12}(\mathbf k,0) \, e^{- i \Delta\omega t} \\
f_{21}(\mathbf k,0) \, e^{i \Delta\omega t} & f_{22}(\mathbf k,0) \ea \right) \; .
\label{eq:vacosc1}
\ee
To be concrete, let us compute the oscillation probability for $\Phi_L \to \Phi_R$, given a pure $\Phi_L$ initial state.  
We define the flavor projection operators $P_L = \textrm{diag}(1,0)$ and $P_R = \textrm{diag}(0,1)$.  
In terms of the mixing matrix $U$, defined in Eq.~\eqref{eq:diagonalization}, 
the pure $L$ initial state corresponds to the mass-basis density matrix $f(\mathbf k, 0) = U^\dagger \, P_L \, U$.  The oscillation probability is
\be
\mathcal{P}_{L\to R}(t) = \textrm{Tr}[ P_R \, U \, f(t) \, U^\dagger] = \sin^2 2\theta \, \sin^2 \left(\frac{\Delta\omega \, t}{2} \right) \;,
\ee
as expected.  

\subsection{Interactions and Collisions}

\label{sec:collisions}

\begin{figure}[!t]
\begin{center}
\mbox{\hspace*{-1cm}\includegraphics{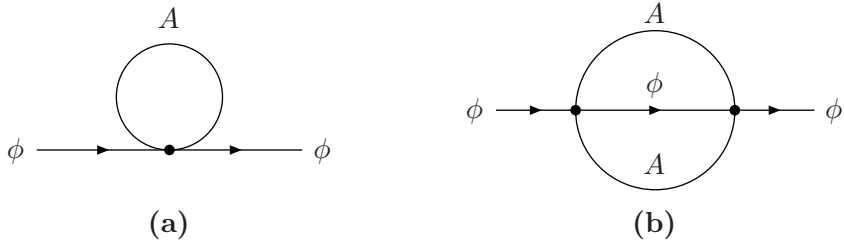}}
\end{center}
\caption{\it\small 
Leading-order self-energy graphs that induce the collision terms in the Boltzmann equations, corresponding to (a) coherent forward scattering, and (b) non-forward scattering ($\phi A \leftrightarrow \phi A$) and annihilation ($\phi \phi^\dagger \leftrightarrow A A$).
}
\label{fig:feyn}
\end{figure}

In this section, 
we evaluate the effects of the interaction
\be
\Lint =  \, - \, \frac{1}{2} \: A^2 \, \phi^\dagger \, Y \, \phi \; ,
\ee
in the kinetic equation \eq{kineticfinal}. 
This simple interaction, described in Sec.~\ref{sec:prelim}, is a toy model for flavor-dependent gauge 
and Yukawa  interactions that are active in the plasma during the early universe.  

\subsubsection{Plasma-corrected mass}

The $\Pi_h$ term in the kinetic equation \eqref{kineticfinal} describes
coherent forward scattering, analogous to the MSW effect for neutrino flavor mixing \cite{MSW}.  The leading contribution to $\Pi_h$ arises from the one-loop diagram shown in Fig.~\ref{fig:feyn}(a).  The self-energy from this diagram is
\be
\widetilde{\Pi}(x,y) = - \frac{i}{2} \, Y(t) \, \delta^4(x-y) \, \left( \ba{cc} G^t_A(x,y) & 0 \\ 0 & G^{\bar t}_A(x,y) \ea \right) \; ,
\ee
where ${G}_A$ is the $A$ boson Green's function.  Assuming that the $A$ bosons are in thermal equilibrium, their Wigner transformed Green's functions are 
\begin{subequations}
\bea
G^<_A(p) &=& (2\pi) \, \delta(p^2 - m_A^2) \, \textrm{sign}(p^0) \, n_B(p^0) \\ 
G^>_A(p) &=& (2\pi) \, \delta(p^2 - m_A^2) \, \textrm{sign}(p^0) \, (1+n_B(p^0)) \; .
\eea
\end{subequations}
Therefore, we have
\bea
\Pi^h(k,t) &=& \, - \, \frac{i}{4} \, Y(t) \, \int \!\! \frac{d^4 p}{(2\pi)^4} \, \left( \, G_A^>(p) + G_A^<(p) \, \right) \notag \\
&=& - \frac{i}{2} \, Y(t) \, \int\!\! \frac{d^3 p}{(2\pi)^3} \, \frac{1}{2 \varepsilon_p} \, \left( 1 + 2 \, n_B(\varepsilon_p) \right)  \; ,
\eea
with energy $\varepsilon_p = \sqrt{p^2 + m_A^2}$.  
The $\Pi_h$ term
effectively shifts the mass in the oscillation term
\be
\left[ \frac{}{} m^2(t) , \, G^\gtrless(k,t) \, \right] \, \longrightarrow \, \left[ \frac{}{} m^2(t) + i \, \Pi^h(k,t), \, G^\gtrless(k,t) \, \right] \;.
\ee
In the $m_A \ll T$ limit, 
this shift is given by
\be
i \, \Pi^h(k,t) = Y(t) \, \left( \, \left[ \int\!\! \frac{d^3 p}{(2\pi)^3} \, \frac{1}{4 \varepsilon_p} \right] + \frac{T^2}{24} \, \right) \; .
\label{eq:thmass} 
\ee
The term in square brackets is the usual zero temperature divergence that must be removed through the renormalization of $m^2$.  The remaining temperature-dependent term is the one-loop thermal mass.  This term is flavor diagonal and induces the shift:
\be
m_{L,R}^2 \  \longrightarrow \ m_{L,R}^2  +  \frac{y_{L,R} T^2}{24}~.  \label{eq:shifty}
\ee
In the limit that $i \Pi^h \gg m^2$  flavor oscillations do not occur since the mass basis coincides with the flavor basis.

\subsubsection{Collision term}

Now we evaluate the collision term \eq{collision}:
\be
\mathcal{C} \equiv \frac{1}{2} \, \left( \, \left\{ \Pi^>(k,t) , \, G^<(k,t) \right\} - \left\{ \Pi^<(k,t) , \, G^>(k,t) \right\} \frac{}{} \right) \; . \label{eq:collterm3}
\ee
This term encodes the effects of the processes of annihilation ($\phi \phi^\dagger \leftrightarrow A A)$, 
(non-forward) scattering ($\phi A \leftrightarrow\phi A$), and emission and absorption ($\phi_i \leftrightarrow \phi_j A A$) on the Green's functions $G^\gtrless$. 
As we observe below, the latter processes are kinematically forbidden 
in the small $\epsilon_\mathrm{osc}$ regime in which we work, 
so we will henceforth retain only the annihilation and scattering terms.  
The leading contribution to Eq.~\eqref{eq:collterm3} is the two-loop graph in Fig.~\ref{fig:feyn}(b).
The resulting self-energy is
\be
\Pi^\lambda(x,y) = - \frac{1}{2} \, Y(x^0) \, G^\lambda(x,y) \, Y(y^0) \, G^\lambda_A(x,y) \, G^\lambda_A(x,y) \; ,
\ee
with $\lambda = <,>$. At linear order in $\epsilon$, its Wigner transformation is
\begin{align}
\Pi^\lambda(k,t) = - \frac{1}{2} \, \int \!\! \frac{d^4 k^\prime}{(2\pi)^4}\int \!\! \frac{d^4 p}{(2\pi)^4} \int \!\! \frac{d^4 p^\prime}{(2\pi)^4} \, & (2\pi)^4 \,  \delta^4(k - k^\prime - p - p^\prime) \label{eq:Pik}\\
& \times \, Y(t) \, G^\lambda(k^\prime,t) \, Y(t) \, G^\lambda_A(p) \, G^\lambda_A(p^\prime) \; . \notag
\end{align}
Through this self-energy, the collision term's contribution to the particle Boltzmann equation \eqref{eq:boltzpart} is, using Eq.~\eqref{eq:Pik}, 
\bea
\int^\infty_0 \frac{dk^0}{2\pi} \, \mathcal{C}(k,t) &=&- \frac{1}{4} \, \int^\infty_0  \frac{dk^0}{2\pi} \int \!\!\! \frac{d^4k^\prime}{(2\pi)^4} \int \!\!\! \frac{d^4p}{(2\pi)^4} \int \!\!\! \frac{d^4 p^\prime}{(2\pi)^4} \; (2\pi)^4 \delta^4(k-k^\prime-p-p^\prime) \notag \\
&\;& \times \left( \frac{}{} \left\{ Y(t) \, G^>(k^\prime,t) \, Y(t) , \, G^<(k,t) \right\} \, G^>_A(p)\, G^>_A(p^\prime) \right. \notag  \\ &\;& \qquad \left. -  \left\{ Y(t) \, G^<(k^\prime,t) \, Y(t) , \, G^>(k,t) \right\} \, G^<_A(p)\, G^<_A(p^\prime) \frac{}{} \right) \; .
\eea
We express this in terms of $f$ and $\bar{f}$ by performing the frequency integrals.
Under our approximations, there are three integration regions receiving nonzero support from the energy-momentum conserving delta function $\delta^4(k-k^\prime-p-p^\prime)$ that correspond to following three physical processes:
\begin{subequations}
\label{processes}
\begin{align}
\text{annihilation:}& & \phi(k) \, \phi^\dagger(k^\prime) &\leftrightarrow A(p) \, A(p^\prime)  \,,& &(k^0,p^0,p'^0 > 0; \; k'^0\!<0) \label{eq:annihproc}\\
\text{scattering:}& &  \phi(k) \, A(p^\prime) &\leftrightarrow \phi(k^\prime)  \, A(p)  \,,& &(k^0,k'^0,p^0 > 0; \; p'^0\!<0)\label{eq:scattproc1}\\
\text{scattering:}& &  \phi(k) \, A(p) &\leftrightarrow \phi(k^\prime)  \, A(p') \,, & &(k^0,k'^0,p'^0 > 0; \;  p^0\!<0)\label{eq:scattproc2}
\end{align}
\end{subequations}
Other regions of integration correspond to processes that are kinematically forbidden, such as $\phi(k) \leftrightarrow \phi(k^\prime)  \, A(p) \, A(p^\prime)$\footnote{This process is forbidden because, at $\mathcal{O}(\epsilon)$, we treat the $\phi$ masses as degenerate within the collision term, consistent with our expansion.  At higher order in $\epsilon$, this process does contribute to the collision term.}.  Only annihilation and scattering are kinematically allowed at $\mathcal{O}(\epsilon)$.

Performing the frequency integrals over the region in \eq{eq:annihproc}, we obtain the contribution to the collision term from annihilation processes,
\begin{align}
 \int^\infty_0 \frac{dk^0}{2\pi} \,& \mathcal{C}_{{\textrm{ann}}}(k,t)   \label{eq:anncol} \\
& \quad = - \frac{1}{4} \, \frac{1}{2\bar\omega_{k}} \, \int \!\!\! \frac{d^3k^\prime}{(2\pi)^3} \, \frac{1}{2\bar\omega_{k^\prime}} \, \int \!\!\! \frac{d^3p}{(2\pi)^3} \, \frac{1}{2\varepsilon_{p}} \, \int \!\!\! \frac{d^3p^\prime}{(2\pi)^3} \, \, \frac{1}{2\varepsilon_{p^\prime}} \, (2\pi)^4 \delta^4(k+k^\prime-p-p^\prime)  \notag \\
&\; \qquad \qquad \times  \left( \frac{}{} \left\{ Y(t) \, \bar{f}(\mathbf k^\prime,t) \, Y(t) , \, f(\mathbf k,t) \right\} \, \left(1+n_B(\varepsilon_{p})\right) \, \left(1+n_B(\varepsilon_{p^\prime})\right) \right. \notag  \\ &\; \qquad \qquad \qquad \left. -  \left\{ Y(t) \, (1+\bar{f}(\mathbf k^\prime,t)) \, Y(t) , \, \left(1+f(\mathbf k,t)\right) \right\} \, n_B(\varepsilon_{p}) \, n_B(\varepsilon_{p^\prime}) \frac{}{} \right) \notag \; .
\end{align}
Here, the frequencies that appear in the delta function above are all positive; e.g., $k^{\prime 0} = \omega_{k^\prime}$, $p^0 = \varepsilon_{p}$, etc. We have assumed the $A$ bosons are in equilibrium.

Similarly, performing frequency integrals over the regions in \eqs{eq:scattproc1}{eq:scattproc2}, we obtain the contribution to the collision term from scattering,
\begin{align}
 \int^\infty_0 \frac{dk^0}{2\pi} &\, \mathcal{C}_{{\textrm{scat}}}(k,t)   \label{eq:scatcol} \\
&\quad = - \frac{1}{2} \, \frac{1}{2\bar\omega_{k}} \, \int \!\!\! \frac{d^3k^\prime}{(2\pi)^3} \, \frac{1}{2\bar\omega_{k^\prime}} \, \int \!\!\! \frac{d^3p}{(2\pi)^3} \, \frac{1}{2\varepsilon_{p}} \, \int \!\!\! \frac{d^3p^\prime}{(2\pi)^3} \, \, \frac{1}{2\varepsilon_{p^\prime}} \, (2\pi)^4 \delta^4(k-k^\prime+p-p^\prime)  \notag \\
&\; \qquad \qquad \times  \left( \frac{}{} \left\{ Y(t) \, \left(1+f(\mathbf k^\prime,t)\right) \, Y(t) , \, f(\mathbf k,t) \right\} \, n_B(\varepsilon_{p}) \, \left(1+n_B(\varepsilon_{p^\prime})\right) \right. \notag  \\ 
&\; \qquad \qquad \qquad \left. -  \left\{ Y(t) \, f(\mathbf k^\prime,t) \, Y(t) , \, \left(1+f(\mathbf k,t)\right) \right\} \, \left(1+ n_B(\varepsilon_{p}) \right) \, n_B(\varepsilon_{p^\prime})  \frac{}{} \right) \; . \notag
\end{align}
If we take the one-flavor limit of these collision terms, they reduce to  
the usual semi-classical collision terms~\cite{Calzetta:1986cq}.

We can evaluate these collision terms further.  If we assume that the distribution functions $f$ and $\bar{f}$ are 
independent of direction of $\mathbf k$~\footnote{
In the case of time-dependent external perturbation considered here, assuming  isotropy of the 
density matrices is well justified.    This assumption will have to be dropped in the realistic case of space-time dependent external field. 
In that case, one has to keep in $f (\mathbf{k})$ the full dependence on $|\mathbf{k}|$ and the cosine of the angle 
between $\mathbf{k}$  and a unit vector normal to the planar propagating bubble wall.
}, these collision terms take the form  
\begin{align}
C[f,\bar f] \equiv&  \int^\infty_0 \frac{dk^0}{2\pi} \, \mathcal{C}(k,t) \label{eq:collexp} \\
=& - \, \frac{1}{2} \, \int^\infty_0 dk^\prime \, \Big[ \Gamma_s(k,k^\prime) \,
\left( 1+ n_B(\bar{\omega}_{k} -\bar{\omega}_{k^\prime}) \right)  \times \nonumber  \\
&   
\left( \left\{ \frac{}{} Y \, (1 + f({k^\prime},t)) \, Y, \, f(k,t)  \, \right\} \,    
 -   e^{- \frac{\bar{\omega}_k -\bar{\omega}_{k^\prime}}{T}}     \,     \left\{ \frac{}{} Y \, f({k^\prime},t) \, Y, \, (1+f(k,t)) \, \right\} \,   \right)    \notag \\
& \qquad  \qquad \quad  + \Gamma_a(k,k^\prime) \,  \left( 1+ n_B(\bar{\omega}_k +\bar{\omega}_{k^\prime}) \right)  \times \notag \\
&\left( \left\{ \frac{}{} Y \,  \bar{f}({k^\prime},t) \, Y, \, f(k,t)  \, \right\}   -    
 e^{- \frac{\bar{\omega}_k +\bar{\omega}_{k^\prime}}{T}}   
\left\{ \frac{}{} Y \, (1+\bar{f}({k^\prime},t)) \, Y, \, (1+f(k,t)) \, \right\} \,  \right) \, \Big]  \notag \; , 
\end{align}
with $k \equiv |\mathbf{k}|$ and $k' \equiv |\mathbf{k}'|$. 
The term with $\Gamma_a$ corresponds to the annihilation collision term in Eq.~\eqref{eq:anncol}, while the term with $\Gamma_s$ corresponds to the scattering collision term in Eq.~\eqref{eq:scatcol}.  
For both annihilation and scattering processes, 
we have cast the collision integral in a form that emphasizes the relation between loss and gain terms implied by 
the principle of detailed balance. 
The anti-particle collision term is simply
\be
- \int_{-\infty}^0 \frac{dk^0}{2\pi} \, \mathcal{C}(k,t) = C[\bar f,f] \; ,
\ee
i.e., given by the same expression as Eq.~\eqref{eq:collexp}, but with $f \leftrightarrow \bar{f}$.  The rates $\Gamma_{a,s}$ are given by
\begin{subequations}
\bea
\Gamma_a(k,k^\prime) &=& \frac{k^\prime \, T}{256 \pi^3 \,\bar{\omega}_k \, \bar{\omega}_{k^\prime} k} \, \int^{k+k^\prime}_{|k-k^\prime|} ds \, \log\left[ \frac{ n_B( s_- ) \, n_B( - s_- ) }{n_B( s_+ ) \, n_B( - s_+ )  } \right] \, \theta(s_0^2 - s^2 - 4m_A^2) \quad \\
\Gamma_s(k,k^\prime) &=& \frac{k^\prime \, T}{128 \pi^3 \, \bar{\omega}_k \, \bar{\omega}_{k^\prime} k} \, \int^{k+k^\prime}_{|k-k^\prime|} dt \, \left( \, \log \left[\frac{ n_B( t_- ) }{ n_B( t_+ ) } \right]\, - \, \frac{t_0}{T} \,  \right) \;,
\eea
\label{eq:sakernels}
\end{subequations}
where
\begin{subequations}
\begin{align}
s_0 \; \equiv \; &\bar\omega_k + \bar\omega_{k^\prime} \, , & s_\pm \; \equiv& \; \frac{s_0}{2} \pm \frac{s}{2} \sqrt{ 1- 4 \, m_A^2/(s_0^2 - s^2) } \\
t_0 \; \equiv \; &\bar\omega_k - \bar\omega_{k^\prime} \, , & t_\pm \; \equiv& \; \pm  \frac{t_0}{2} + \frac{t}{2} \sqrt{ 1- 4 \, m_A^2/(t_0^2 - t^2) } \; . 
\end{align}
\end{subequations}


\section{Solving the Quantum Boltzmann Equations}
\label{sec:numerical}

In the preceding section, we derived the quantum Boltzmann equations for particle and antiparticle density matrices 
($f( k,t)$ and $\bar f( k  ,t)$, respectively). 
The diagonal entries correspond to the distributions of the individual mass eigenstates, 
while the off-diagonal elements measure the quantum coherence between them. 
As we have emphasized earlier, it is only by evolving the density matrix as a whole 
that one can properly study the coherent $CP$-violating oscillations (responsible for generating $CP$ asymmetry)  and the collisions that tend to wash out this coherence.
To summarize, our results are: 
\begin{subequations}
\label{eq:boltzfinal}
\bea
\frac{\partial f(k,t)}{\partial t} &=& \, - \, i \, \left[ \, \omega_k(t) + \delta \omega_k(t) - i \, \Sigma(t), \, f(k, t) \, \right] +   \, g_{\rm eff}  \, C[f,\bar f] \\
\frac{\partial \bar f(k,t)}{\partial t} &=& \,  \, i \, \left[ \, \omega_k(t) + \delta \omega_k(t) + i \, \Sigma(t), \, \bar f( k, t) \, \right] + \,  g_{\rm eff}  \, C[\bar f, f] \; .
\eea
\end{subequations}
The various contributions to these equations are as follows:
\begin{itemize}
\item the flavor oscillation terms $[\omega, \,f]$ and $[\omega, \, \bar f]$, as discussed in Eq.~\eqref{eq:freeboltz}; 
\item the forward scattering terms, giving rise to medium-induced mass terms 
(analogue to the  MSW effect),  with 
\be
\label{MSW}
\delta \omega_k(t) \equiv \frac{Y(t)}{2 \, \bar\omega_k} \, \int\!\! \frac{d^3 p}{(2\pi)^3} \, \frac{n_B(\varepsilon_p)}{2 \varepsilon_p}  \; .
\ee
As previously discussed in Eq.~\eqref{eq:shifty}, these terms shift the zero-temperature 
flavor diagonal masses:  $m_{L,R}^2 \to m_{L,R}^2 +  y_{L,R}\, T^2/24$. 
Since  $m_{L,R} \sim O(T)$  and $y_{L,R} \sim O(1)$,  we can safely neglect these contributions in the rest of the analysis. 

\item the terms $[\Sigma, \, f]$ and $[\Sigma, \, \bar f]$ arising from the presence of a time-dependent mixing matrix $U(t)$;
\item and the collision terms $C[f,\bar f]$ and $C[\bar f,f]$, given by Eq.~\eqref{eq:collexp}, 
which arise from scattering and annihilation processes of  the mixing scalars $\Phi_{L,R}$ with a single real degree 
of freedom ($A$). 
In Eqs.~\eqref{eq:boltzfinal} we have 
introduced  the coefficient  $g_{\rm eff}$,  counting the 
number of  $A$-type degrees of freedom present in  the thermal bath.   
In the electroweak plasma there are $\mathcal{O}(100)$  degrees of freedom from which the scalars can scatter, 
so   $g_{\rm eff} \sim \mathcal{O}(100)$ will  provide  a  realistic estimate of the  collision rate. 
In our numerical explorations we will vary $g_{\rm eff}$ 
to dial the relative size of the inverse collision rate   $\tau_{\rm coll}$  versus the other  scales in the problem, namely $\tw$ and $\tau_{\rm osc}$. 
\footnote{Strictly speaking, in this theory, we should then multiply the medium-induced mass correction \eq{MSW} by $g_{\text{eff}}$ as well. However, in a gauge theory like the electroweak theory or QCD, the leading thermal correction to, say, a fermion mass is summed over the number of gauge boson degrees of freedom in the theory, whereas the collision terms receive contributions from every particle with which the fermion can scatter, such as fermions of other flavors and colors as well as all the gauge bosons. This number of degrees of freedom is typically much greater than the number contributing to the leading correction to the thermal mass. We mimic this in our simple toy model by the heuristic device of multiplying the collision term by $g_{\text{eff}}$ but not the thermal mass correction.}

\end{itemize}
In this section, we describe numerical solutions to  these equations in a variety of  regimes.  Treating this as a toy model for the dynamics of electroweak baryogenesis, we wish to study how the time-dependent mixing matrix (encoded by $\Sigma$) leads to the generation of a $CP$-asymmetry in an initially $CP$-symmetric plasma, and how this asymmetry evolves in the presence of flavor oscillations and flavor-dependent interactions.
To illustrate these points, we first  analyze the collisionless case, paying special attention to 
the generation of the $CP$ asymmetry. 
We subsequently consider in detail the effect of collisions.

\subsection{Collisionless flavor oscillations and $CP$ violation}

\subsubsection{Analogy with spin precession in time-dependent magnetic field}

Neglecting the collision terms,  the Boltzmann equations  (\ref{eq:boltzfinal})
describe the independent time  evolution of  $f_{ij} (k, t)$ and $\bar{f}_{ij} (k,t)$,  
with no coupling between different momentum bins or  between  particles and antiparticles.  
Without loss of generality, we can parametrize the $2\times 2$ mass-basis density matrix $f (k, t)$ in terms of four real functions by expanding in the identity matrix and Pauli matrices~\cite{Stodolsky:1986dx}: 
\be
f (k, t) =    I \, p_0( k,t)    \ + \ \vec{\sigma} \cdot  \vec{p} (k,t) \;.
\ee
Here, $p_{0}(k,t)$ is the  total occupation number of momentum mode $k$, and 
the polarization vector $\vec{p} (k) = (p_x, p_y, p_z)$ describes the density matrix for the  ``internal"  (flavor) degrees of freedom.  
Similarly,  we parametrize the anti-particle density matrix $\bar{f}_{ij} (k,t)$
in terms of $\tilde{p}_0 (k,t)$ and   $\vec{\tilde{p}} (k,t)$. 

In the collisionless case, Eqs.~(\ref{eq:boltzfinal}) imply that   $p_{0}(k,t)$ 
and $\tilde{p}_{0}(k,t)$ do not evolve in time, while the polarization vectors 
$\vec{p} (k,t)$  and $\vec{\tilde{p}} (k,t)$ obey equations of motion suggestive of spin precession:
\begin{subequations}
\bea
\frac{d \vec{p} (k,t) }{d t}   &= & \Big( \vec{B}_0(k,t)  +  \vec{B}_\Sigma (t)    \Big) 
 \times  \vec{p} (k,t) \\
\frac{d \vec{\tilde p} (k,t) }{d t}   &= &  -  \Big( \vec{B}_0 (k,t)  - \vec{B}_\Sigma (t) \Big)  \times  \vec{\tilde p} (k,t) ~, 
\eea
\end{subequations}
with  effective time-dependent magnetic field dictated by the form of the mass matrix:  
\begin{subequations}
\bea
\vec{B}_0    (k,t)  &=&   \Big( 0, 0,\omega_1(k,t) - \omega_2 (k,t) \Big) \\
\vec{B}_\Sigma  (t)  &=& \left( 2 \sin \sigma \ \dot{\theta} + \sin 2 \theta \, \cos \sigma \, \dot{\sigma}, 
\ - 2 \cos \sigma \, \dot{\theta}  + \sin 2\theta \, \sin \sigma \dot{\sigma},  \ 2 \sin^2 \theta \, \dot{\sigma} 
\right)~.
\eea
\end{subequations}
The particle polarization vector  $\vec{p} (k,t)$ precesses  around $\vec{B}_0 + \vec{B}_\Sigma$, 
while the anti-particle polarization vector $\vec{\tilde{p}} (k,t)$ 
precesses around  $\vec{B}_0 - \vec{B}_\Sigma$  in the opposite direction: 
this describes the flavor oscillation dynamics. 
The usual  vacuum oscillation is recovered in the case of $t$-independent mass matrix, 
in which the effective magnetic field ($\vec{B}_0$) points in the $z$-direction. 
An initial flavor eigenstate---having $p_{x,y} (k,t) \neq 0$---precesses around $\vec{B}_0$ with period 
$\tau_{\rm osc} = 2 \pi/(\omega_1 - \omega_2)$; see also Eq.~\eqref{eq:vacosc1}. 

\subsubsection{Adiabatic and non-adiabatic regimes}

\begin{figure}[!t]
\begin{center}
\mbox{\hspace*{-1cm}\epsfig{file=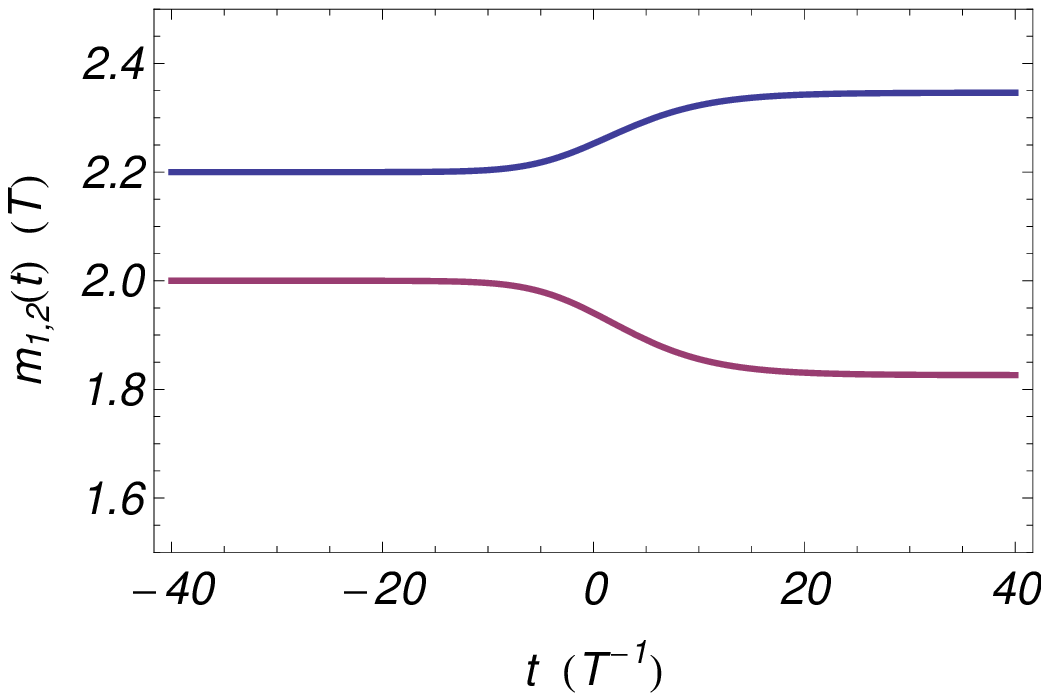,height=4.5cm}}
\mbox{\hspace*{1cm}\epsfig{file=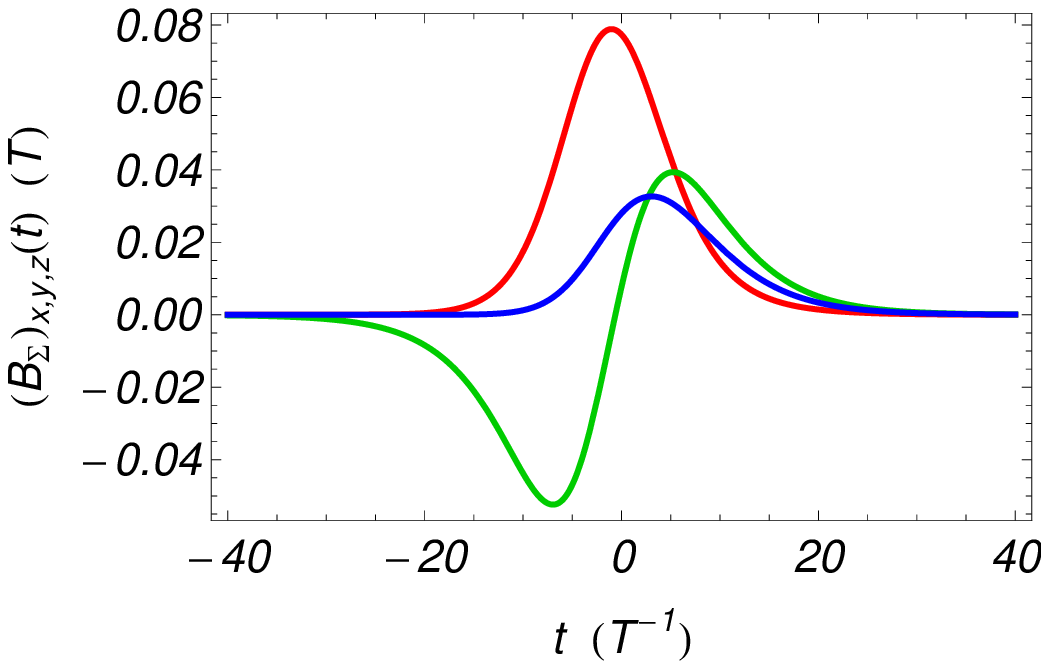,height=4.5cm}} 
\end{center}
\caption{\it\small 
Left panel: mass eigenvalues $m_{1,2}(t)$ as a function of time. 
Right panel:  components of $\vec{B}_\Sigma$ as a function of time: 
$(\vec{B}_\Sigma)_x$ (red),  
$(\vec{B}_\Sigma)_y$ (green),  
$(\vec{B}_\Sigma)_z$ (blue).  
Input parameters are as in Table~\ref{tab:baseline}.
}
\label{fig:Bsigma}
\end{figure}

\begin{table}[b!]
\begin{center}
\begin{tabular}{|ccccc|}
\hline 
 &  & & &   \\
\; \; $\tw = 10/T  \quad$ &       
$v_0 = T^2    \quad $       &  
$a_0=\pi/2 \quad $  &   
$m_L = 2.2\, T \quad $ & 
$m_R= 2 \,  T \quad $   \\
 &  & & &   \\
\hline
\end{tabular}
\end{center}
\caption{Baseline input parameters for the toy model.  All dimensionful parameters are expressed in 
units of the temperature $T$ or its inverse.  In addition, we take $m_A/T \ll 1$.}
\label{tab:baseline}
\end{table}

In the time dependent case, the qualitative behavior of the solution depends on the ratio of 
 the oscillation time  $\tau_{\rm osc} = 2 \pi/(\omega_1 - \omega_2)$ 
and $\tw$, which controls the time variation of the effective magnetic field 
through $m_{1,2}$,   $\theta$,   and $\sigma$.
In Fig.~\ref{fig:Bsigma},  we show the time-dependence of $m_{1,2}(t)$ (determining $\vec{B}_0(t)$ and $\tau_{\rm osc}$) and 
$\vec{B}_\Sigma (t)$ for our baseline choice of parameters, reported in Table~\ref{tab:baseline}.
Let us consider the evolution of a $CP$ invariant, purely L-handed initial state, given by 
$(p_0, \vec{p}) = (\tilde p_0, \vec{\tilde p} ) = (1/2,0,0,1/2)$ at initial time $t_{\rm in} < - \tw$. 
Before the external  wall turns on ($t <  - \tw$),  both $\vec{B}_0$ and $\vec{p}$
point  along the $z$ axis and there is no precession.  
As the ``wall'' turns on,   the non-vanishing $\vec{B}_\Sigma$ tends to push $\vec{p}$ 
out of its original  stationary state, triggering the precession around the 
time-dependent  field  $\vec{B}_0 + \vec{B}_\Sigma$. 
In the adiabatic regime ($\tw \gg \tau_{\rm osc}$), the polarization vector $\vec{p}$  
effectively tracks the magnetic field (with a small precession amplitude that vanishes in the $\tau_{\rm osc}/ \tw \to 0$ limit).   
On the other hand, in the non-adiabatic regime ($\tw \leq  \tau_{\rm osc}$), when the magnetic field changes on time scales comparable to or faster than 
the oscillation time scale,
the polarization vector will lag behind the magnetic field and begin precessing 
with a large amplitude. The precession persists
at late time ($t > \tw$) around the final constant magnetic field.

%

\begin{figure}[!t,p]
\begin{center}

\ \qquad \boxed{\text{Particle polarization}}  \  \ \ \quad\quad  \boxed{\text{Anti-particle polarization}} \quad \quad \ \boxed{\text{\scriptsize Flavor-diagonal $CP$ asymmetry}}

\bigskip

\mbox{\hspace*{0cm}\epsfig{file=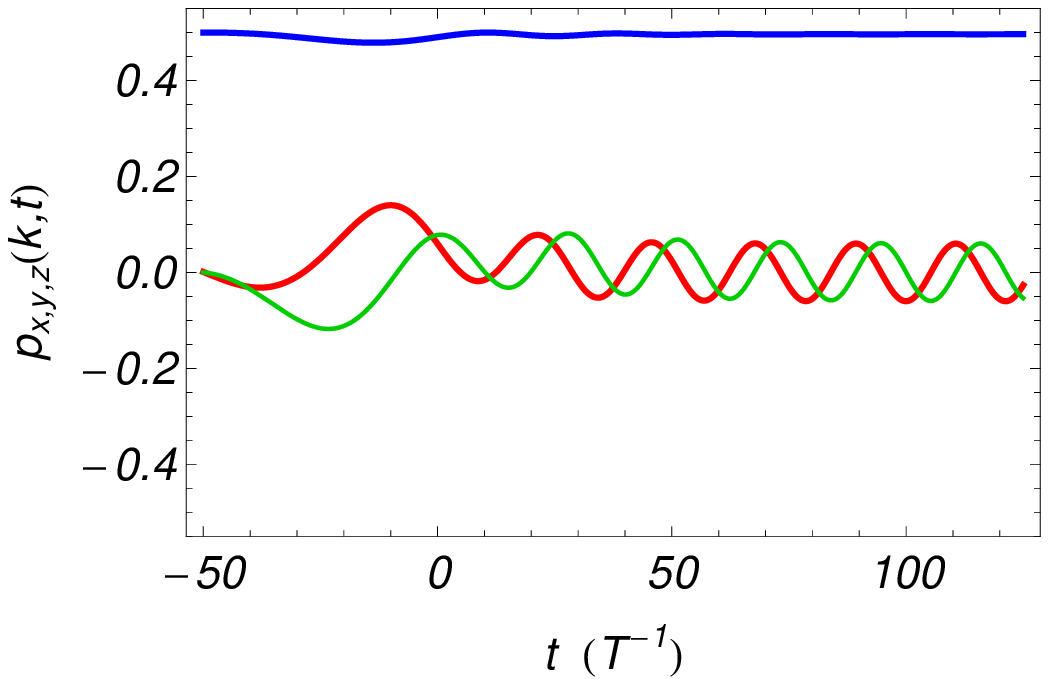,height=3.3cm}} \mbox{\hspace*{0cm}\epsfig{file=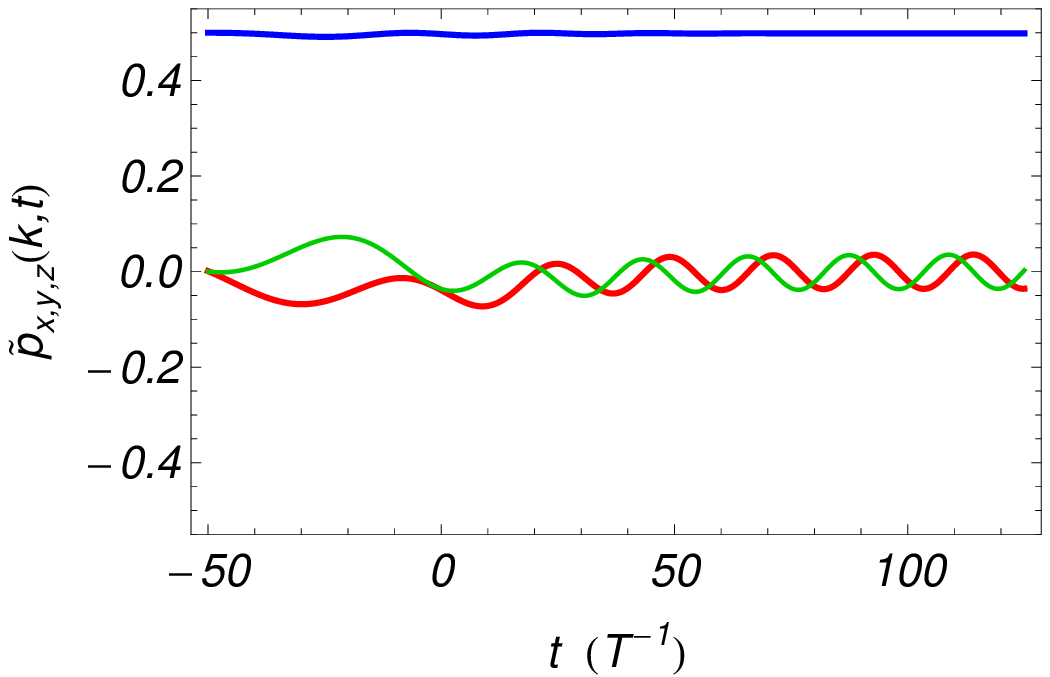,height=3.3cm}} \mbox{\hspace*{0cm}\epsfig{file=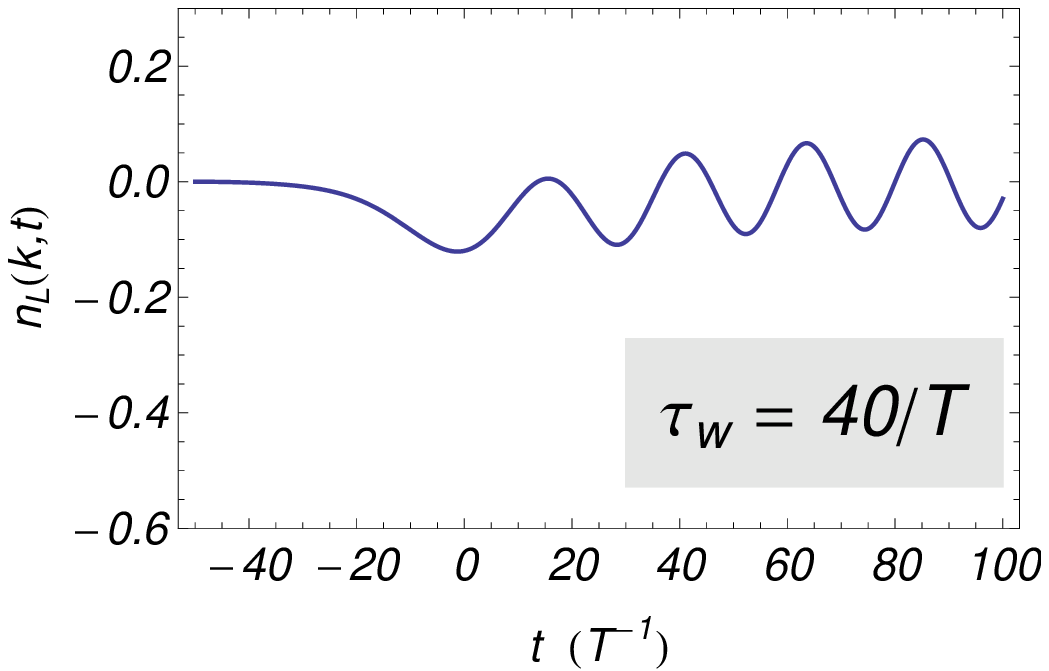,height=3.3cm}}
\vspace{0.7cm}

\mbox{\hspace*{0cm}\epsfig{file=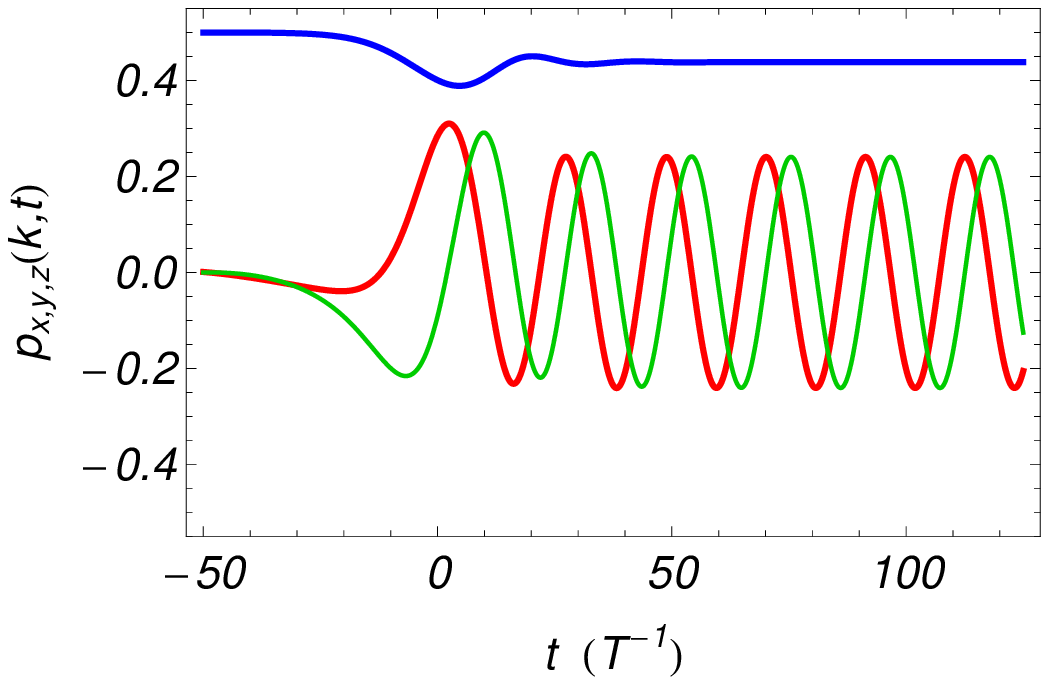,height=3.3cm}} \mbox{\hspace*{0cm}\epsfig{file=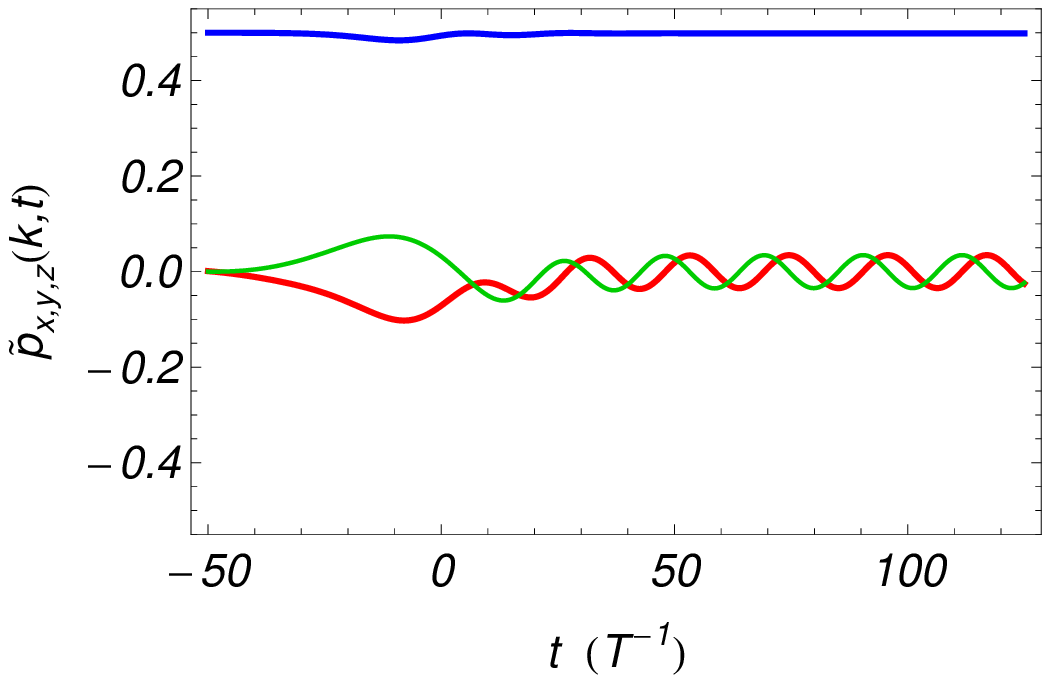,height=3.3cm}} \mbox{\hspace*{0cm}\epsfig{file=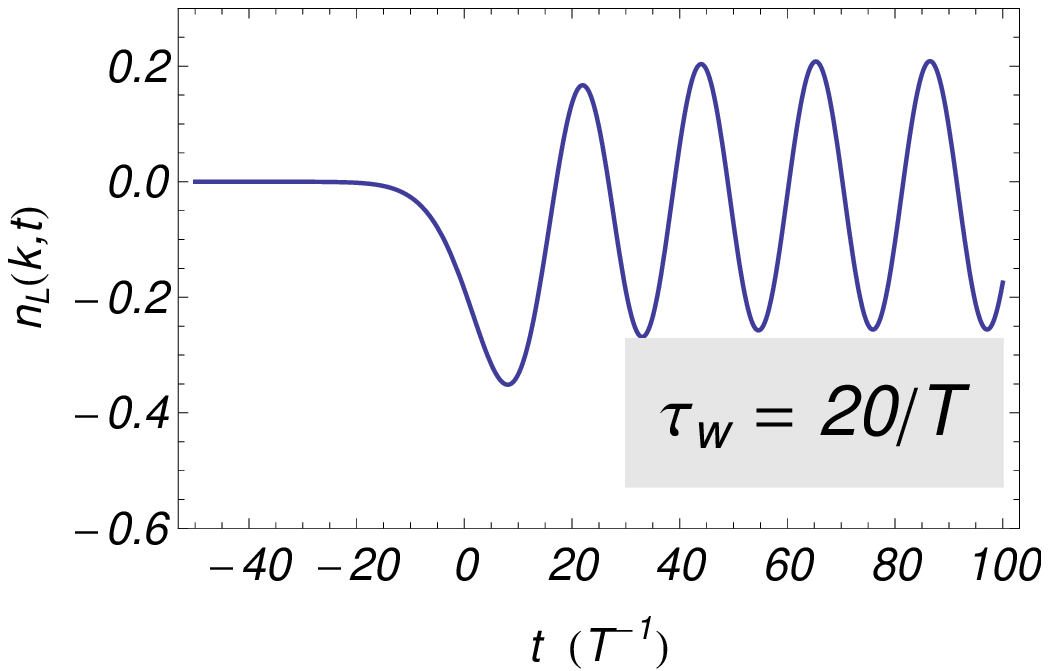,height=3.3cm}}
\vspace{0.7cm}

\mbox{\hspace*{0cm}\epsfig{file=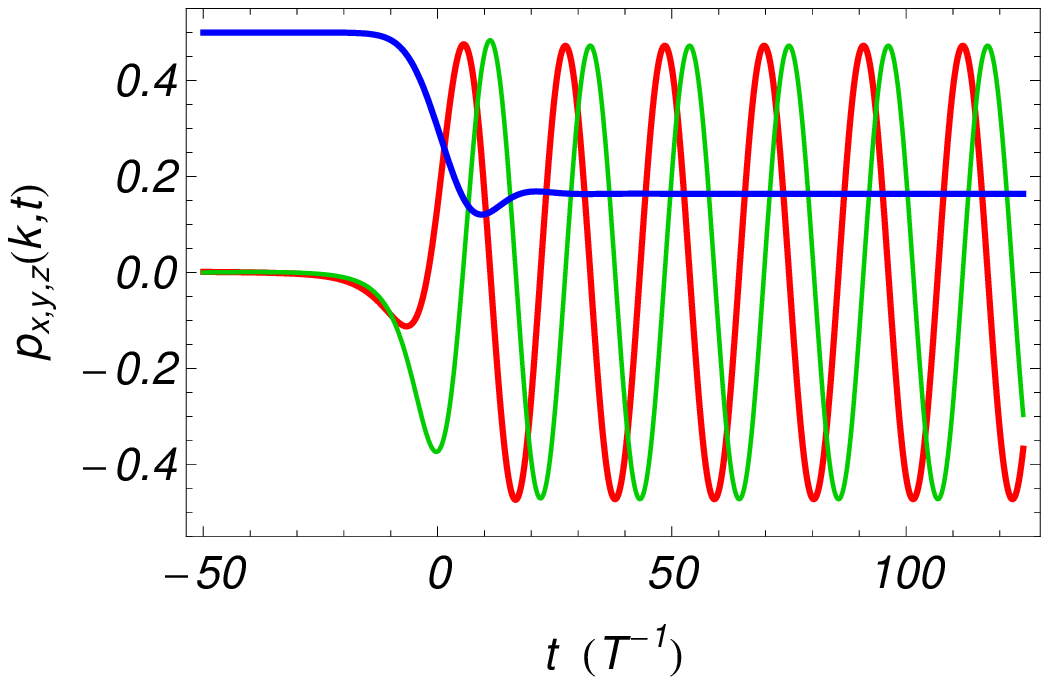,height=3.3cm}} \mbox{\hspace*{0cm}\epsfig{file=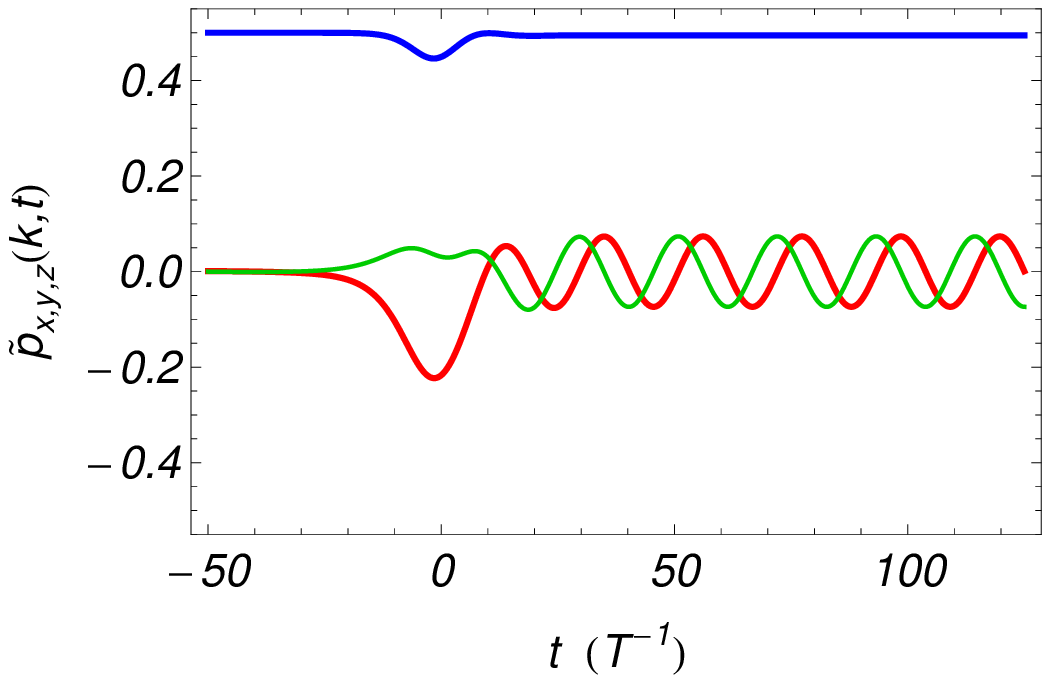,height=3.3cm}} \mbox{\hspace*{0cm}\epsfig{file=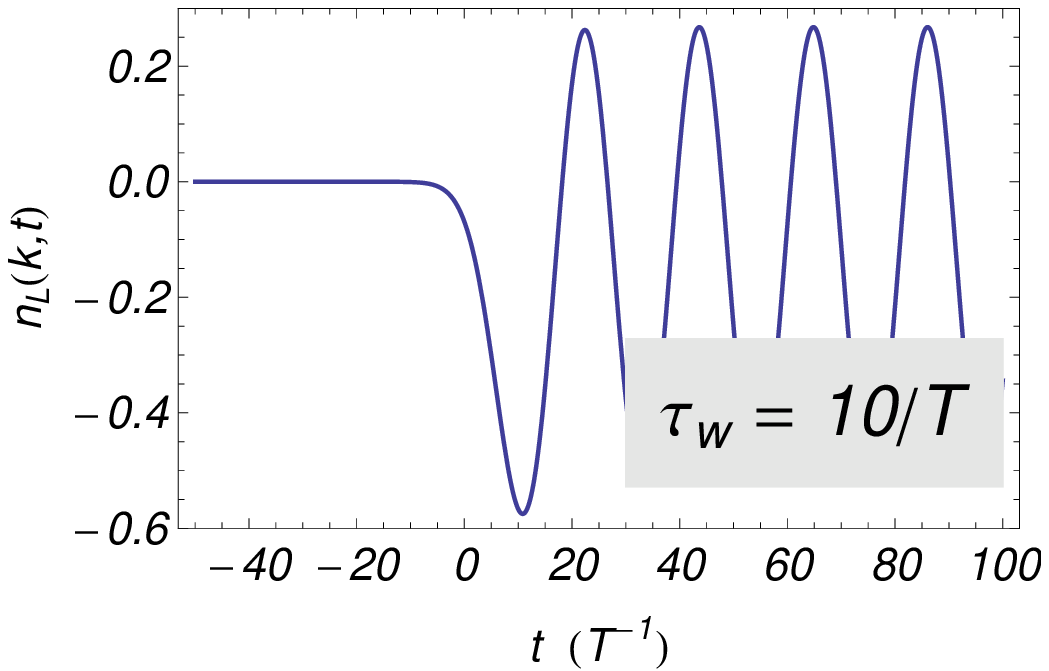,height=3.3cm}}
\vspace{0.7cm}

\mbox{\hspace*{0cm}\epsfig{file=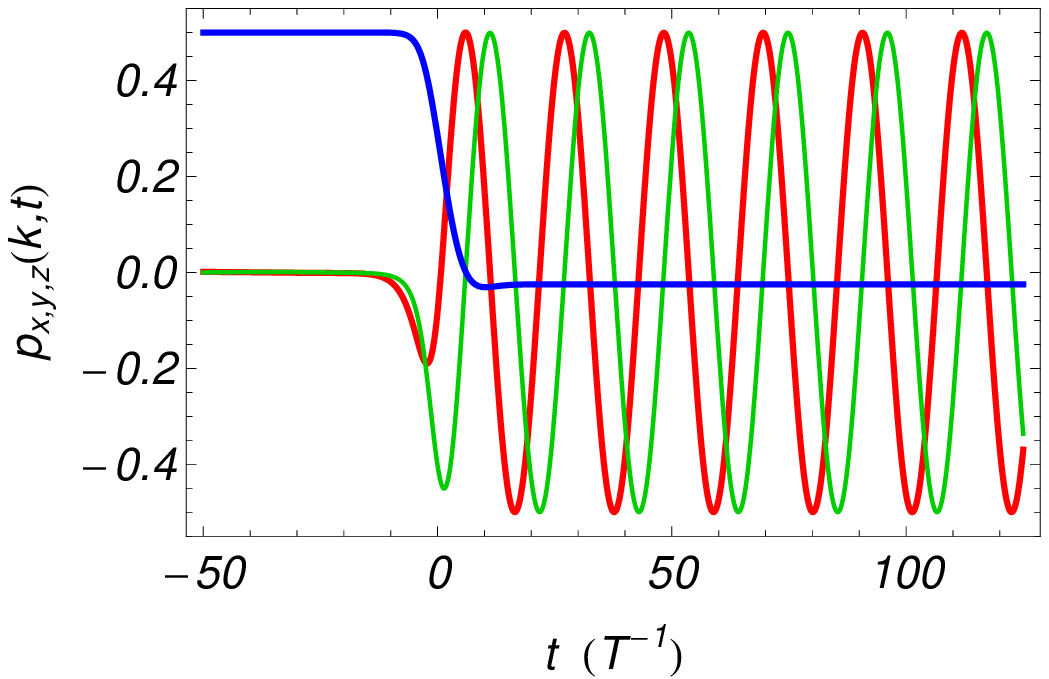,height=3.3cm}} \mbox{\hspace*{0cm}\epsfig{file=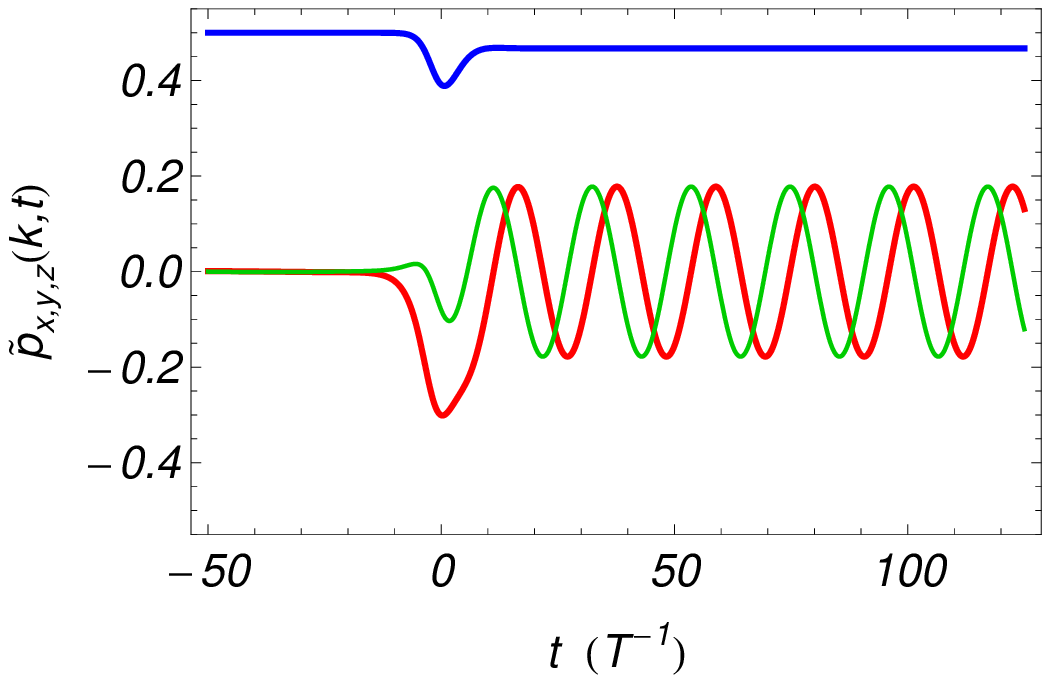,height=3.3cm}} \mbox{\hspace*{0cm}\epsfig{file=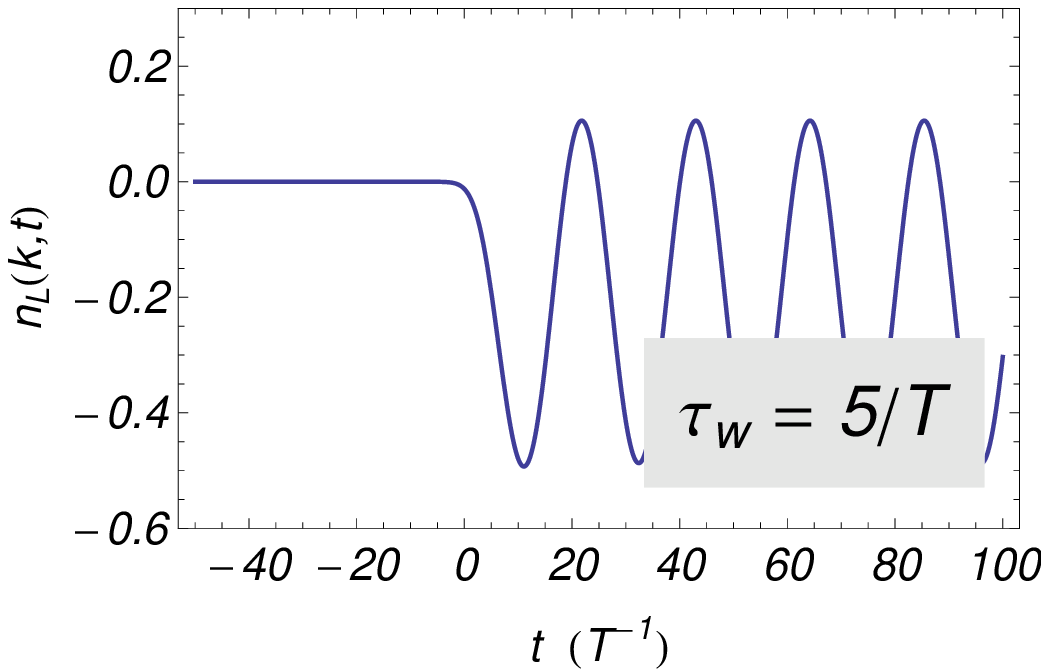,height=3.3cm}}
\vspace{0.7cm}

\end{center}
\vspace{-1cm}
\caption{\it\small 
Evolution of flavor polarization and CP asymmetry in the absence of collisions, 
starting from a $CP$-invariant, pure $L$-handed initial state. 
Left column:   time dependence of the particle polarization vector components $p_x$ (red), $p_y$ (green), and $p_z$ (blue),  
for different values  of   $\tw=40/T,20/T,10/T,5/T$ (from top to bottom). 
Middle column:   time dependence of the anti-particle polarization vector components $\tilde{p}_x$ (red), $\tilde{p}_y$ (green), 
and $\tilde{p}_z$ (blue),   for different values  of   $\tw=40/T,20/T,10/T,5/T$ (from top to bottom). 
Right column:   time dependence of the flavor diagonal $CP$ asymmetry $n_L(k,t)$ 
 for different values  of  $\tw=40/T,20/T,10/T,5/T$ (from top to bottom). 
In all cases $k=3T$ and all other input parameters (except $\tw$) are as in Table~\ref{tab:baseline},
corresponding to $\tau_{\rm osc} \simeq 35/T$ (at $t=0$). 
}
\label{fig:nocoll}
\end{figure}

The onset of the non-adiabatic regime 
is illustrated in the first column of Fig.~\ref{fig:nocoll}, in which we plot the  
the time-dependence of the particle polarization  $\vec{p} (k,t)$ with  $k=3 T$
(corresponding to $\tau_{\rm osc} \simeq 35/T$, at $t=0$).  We show four different values of 
$\tw=40/T,20/T,10/T,5/T$ (from top to bottom), in order of increasing non-adiabaticity, resulting in increasing precession amplitudes.
Within any realistic model, $\tw$ is fixed and the 
adiabaticity is controlled by $k$ and $m_{1,2}$. 
Larger values of $k$ and smaller mass splittings increase $\tau_{\rm osc}$, 
thus leading to increasingly non-adiabatic evolution.

\subsubsection{$CP$ violation}

Having identified the basic features of the particle density matrix evolution, we 
can now turn to discuss the effects of $CP$ violation. 
As discussed in Section~\ref{sec:prelim}, the dynamics are $CP$ invariant if and only if $\dot{\sigma} = 0$. 
In this case, with a time-independent phase redefinition of $\Phi_{L,R}$, one can set $\sigma = 0$. 
Therefore, $\vec{B}_0 + \vec{B}_\Sigma$ and $\vec{B}_0 -  \vec{B}_\Sigma$ lie on the $y-z$ plane; 
for any $CP$-invariant initial condition (such that 
$\vec{p} (t_{\rm in})= \vec{\tilde{p}} (t_{\rm in})$), the evolution implies  that 
\bea 
\label{eq:CPconditions2}
\tilde{p}_x (k ,t)  = p_x ( k  ,t) \, , \qquad \tilde{p}_y (k   ,t)  =  - p_y (k  ,t) \, ,  \qquad 
\tilde{p}_z ( k  ,t)  = p_z (k  ,t)  \; ,
\eea 
or, equivalently, $f(k,t) = \bar{f}^T(k,t)$.
Geometrically, this means that the angle between  $\vec{p}$  and $\vec{B}_0 + \vec{B}_\Sigma$ 
is equal to the angle between   $\vec{\tilde{p}}$  and $\vec{B}_0 - \vec{B}_\Sigma$. 
On the other hand, if $\dot{\sigma} \neq 0$ the evolution violates $CP$. Geometrically,  the effective fields 
$\vec{B}_0 + \vec{B}_\Sigma$  and $\vec{B}_0 - \vec{B}_\Sigma$ are not confined to the $y-z$ plane 
but develop opposite $x$ components,  so that the 
angle between  $\vec{p}$  and $\vec{B}_0 + \vec{B}_\Sigma$ differs from 
the angle between   $\vec{\tilde{p}}$  and $\vec{B}_0 - \vec{B}_\Sigma$. 
At late times ($t \gg \tw$, such that $\vec{B}_\Sigma  \to  0$), the polarization vectors $\vec{p}$ and $\vec{\tilde{p}}$ 
precess in opposite directions and with different angles around $\vec{B}_0$ (parallel to the $z$ axis); therefore, $CP$ is violated since Eqs.~(\ref{eq:CPconditions2}) are not satisfied. 
The above qualitative discussion is confirmed by the explicit numerical solution, 
as  illustrated in the second column of Fig.~\ref{fig:nocoll}, in which we plot the  
the time-dependence of  $\tilde{\vec{p}} (k,t)$ for $k=3 T$ 
(with $\tw=40/T,20/T,10/T,5/T$, from top to bottom).

In light of future applications of this formalism to weak-scale baryogenesis, it is highly relevant 
to quantify the $CP$ asymmetry generated in the diagonal elements of the density matrix in flavor basis. 
In particular, we are interested in the L-handed charge density, defined by 
\be
n_L (k,t) \equiv 
\Tr\left[ \,  P_L \, U(t) \, \left(  f( k, t) -  \bar{f}( k, t) \right)   \, U^\dagger(t) \, \right]
\ee
where  $P_L = \textrm{diag}(1,0)$.
We plot $n_L (k=3T,t)$ in the third column of Fig.~\ref{fig:nocoll}, 
as usual for different values of $\tw=40/T,20/T,10/T,5/T$ (from top to bottom). 
The flavor-diagonal $CP$ asymmetry is small in the quasi-adiabatic regime (top panel) 
and increases with $\tau_{\rm osc}/\tw$   (the evolution becomes more non-adiabatic).

\begin{figure}[!t]
\begin{center}
\mbox{\hspace*{-1cm}\epsfig{file=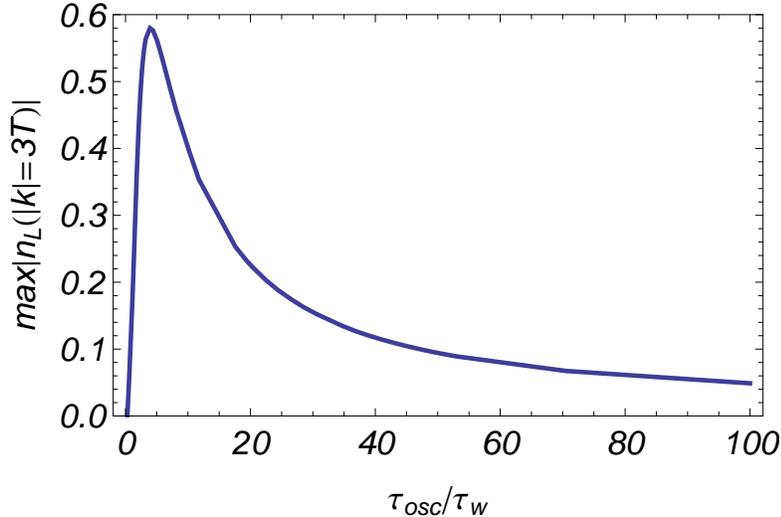,height=7cm}} 
\end{center}
\caption{\it\small 
Maximum value of $CP$ asymmetry $|n_L(k,t)|$ for $k=3 T$  as a function of $\tau_{\rm osc}(t=0)/\tw$. 
Except for $\tw$ that is varied, all other input parameters are as in Table~\ref{tab:baseline}. 
}
\label{fig:nLvslw}
\end{figure}

However, as one increases the non-adiabaticity, a maximal $CP$ asymmetry is reached.  This is illustrated in Fig.~\ref{fig:nocoll} by the fact that the $CP$ asymmetry is larger for $\tw = 10/T$ than for $\tw = 5/T$.
As a measure of this effect, we plot in Fig.~\ref{fig:nLvslw} the maximum  of $|n_L (k=3T, t)|$ 
versus the ratio $\tau_{\rm osc}(t=0)/\tw$.   The $CP$ asymmetry vanishes in both the adiabatic $(\tau_{\rm osc}/\tw \ll 1)$ and extreme non-adiabatic $(\tau_{\rm osc}/\tw \gg 1)$ limits.  A maximal $CP$-asymmetry occurs for $\tau_{\rm osc} /\tw \sim 4$.  This result is independent of the particular choice of momentum bin $k = 3 T$. 

The behavior of these limiting cases can be simply understood on a qualitative level.  In the adiabatic limit  ($\tau_{\rm osc}/\tw \ll 1$), 
both  $\vec{p}$ and   $\vec{\tilde{p}}$ track their effective magnetic fields $\vec{B}_0 \pm \vec{B}_\Sigma$.
Since at late times these effective magnetic fields coincide ($\vec{B}_\Sigma = 0$), the resulting $\vec{p}$ and   $\vec{\tilde{p}}$ are equal and no $CP$ asymmetry is present. 
On the other hand, in the extreme non-adiabatic regime ($\tau_{\rm osc}/\tw \gg 1$),  
the time behavior of the effective magnetic field is singular:  $\vec{B}_\Sigma (t) \sim \delta(t) \ \vec{b}_\Sigma$. 
This  produces a common shift $\vec{b}_\Sigma \times \vec{p}(0)$ in both $\vec{p}$ and $\vec{\tilde{p}}$ at $t=0$.
After this modified initial condition,  $\vec{p}$ and   $\vec{\tilde{p}}$ precess around the constant 
$\vec{B}_0$ in exactly opposite directions, so again no $CP$ asymmetry can arise.~\footnote{Strictly speaking, a time-independent phase redefinition 
has to be performed to satisfy conditions~\eqref{eq:CPconditions2}.}
So  for the system to feel the $CP$-violating effects,  $\vec{B}_\Sigma$ has to vary not too 
fast compared to the  oscillation scale. 

In summary, we have shown how a $CP$-asymmetry can arise in an intially $CP$-symmetric plasma in equilibrium through 
time-dependent, $CP$-violating flavor oscillations.  This provides a conceptually simple picture for the EWB mechanism.  We have also identified and discussed  the regime in which the $CP$ asymmetry is largest, namely 
$\tau_{\rm osc} \sim \tw$. 
We discuss next  how interactions between the mixing scalars and the thermal bath affect this picture.

\subsection{Full Boltzmann equations}

\subsubsection{General considerations}

The primary technical challenge in solving the full  Boltzmann equations is that they  are  coupled 
integro-differential equations.  We surmount this obstacle by discretizing $f$ and $\bar{f}$ into $N$ momentum bins, each of width $\Delta$.  Whereas Eqs.~\eqref{eq:boltzfinal} are 8 coupled integro-differential equations (each $f$ and $\bar f$ are composed of 4 real functions), discretization converts the Boltzmann equations into $8\,N$ coupled ordinary differential equations, 
which can be handled numerically. 
We have explored several choices of binning, keeping in mind two criteria: (i) the need to cover 
the entire thermal spectrum  up to sufficiently large momentum (we have included up to $k \sim 8 T$); (ii) the need to ensure thermalization,    
which constrains the bin width  to be not much larger than   the width of the 
scattering and annihilation kernels  $\Gamma_{s,a}(k,k')$  (see Eqs.~(\ref{eq:sakernels})). 
For our baseline numerical analysis we have adopted the following   choice:     
\begin{equation}
f(k,t)  \to    f_{n} (t) \equiv  f(k_n,t)   \quad n=1, N=8\ ;    \qquad k_n = \left( n- \frac{1}{2}\right) \Delta  \ ;  \qquad \Delta =  1.2 \,  T  \ , 
\end{equation}
and we have verified that the main results are stable at the percent level against 
changes of binning that satisfy criteria (i) and (ii) above.

Let us now turn to the physics.   At late negative times $t \ll  - \tw$, we assume the system to be in 
thermal equilibrium with zero total charges.   Hence, our initial condition is 
\be
f_{ij} (k,t_{\rm in}) =  \bar{f}_{ij} (k,t_{\rm in}) = 0      \qquad  i \neq j \ , \qquad \qquad  
f_{ii}(k,t_{\rm in}) = \bar{f}_{ii}(k,t_{\rm in}) = n_B (\omega_{i} (k)) ~. 
\ee
As shown in the previous section, in the non-adiabatic regime  the turning on of the wall kicks the system out 
of equilibrium and induces flavor oscillations.  
The inclusion of   $C[f,\bar{f}]$ in the kinetic equations (\ref{eq:boltzfinal})
has two important effects on the evolution of the system:
\begin{itemize}
\item   Collisions tend to destroy the quantum coherence characterizing 
the evolution of  linear combinations  of different mass (or flavor) eigenstates. 
This is highly relevant to our problem,  as  
coherence is essential for flavor oscillations and hence $CP$ violation to emerge.  
Even iso-energetic collisions would stop the coherent development of the state, 
as long as the two mixing particles  have different couplings to the thermal bath 
($y_L \neq  y_R$ in our model)~\cite{Stodolsky:1986dx,Sigl:1992fn}. 
So on general grounds we can conclude that  collisions will push the off-diagonal elements of the density 
matrices  $f(k,t)$ and $\bar{f}(k,t)$ (hence $p_{x,y}(k,t)$ and  $\tilde{p}_{x,y}(k,t)$)  to zero.

\item More generally, collisions allow for energy exchange between the scatterers and the thermal bath, 
thus pushing  the system towards its thermal equilibrium state. As a consequence,  the 
diagonal entries of the density matrices  tend to the appropriate thermal form consistent with 
quantum statistics and the conserved charges in the system.   

\end{itemize}

\subsubsection{Reaching equilibrium} 

In the toy model under study,  the form of the  late-time equilibrium state depends on the interaction coupling constants 
$y_{L,R}$.
By analyzing the collision term, one can easily verify that  the general form of the equilibrium distributions 
will be
\begin{subequations}
\label{eq:latetime}
\begin{align}
f_{11}(k) = \ & n_B(\bar \omega _k- \mu_1) & \bar f_{11}(k) = \ & n_B(\bar\omega_k+ \mu_1) \\
f_{22}(k) = \ & n_B(\bar\omega_k- \mu_2) & \bar f_{22}(k) = \ & n_B(\bar\omega_k+ \mu_2) ~,
\end{align}
\end{subequations}
with vanishing off-diagonals. 
In the case of flavor-sensitive interactions  ($y_L \neq  y_R$),  one also has the condition 
$\mu_2 = \mu_1$,  since  there is only one conserved charge in the system, the total charge $Q_1 + Q_2$. 
On the other hand, flavor-blind interactions ($y_L = y_R$) can lead to equilibrium with $\mu_1 \neq \mu_2$, since at late times $Q_1$ and $Q_2$ are separately conserved. 
In addition, if we begin with a $CP$-symmetric initial state (with $Q_1 + Q_2 = 0$), then at late times we must have $\mu_1 = - \, \mu_2$.  Therefore, we expect our numerical solutions to reach the form given in Eq.~\eqref{eq:latetime}, with (i) $\mu_1 = \mu_2 = 0$ for flavor-sensitive interactions, or (ii) $\mu_1 = - \, \mu_2 \ne 0$ for flavor-blind interactions.  

In the flavor-blind case ($y_L = y_R$), we can compute the individual charge generated at late times by the discretized momentum integral
\be
Q_1(t) = \Delta \, \sum_{i=1}^N \, \frac{k^2_i}{2\pi^2} \, \left( \, f_{11}(k,t) - \bar{f}_{11}(k,t) \, \right) \; , 
\ee
and, using Eq.~\eqref{eq:latetime}, the corresponding chemical potential (working to linear order in $\mu_1/T$): 
\be
\frac{\mu_1(t)}{T} = 
\label{eq:mu1}
Q_1(t)    \left[ \frac{\Delta}{\pi^2}  \sum_{i=1}^N \,  k^2_i \, \frac{e^{\bar\omega/T}}{(e^{\bar\omega/T}-1)^2} \right]^{-1}\; .
\ee
We have explicitly verified that our numerical solutions converge at late time to these expected equilibrium results,  
turning  the  above relations into a nontrivial check of our numerical codes.

\begin{figure}[h]
\begin{center}
\mbox{\hspace*{-0.cm}\epsfig{file=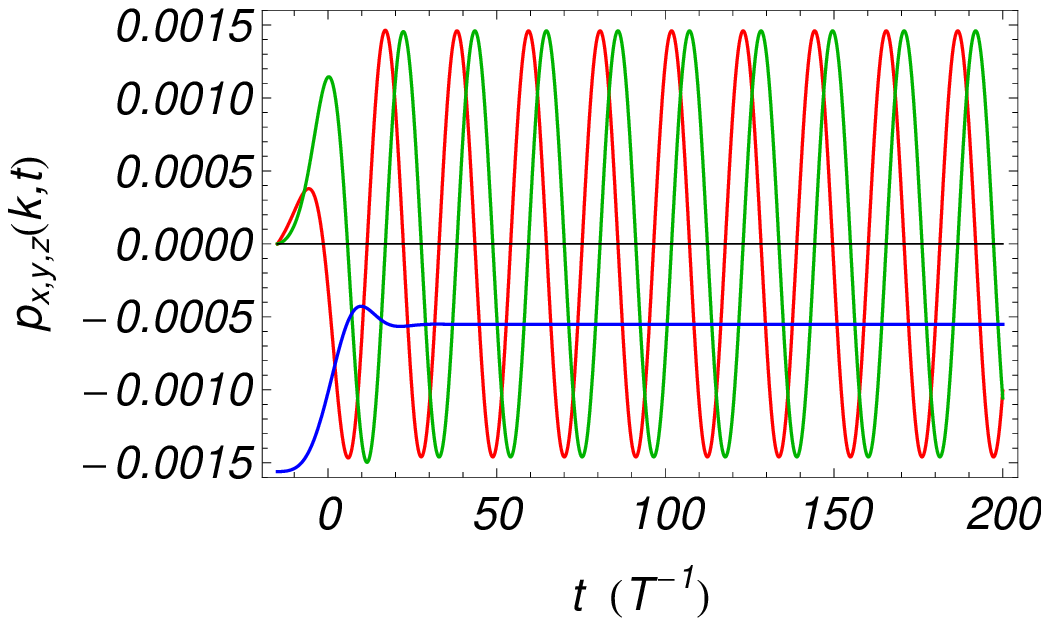,height=4.5cm}} \mbox{\hspace*{0.5cm}\epsfig{file=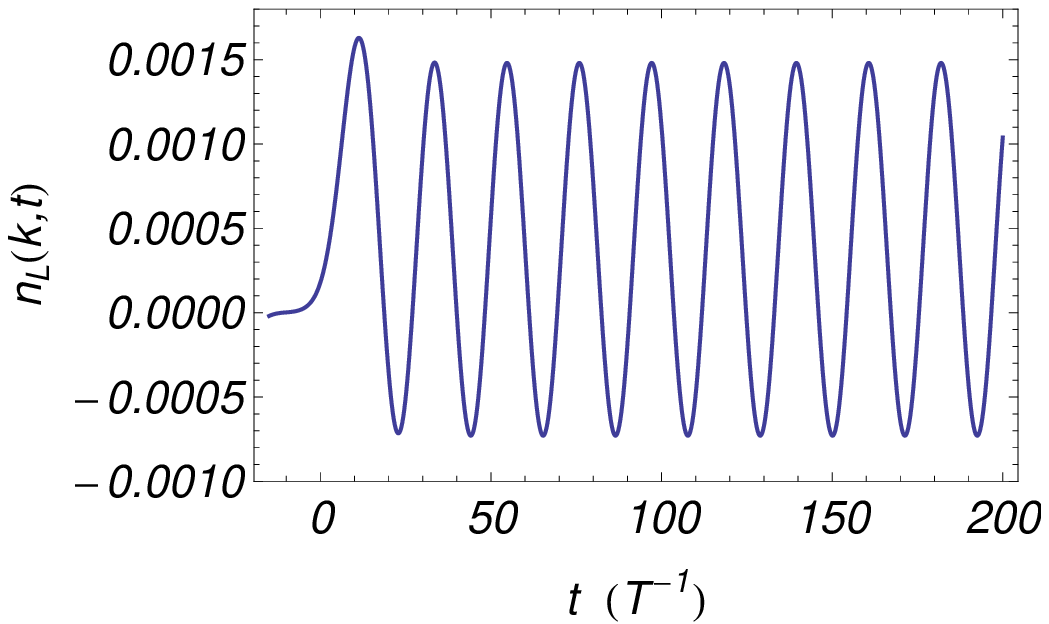,height=4.5cm}}
\vspace{0.0cm}

\mbox{\hspace*{-0.cm}\epsfig{file=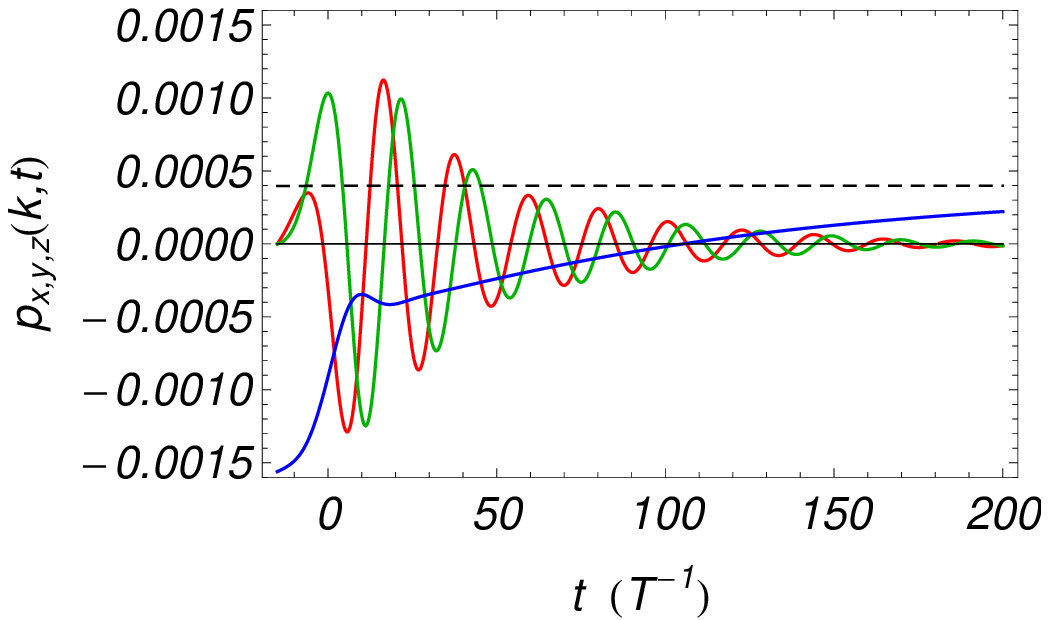,height=4.5cm}} \mbox{\hspace*{0.5cm}\epsfig{file=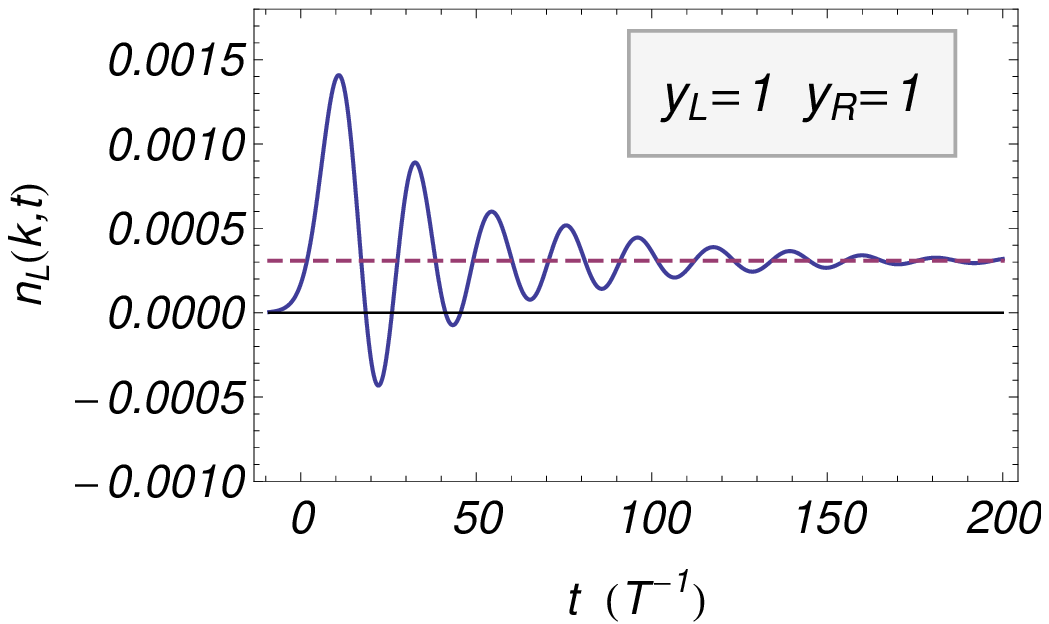,height=4.5cm}}
\vspace{0.0cm}

\mbox{\hspace*{-0.cm}\epsfig{file=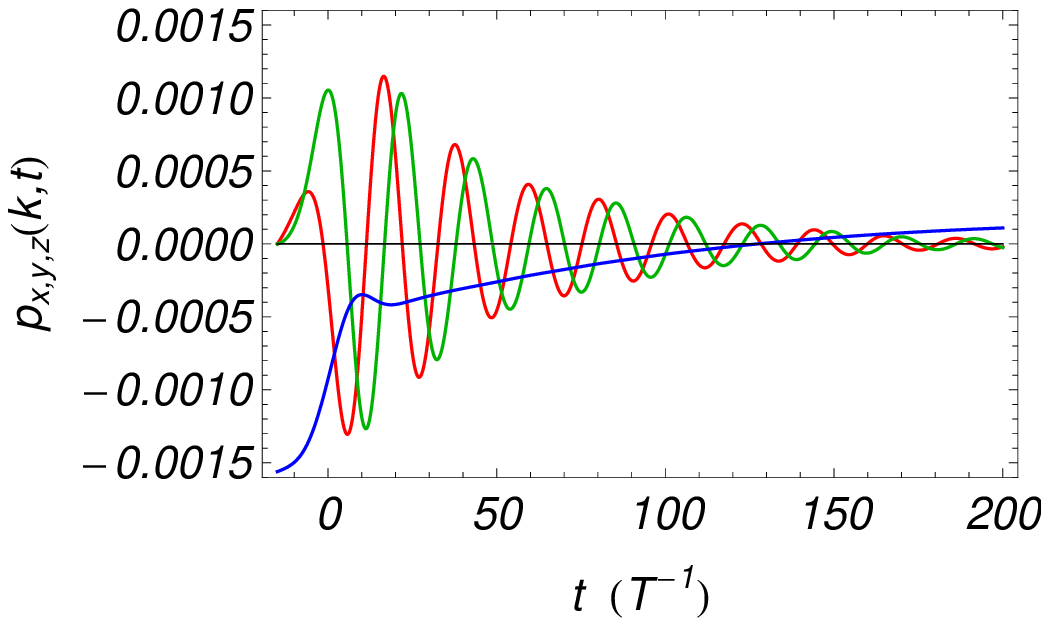,height=4.5cm}} \mbox{\hspace*{0.5cm}\epsfig{file=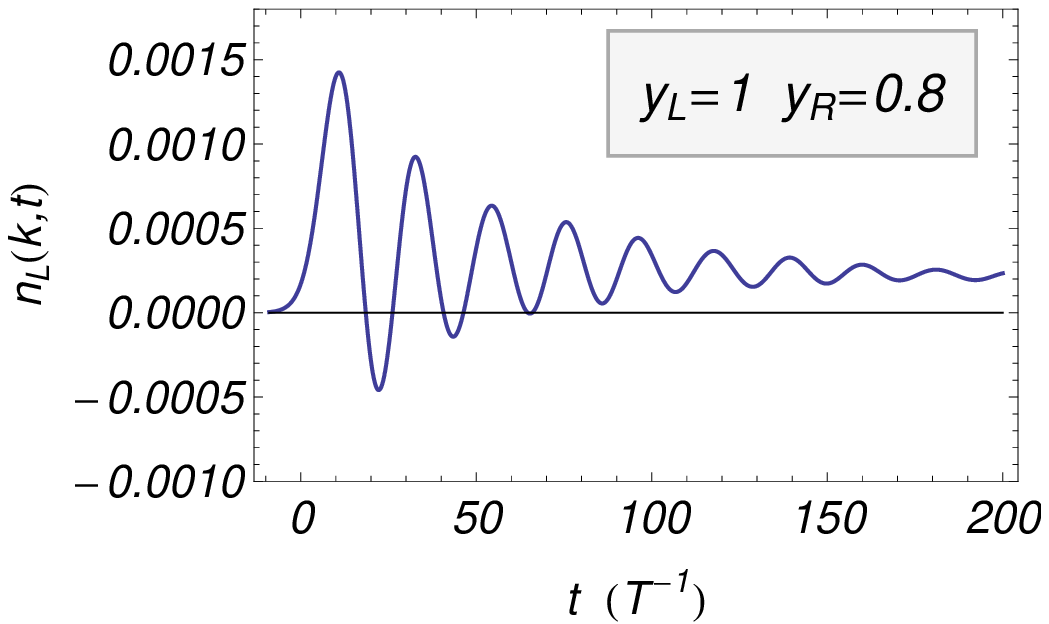,height=4.5cm}}
\vspace{0.0cm}

\mbox{\hspace*{-0.cm}\epsfig{file=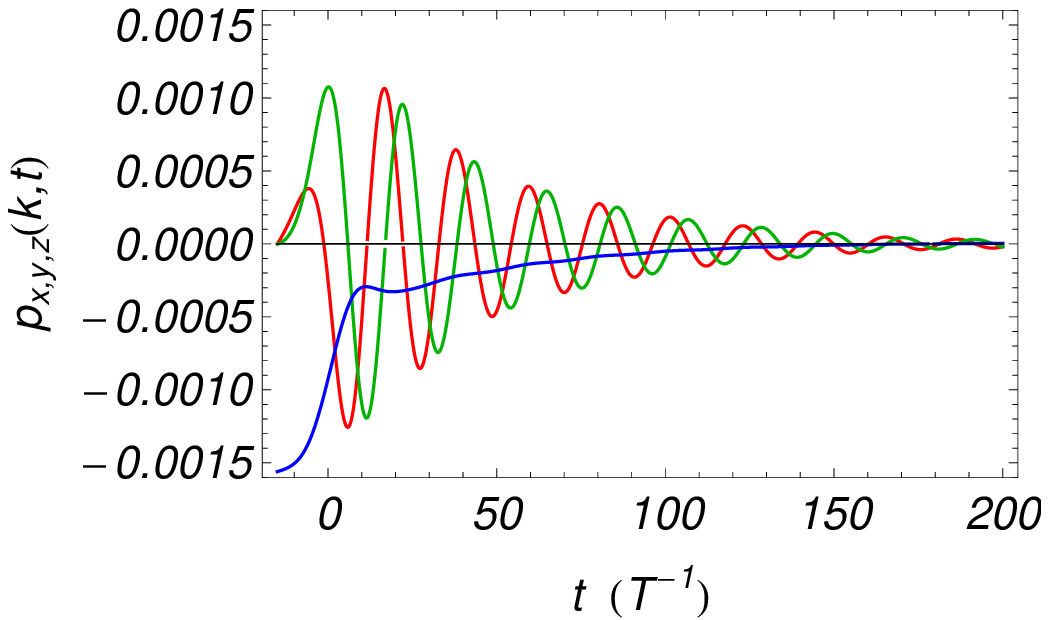,height=4.5cm}} \mbox{\hspace*{0.5cm}\epsfig{file=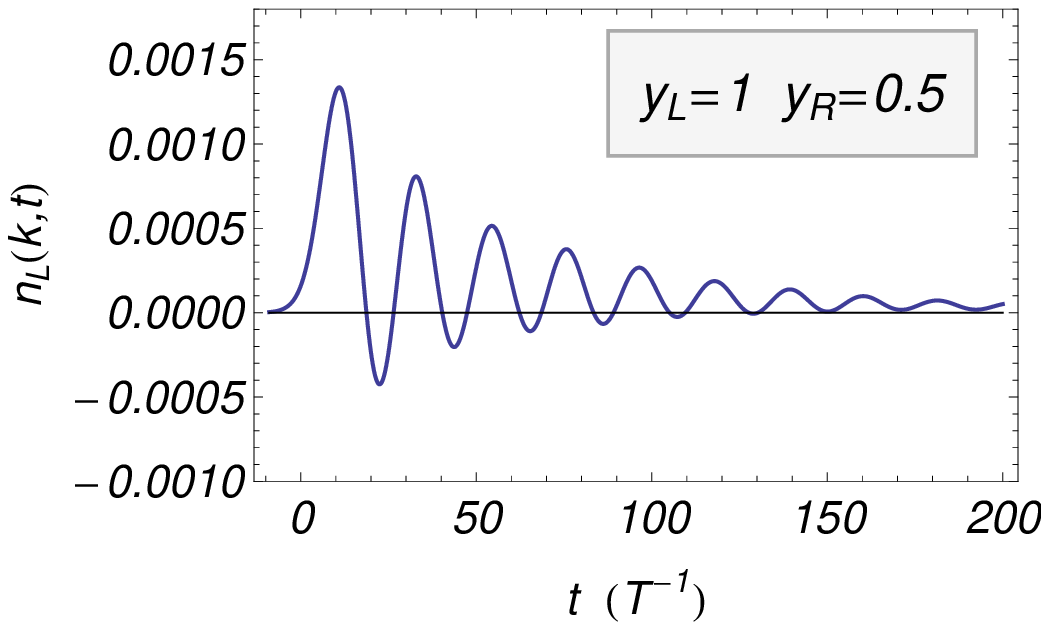,height=4.5cm}}

\end{center}
\vspace{-15pt}
\caption{\it\small 
Evolution of flavor polarizations with no collisions, flavor-blind collisions, and flavor-sensitive collisions. 
We plot the components  $p_x(k,t)$ (red), $p_y(k,t)$ (green), and $p_ z (k,t)$ (blue) [left panels] and $n_L(k,t)$ [right panels], 
for $k=3 T$, as a function of time in several regimes: 
no collisions  (1st row); 
flavor-blind collisions,   $y_L = y_R=1$, $g_{\textrm{eff}} = 200$   (2nd row); 
flavor-sensitive collisions with $y_L = 1$, $y_R=0.8$, $g_{\textrm{eff}} = 200$ (3rd row); 
flavor-sensitive collisions with $y_L = 1$, $y_R=0.5$, $g_{\textrm{eff}} = 200$ (4th row). 
In all cases we use equilibrium initial conditions at $t_{\rm in}= - 15/T$. 
All other input parameters are as in Table~\ref{tab:baseline}. The dotted line in the 2nd row represents the nonzero density $n_L$ surviving at late times corresponding to a non-vanishing chemical potential $\mu_L$. In the 3rd and 4th rows, the late-time density goes to zero since $y_L\not=y_R$. 
See text for additional details. 
}
\label{fig:collision1}
\end{figure}

\subsubsection{The effect of collisions:  $\tau_{\rm coll} \gg \tw$ and  $\tau_{\rm coll} \leq \tw$}

\begin{figure}[!t]
\begin{center}

\mbox{\hspace*{-0.cm}\epsfig{file=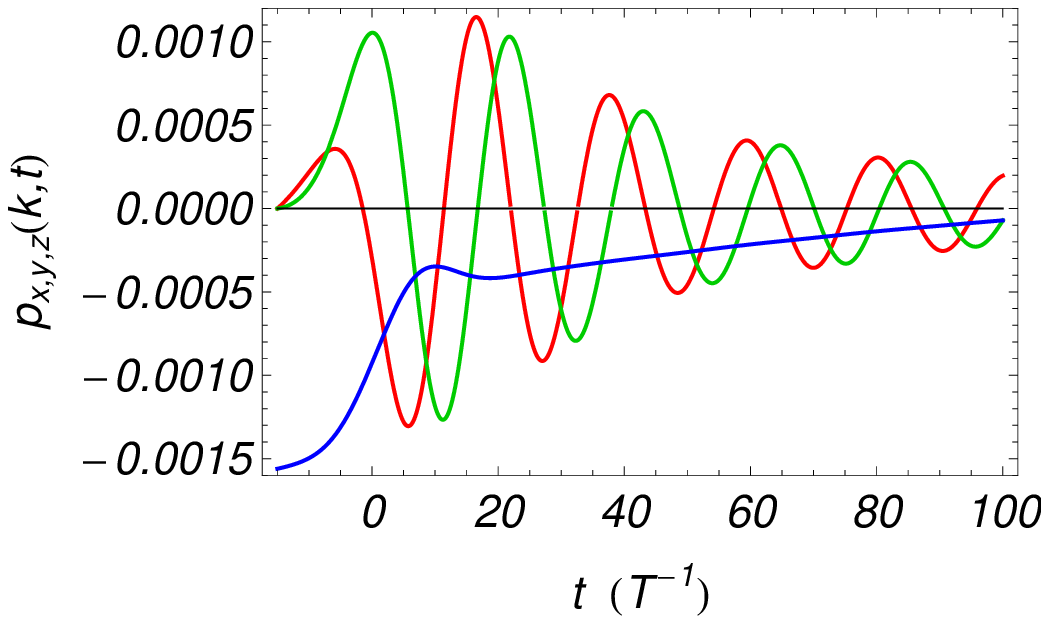,height=4.5cm}} \mbox{\hspace*{0.5cm}\epsfig{file=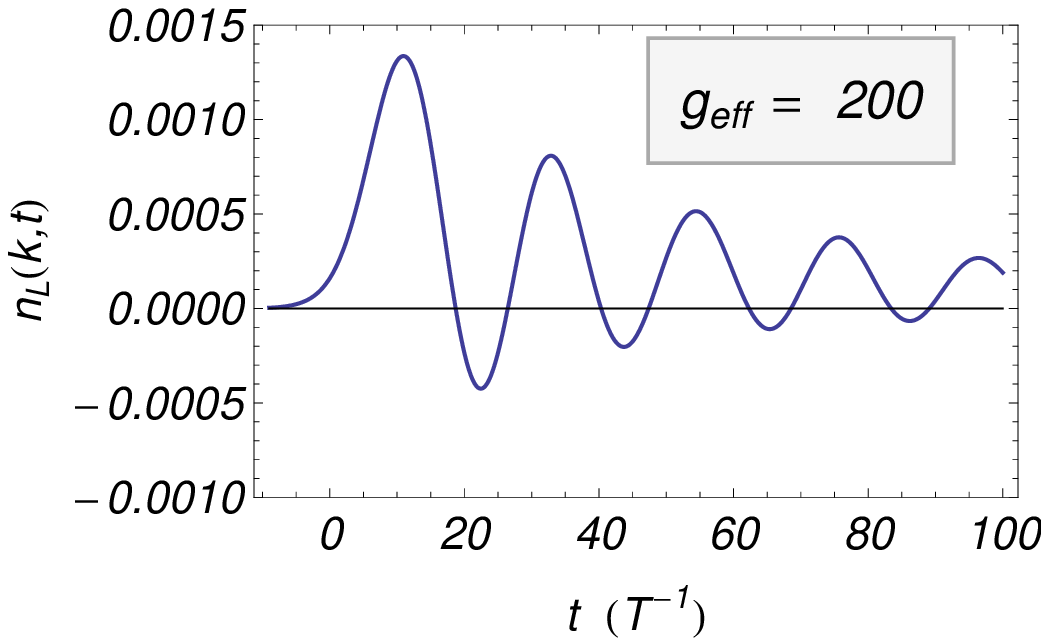,height=4.5cm}}
\vspace{0.0cm}

\mbox{\hspace*{-0.cm}\epsfig{file=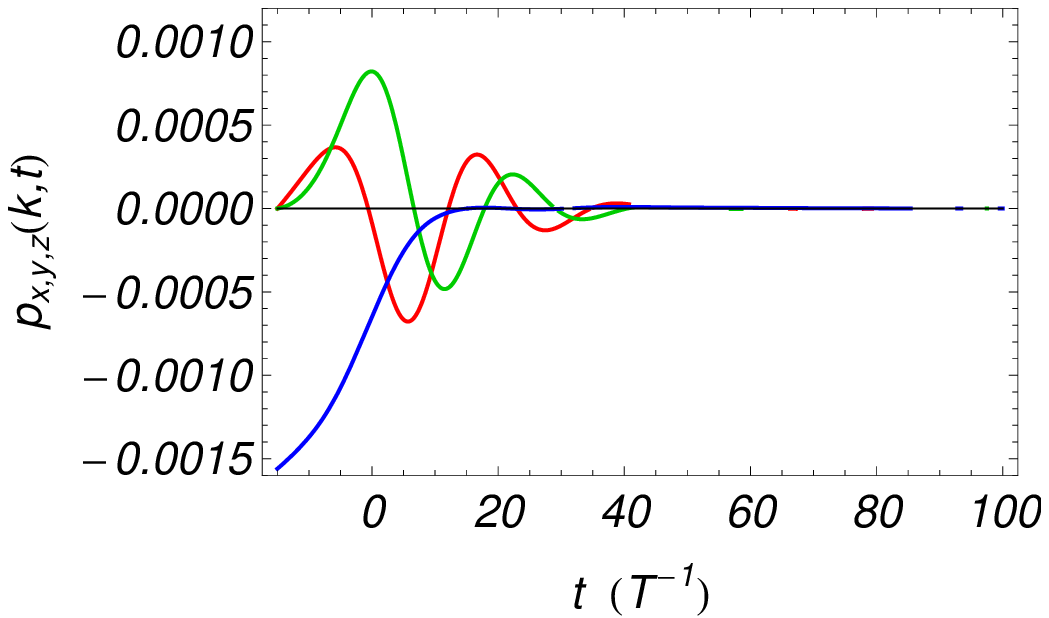,height=4.5cm}} \mbox{\hspace*{0.5cm}\epsfig{file=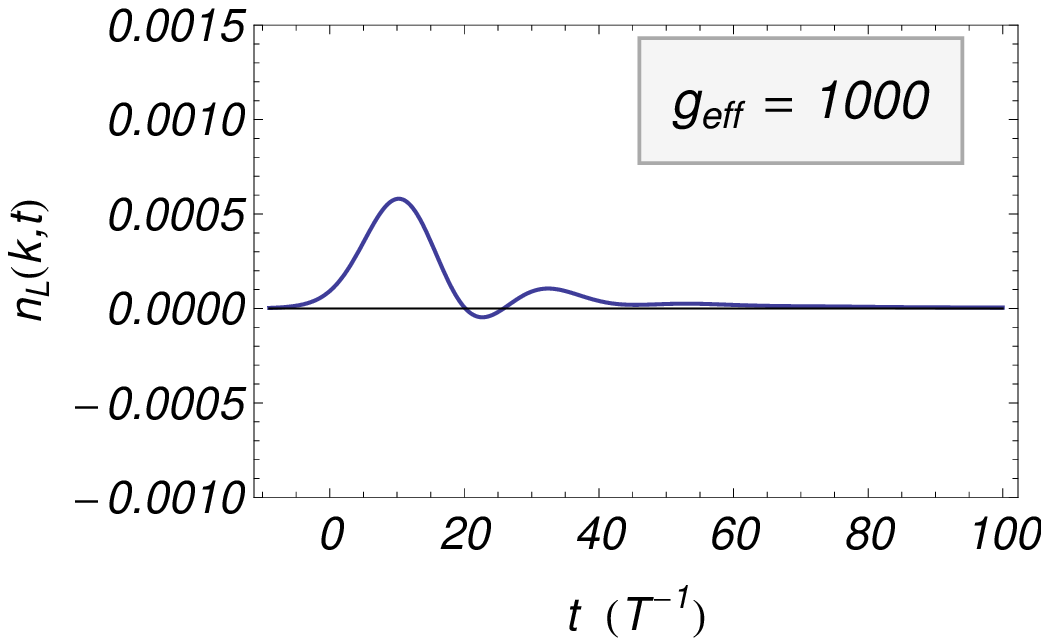,height=4.5cm}}
\vspace{0.0cm}

\mbox{\hspace*{-0.cm}\epsfig{file=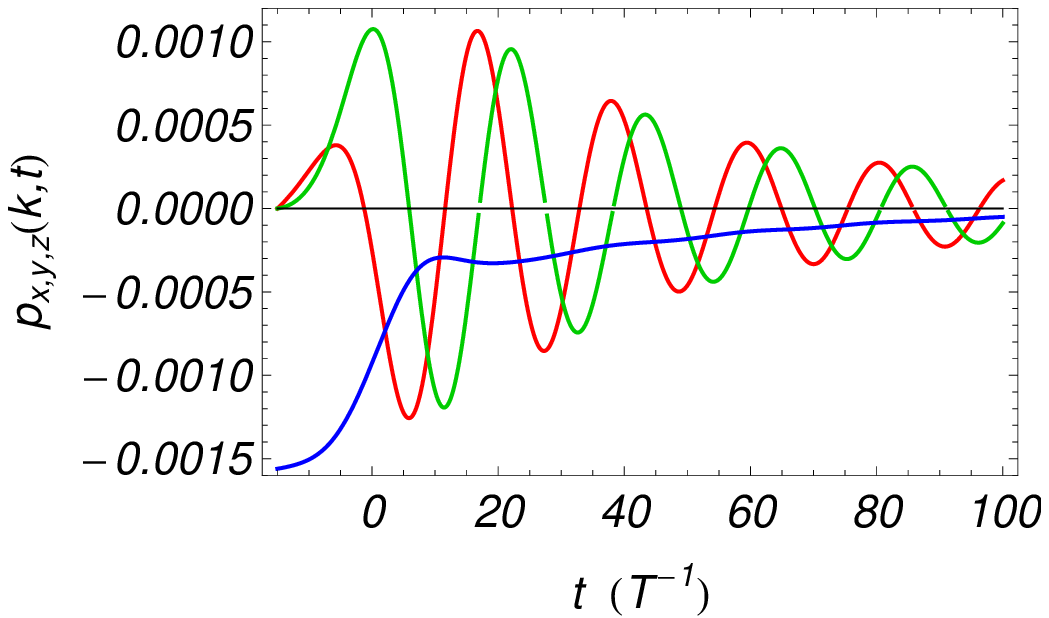,height=4.5cm}} \mbox{\hspace*{0.5cm}\epsfig{file=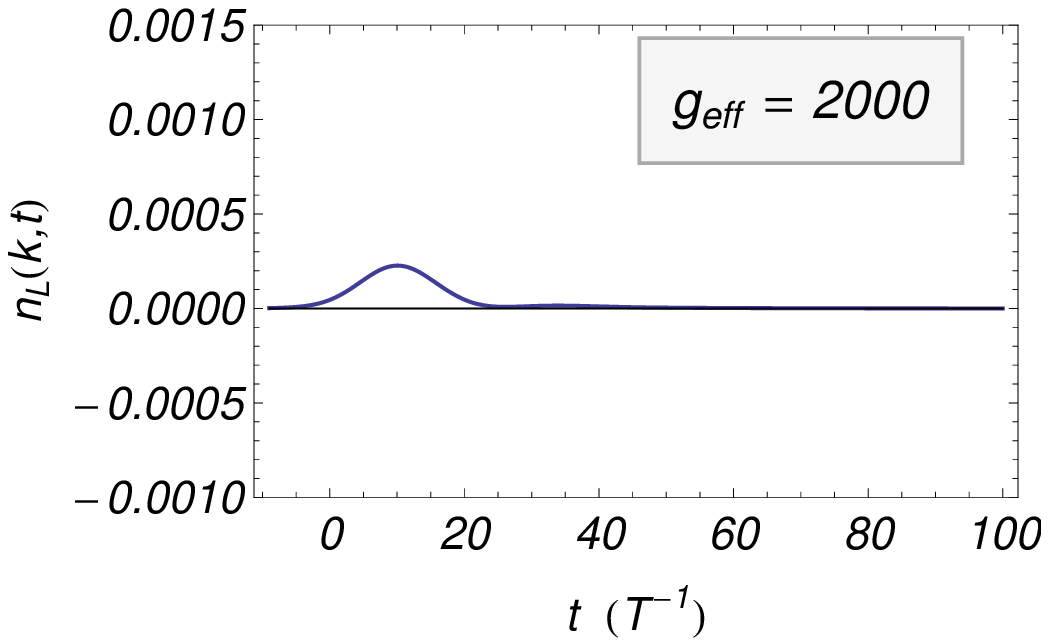,height=4.5cm}}

\end{center}
\caption{\it\small 
Dependence of solutions on the relative size of  $\tau_{\rm w}$ and $\tau_{\rm coll}$. 
Keeping $\tau_{\rm w}$ fixed, we control $\tau_{\rm coll}$ by dialing 
the  effective number of degrees of freedom $g_{\rm eff}$ in thermal bath. 
We plot the  polarization vectors $p_x(k,t)$ (red), $p_y(k,t)$ (green), and $p_ z (k,t)$ (blue) [left panels] and $n_L(k,t)$ [right panels], 
for $k=3 T$, as a function of time with  $y_L = 1$, $y_R=0.5$ and all other input parameters are as in Table~\ref{tab:baseline}, 
for different values of $g_{\rm eff}$. 
$g_{\rm{eff}} = 200$   (1st row); 
$g_{\rm{eff}} = 1000$ (2nd  row); 
$g_{\rm{eff}} = 2000$ (3rd  row). 
In all cases we use equilibrium initial conditions at $t_{\rm in}= - 15/T$. 
See text for additional details. 
}
\label{fig:collision2}
\end{figure}

In this section, we study quantitatively the impact of interactions with the thermal bath through 
a set of simulations whose results are shown in  Figs.~\ref{fig:collision1} and \ref{fig:collision2}.  Our key results are:
\begin{itemize}
\item Collisions (both flavor-blind and flavor-sensitive) lead to damping of flavor oscillations, such that $p_{x,y}, \, \tilde{p}_{x,y} \to 0$.
\item For flavor-sensitive collisions, all charge induced by the wall is damped away at late times ($p_{z}, \, \tilde{p}_z \to 0$).
\item For flavor-blind collisions, charge induced by the wall is not damped away at late times ($p_z \ne \tilde{p}_z \ne 0$).  This charge can be interpreted as a chemical potential.
\item Fast collisions $(\tau_{\textrm{coll}} \leq \tw)$ suppress the generation of $CP$ asymmetry.
\end{itemize}

Fig.~\ref{fig:collision1} illustrates the impact of flavor-blind versus flavor-sensitive interactions; it is organized as follows:
In the left column of  Fig.~\ref{fig:collision1} we plot the time evolution of the particle polarization vector $\vec{p} (k=3T,t)$, 
while in the right column we show the evolution of the flavor-diagonal $CP$ asymmetry $n_L (k=3T,t)$. 
We take different choices for $y_{L,R}$, illustrative of several regimes: (i) top row: collisionless case $y_L = y_R = 0$; (ii) second row: flavor-blind case $y_L = y_R = 1$; (iii) third row: flavor-sensitive case $y_L = 1$, $y_R = 0.8$, illustrative of a squark-driven EWB scenario, where the two flavor eigenstates share a common strong interaction and are distinguished only by electroweak interactions; and (iv) fourth row: flavor-sensitive case $y_L = 1$, $y_R = 0.5$, illustrative of
the chargino-like  scenario in the MSSM, in which the two flavor eigenstates have interaction strengths that might 
differ by a factor of $O(1)$. 
We have  adopted  $g_{\rm eff} = 200$,  
corresponding to an inverse collision rate $\tau_{\rm coll}$  considerably larger than 
than  the characteristic time scale of the external field $\tw = 10/T$.
We evolve the density matrix from equilibrium initial conditions at $t_{\rm in} = -15/T$,  
using  input parameters  as in Table~\ref{tab:baseline}.

The panels  in Fig.~\ref{fig:collision1}  clearly illustrate the impact of collisions on flavor oscillation dynamics: 
in all cases  the expected late-time thermalization is reached. 
In the case $y_L = y_R$  (second row), while the coherences decay ($p_{x,y} (k,t) \to 0$) at late times, 
we find $p_z (k,t) \to  ( n_B (\bar{\omega}_k - \mu_1) -  n_B (\bar{\omega}_k + \mu_1))/2 $, 
with $ \mu_1$ given by Eq.~\eqref{eq:mu1}.
Similarly, the late time behavior of $n_L (k,t) \equiv f_{LL} (k,t) - \bar{f}_{LL} (k,t)$  is consistent with 
the equilibrium form for the flavor diagonal density  $f_{LL} (k,t) = n_B (\bar{\omega}_k - \mu_L)$, 
with effective  L-handed chemical potential  given  by $\mu_L = (\cos^2  \theta - \sin^2 \theta)\,  \mu_1$~
\footnote{At late time $f_{LL}$ can be expressed in terms of $ f_{11}$ and
$f_{22}$ via  $f_{LL} = \cos^2 \theta  \, f_{11} + \sin^2 \theta \, f_{22}$. ÊUsing
these relations, and expanding to linear order in $\mu$, one can relate $\mu_L$
to $\mu_1$ as described.}.
The equilibrium value of $n_L$ corresponding to this $\mu_L$ is shown as the dotted line in 
the second row of Fig.~\ref{fig:collision1}.
Introducing flavor-sensitive interactions ($y_L \neq y_R$, third and fourth rows) induces 
no big differences at early times ($t \sim \tw$ for $g_{\rm eff} = 200$),   while it modifies the late time behavior 
as now $p_z (k,t) \to 0$ and $n_L (k,t) \to 0$. 
By comparing the behavior in the third and fourth rows of Fig.~\ref{fig:collision1}, 
one can also observe that as $|y_L - y_R|$ grows,  the chemical equilibrium 
($\mu_1 = \mu_L =  0$) is reached faster, in accord with intuitive expectations. 

So far we have used  values of the input parameters so that $\tau_{\rm coll} \gg  \tw, \tau_{\rm osc}$.
In this case the dynamics divides into three regimes.
(i) Early times $t = \mathcal{O}(\tw) \sim 10/T$.  This is the time scale over which the $CP$-asymmetry is generated through flavor 
oscillations,  and to a first approximation one can neglect the effect of collisions.   
(ii) Intermediate times $t \sim 100/T$.  This is the time scale when collisions damp away flavor oscillations.  Over these times, it appears irrelevant if the interactions are flavor-dependent or flavor-blind.
(iii) Late times $t \sim 1000/T$.  This the time scale over which chemical equilibrium is reached.  Here it 
matters crucially if the interactions are (nearly) flavor-blind or flavor-dependent.

If the underlying parameters are such that $\tau_{\rm coll} \leq  \tw , \tau_{\rm osc}$, one expects 
a strong damping of flavor oscillations and a strong suppression of the $CP$ asymmetry. In this case
the role of collisions is not merely  to relax to equilibrium the asymmetry generated by turning on the wall. 
Collisions are now so frequent that they 
prevent the system from going sufficiently out of equilibrium; oscillations cannot play a significant role in generating a $CP$ asymmetry. 
We illustrate this effect quantitatively  in  Fig.~\ref{fig:collision2},  in which we change  $\tau_{\rm coll}$ 
by dialing the parameter $g_{\rm eff}$ representing  the effective number of degrees of freedom present in the thermal bath. 
Again, we adopt equilibrium initial conditions at $t_{\rm in} = -15/T$,  
use  input parameters  as in Table~\ref{tab:baseline}, 
and  adopt  $y_L = 1, y_R = 0.5$.  
In the left column of  Fig.~\ref{fig:collision2} we plot the time evolution of the particle polarization vector $\vec{p} (k=3T,t)$, 
while on the right column we show the evolution of the flavor-diagonal $CP$ asymmetry $n_L (k=3T,t)$. 
From top to bottom we have 
$g_{\rm eff} = 200$,  
$g_{\rm eff} = 1000$,  
$g_{\rm eff} = 2000$,  
respectively.

\subsubsection{Is there a resonance?}

\begin{figure}[!t]
\begin{center}
\mbox{\hspace*{-1cm}\epsfig{file=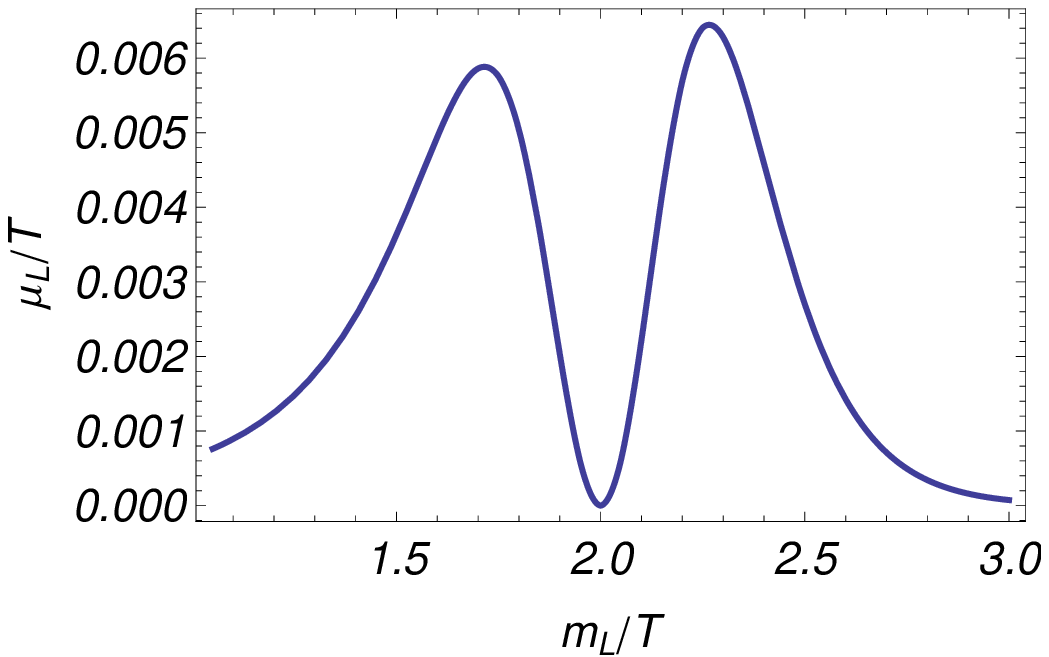,height=7cm}} 
\end{center}
\caption{\it\small 
Dependence of the  late time L-handed chemical potential $\mu_L$  on 
the mass parameter $m_L$,  with  $m_R/T=2$ and all other input parameters as in 
Table~\ref{tab:baseline}, with flavor-blind interaction $y_L=y_R = 1$. 
}
\label{fig:resonance}
\end{figure}

 In the realistic MSSM case,  the baryon asymmetry depends on the amount of L-handed charge asymmetry
 that survives in the unbroken electroweak phase after being   generated  within  the bubble wall.  
 While the  toy model, with only a time-dependent mass matrix, does not allow us to obtain a realistic estimate of the 
space-dependent  L-handed charge distribution,  it does allow us to study how the generated L-handed 
density depends on the  underlying parameters of the mass matrix. 
In particular,  our toy model can reveal whether
the so-called ``resonant" enhancement  
found in Refs.~\cite{Carena:1997gx,Riotto:1998zb,
Carena:2000id,
Carena:2002ss,
Cirigliano:2006wh}  for $m_L \sim m_R$ survives  within  the full particle mixing treatment. 
To address this question, we need a measure of the total charge asymmetry generated. 
For this purpose, we find it most convenient to work in the flavor-blind limit $y_L = y_R$ and  
evaluate the effective L-handed chemical potential $\mu_L = \mu_1  (\cos^2 \theta - \sin^2 \theta)$ 
that emerges in the  late-time solution,  starting from equilibrium initial conditions 
at some early initial time $t_{\rm in}  < - \tw$.  
In Fig.~\ref{fig:resonance} we plot the behavior of $\mu_L/T$  versus $m_L/T$, 
with all other input parameters fixed to the values of Table~\ref{tab:baseline}  (in particular $m_R/T = 2$).  
The behavior shown in Fig.~\ref{fig:resonance} results from the competition of several effects: 
\begin{itemize}
\item 
For $m_L = m_R$, there is no generation of $CP$ asymmetry at all,  
because the equilibrium initial condition with equal masses implies that $\vec{p} (k, t_{\rm in}) = (0,0,0)$, 
so there is no effective source (the commutator terms  $[\Sigma, f ]$ and $[\Sigma,\bar{f}]$ vanish).   
As long as $m_L \neq m_R$, the system gets pulled out of equilibrium when the wall turns on. 
However,  for  $m_L \to  m_R$ the initial condition $p_z (k, t_{\rm in}) \to 0$, thus  suppressing the 
final asymmetry.   

\item On the other hand, for given momentum  $k$,   as    $m_L \to  m_R$ 
the ratio $\tau_{\rm osc}/\tw$  increases and so one enters the non-adiabatic 
regime. This tends to increase  the $CP$ asymmetry, with 
maximal $CP$-violating effects obtained  for  $\tau_{\rm osc}/\tw \sim 4$  (see Fig.~\ref{fig:nLvslw}). 

\item The chemical potential $\mu_1$ is enhanced when the mixing angle is maximal $(\theta = \pi/4)$ and suppressed for $\theta = 0$.  On the other hand, the resulting $\mu_L$ is suppressed when $\theta \to \pi/4$, due to the $\cos 2\theta$ prefactor.

\end{itemize} 
It will be intriguing to study the extent to which this double-peaked structure will persist beyond leading 
non-trivial order in $\epsilon_{\rm coll}$ (see, e.g., \cite{Riotto:1998zb,Cirigliano:2006wh}) or when the bubble wall evolves not only in time but also in space. \\

In summary, in this section we have studied the effect of collisions on the generation and evolution of 
flavor-diagonal  $CP$ asymmetries. 
In preparation for  more realistic studies in supersymmetric scenarios,  
we have explored  different regimes, by varying: 
(i)  the degree to which interactions distinguish the two mixing flavors ($y_L$ versus $y_R$); 
(ii) the ratio $\tw/\tau_{\rm coll}$  of the wall time scale over the typical collision time. 
In the supersymmetric EWB scenarios, several mixing species can potentially contribute to the baryon asymmetry (squarks, staus, charginos, neutralinos, or Higgs scalars) and clearly their interactions fall into different regimes.  
While here we do not attempt to make predictions  for the realistic MSSM case,  
we can draw two general conclusions: 
(i)  a larger asymmetry is generated as long as   $\tw/\tau_{\rm coll} < 1$; 
(ii)  the generated  asymmetry persists longer  if  $y_L/y_R$ is close to  unity 
(the case of roughly flavor insensitive interactions).

\section{Conclusions}
\label{sec:conclude}

In the  present work  we have presented  the first step towards a consistent analysis of 
the  transport of  mixing particles in a $CP$-violating external background. 
While our ultimate goal is to apply the insight and techniques to the problem of 
weak scale baryogenesis,  here 
we focused on a simplified model of  mixing scalars $\Phi_{L,R}$ with time-dependent mass matrix, 
capturing the central ingredients  of the full baryogenesis problem. 
We derived quantum kinetic equations  for mixing scalars from first principles in non-equilibrium quantum field theory in the regime in which the oscillation 
time scale $\tau_{\rm osc}$ is comparable to or longer than the external  ``wall''  time-scale $\tw$. 
This is the physically interesting regime of non-adiabatic evolution, in which the 
non-equilibrium and $CP$-violating effects are largest. 

Our analysis provides a novel,  simple physical picture for the 
generation of  flavor diagonal  $CP$ asymmetries in
 the weak scale baryogenesis scenario: 
starting from a $CP$-conserving equilibrium initial state, 
a $CP$ asymmetry arises through coherent flavor oscillations. 
We have shown that, fixing the underlying $CP$-violating phases, 
the $CP$ asymmetry is essentially determined by the ratio $\tau_{\rm osc}/\tw$, 
reaching a maximum for $\tau_{\rm osc}/\tw\sim 4$. 
 
We have also  studied  in detail the effect of collisions, 
by solving the coupled kinetic equations for particle and anti-particle density matrices 
in a number of different regimes.  
As expected, collisions lead to decoherence and hence tend to  suppress the $CP$-violating asymmetry. 
From our analysis, two general lessons can be drawn:   
(i)  a larger asymmetry is generated as long as   $\tw/\tau_{\rm coll} < 1$, 
i.e.  for mean free path longer than the time scale over which the wall turns on;  
(ii)  the generated  asymmetry persists longer  if collisions are nearly flavor insensitive 
(that is, if $y_L/y_R$ is close to  unity in the toy model). 

To make contact with the existing baryogenesis phenomenology, 
we have  studied the dependence of the $CP$ asymmetry on the underlying 
mass matrix parameters, in particular on the difference of the flavor-diagonal 
mass entries $m_L$ and $m_R$.  
Working to linear order in the (small) ratios of physical timescales, 
we found that  while the  non-adiabaticity 
condition tends to maximize the asymmetry for $m_L \sim m_R$, 
in the case of exact degeneracy $m_L = m_R$ the asymmetry vanishes. 
So some sort of  ``resonant behavior" persists but with a double peak 
shifted above and below the $m_L = m_R$ point.  

As already emphasized, the results presented here represent  a first  step towards a realistic treatment 
of transport  phenomena involving  mixing particles at the weak phase transition.  
A number of outstanding issues have to be clarified before more realistic BAU 
calculations can be performed.  In particular,  work is in progress  to: 

\begin{itemize}

\item  Include a spacetime-dependent mass matrix. In this case  the main novel ingredient is  
the generation of spatial $CP$-violating currents. 
We plan to  study them numerically in the toy model and compare to the diffusion approximation. 
A proper understanding of spatial currents (whether or not in the diffusion limit)   
is crucial for the application to baryogenesis, as currents are known 
to enhance  the BAU~\cite{Cohen:1994ss} 
by transporting  $CP$ asymmetries in the unbroken phase where sphalerons are active. 

\item Formulate kinetic equations including effects beyond linear  order in $\epsilon$.

\item   Include in this formalism inelastic  (particle-number changing) 
interactions~\cite{Chung:2008aya}.

\item  Extend the formalism to fermions and  extend the network of kinetic equations 
to realistic cases in the MSSM and other extensions of the SM~\cite{Joyce:1994zt,Joyce:1994zn,Huet:1995sh,Kang:2004pp,Kang:2009rd}. 

\end{itemize}

\begin{acknowledgments}

We acknowledge useful discussions at various stages of this project with Alex Friedland, 
Bj\"orn Garbrecht, Boris Kayser, Thomas Konstandin, Emil Mottola and Petr Vogel.   
The work of VC  is supported by the Nuclear Physics Office of the U.S.
Department of Energy under Contract No.~DE-AC52-06NA25396 and by the LDRD program at Los Alamos National Laboratory. CL was supported in part by the U.S. Department of Energy under Contract DE-AC02-05CH11231, and in part by the National Science Foundation under grant  PHY-0457315.  
MJRM  and ST  were  supported in part by U.S. Department of Energy
contract DE-FG02-08ER41531 and by the Wisconsin Alumni Research Foundation.
ST is kindly supported by the Natural Sciences and Engineering Research Council of Canada.  For hospitality during significant portions of this work, we collectively thank the particle and nuclear theory groups at the Berkeley Center for Theoretical Physics and Lawrence Berkeley National Laboratory, the Institute for Nuclear Theory at the University of Washington, the University of Wisconsin-Madison, the California Institute of Technology, and Los Alamos National Laboratory.
\end{acknowledgments}

\appendix

\section{Review: Closed Time Path Formalism}
\label{appx:CTP}

In this section, we provide a brief introduction to the basics of the Closed Time Path (CTP) formalism.

Let us begin by considering a single complex scalar field $\varphi$, governed by the Lagrangian
\begin{equation}
\label{1flLag}
\mathcal{L} = \partial_\mu\varphi^*\partial^\mu\varphi  - m^2\varphi^*\varphi + \mathcal{L}_{\text{int}}\,.
\end{equation}
At zero temperature, perturbation theory is essentially the study of time-ordered propagators, such as
\be
G^t(x,y) = \left\langle \, \mathcal{T} \left\{ \varphi_h(x) \frac{}{}  \varphi_h^\dagger(y) \right\} \, \right\rangle \;, \label{eq:Gt}
\ee
where $\varphi_h$ is the Heisenberg-picture field.  The key difference when moving to finite temperature is that the expectation value in Eq.~\eqref{eq:Gt} is calculated not in the vacuum but in an ensemble of states defined by a density matrix 
\be
\hat\rho \equiv \sum_n w_n |n_h\rangle \langle n_h| \;,
\ee
where the time-independent Heisenberg-picture states $|n_h\rangle$ each have weight $w_n$.

Now, let us move to the interaction picture.  First, the interaction-picture states $|n(t)\rangle$ are functions of time; we define the interaction states at time $t=-\infty$ to coincide with the Heisenberg states: $|n_-\rangle \equiv |n(-\infty)\rangle = |n_h \rangle$.  The density matrix is $\hat\rho = \sum_n w_n |n_-\rangle \langle n_-|$.
Second, the interaction fields $\varphi$ are related to their Heisenberg counterparts by the time-evolution operator $\hat U$
\be
\varphi_h(x) = \hat U(x_0, -\infty)^\dagger \, \varphi(x) \, \hat U(x^0, -\infty) \;.
\ee
The operator $\hat U$ obeys the usual relations:
\be
\hat U(t_1, t_2) = \hat U(t_2, t_1)^\dagger =  \hat U(t_2, t_1)^{-1}
\ee
and
\be
\hat U(t_1,t_2) = \mathcal{T} \left\{ \exp \left( i \int_{t_1}^{t_2} dz^0 \int d^3 z \; \Lint(z) \right) \, \right\} \;.
\ee
With these relations, Eq.~\eqref{eq:Gt} becomes
\bea
G^t(x,y) &=& \sum_n \, w_n \left\langle \frac{}{} n_- \right| \left(\mathcal{T} \left\{ \exp \left[ i \int d^4 z \, \Lint(z) \right] \right\}\right)^{\dagger} \label{eq:Gtint} \\ 
&\;& \qquad \qquad \qquad \times \; \mathcal{T} \left\{ \varphi(x) \frac{}{}  \varphi^\dagger(y) \exp \left[ i \int d^4 z \, \Lint(z) \right] \right\} \,\left| \frac{}{} n_- \right\rangle \;,  \notag
\eea
where $\int d^4 z = \int^{\infty}_{-\infty} d z^0 \int d^3 z$.
Reading from right to left, Eq.~\eqref{eq:Gtint} corresponds to starting with the ``in''-state $|n_-\rangle$, then time-evolving from $-\infty$ to $+\infty$, acting with the field operators at times $x^0$ and $y^0$ along the way, and lastly time-evolving from $+\infty$ back to $-\infty$, returning to the ``in''-state.  This time-contour, denoted $\mathcal C$, is the ``closed time path''; it is closed in the sense that the contour begins and ends at $t=-\infty$, connecting ``in''-states with ``in''-states.
Eq.~\eqref{eq:Gtint} can then be succinctly written as
\bea
G^t(x,y) &=& \left\langle \mathcal{P} \left\{ \varphi_+(x) \varphi^\dagger_+(y) \exp \left[ i \int_{\mathcal C} d^4 z \, \Lint(z) \right] \right\} \right\rangle \notag \\
&=& \left\langle \mathcal{P} \left\{ \varphi_+(x) \varphi^\dagger_+(y) \exp \left[ i \int d^4 z \, \left( \Lint^{(+)}(z) - \Lint^{(-)}(z) \frac{}{} \right) \right] \right\} \right\rangle \label{eq:porder}
\eea
where $\mathcal P$ means path-ordering of fields along $\mathcal C$.  In the second line, we have reverted to a time integral from $-\infty$ to $+\infty$,  splitting $\Lint$ into two pieces corresponding to the forward and backward branches along the closed contour $\mathcal C$.  The notation $\varphi_{\pm}(x)$ and $\Lint^{(\pm)}(x)$ --- itself a function of $\varphi_\pm(x)$ --- denotes whether $x^0$ is on the time-increasing (+), or time-decreasing (--) branch of $\mathcal C$.  The path-ordering prescription is to time-order the (+) fields, to anti-time-order ($\mathcal{T}^\dagger$) the (--) fields, and lastly to put all the (--) fields to the left of the (+) fields.

A perturbative evaluation of $G^t(x,y)$ proceeds similarly to zero-temperature field theory.  Wick's theorem applies as usual, but with $\mathcal{P}$-ordering instead of $\mathcal{T}$-ordering.  Therefore, we must consider not one but four different propagators, corresponding to all possible path-ordering of $x^0$ and $y^0$ in $\langle \varphi(x) \varphi^\dagger(y) \rangle$:
\begin{subequations}
\bea
G^>(x,y) &\equiv& \left\langle \, \mathcal{P} \left\{ \varphi_-(x) \, \varphi_+^\dagger(y) \right\} \, \right\rangle = \left\langle \,  \varphi(x) \,  \varphi^\dagger(y) \, \right\rangle  \\
G^<(x,y) &\equiv& \left\langle \, \mathcal{P} \!\left\{ \varphi_+(x) \,  \varphi_-^\dagger(y) \right\} \, \right\rangle = \left\langle \,  \varphi^\dagger(y) \,  \varphi(x) \, \right\rangle    \\
G^t(x,y) &\equiv& \left\langle \, \mathcal{P} \!\left\{ \varphi_+(x) \,  \varphi_+^\dagger(y) \right\} \, \right\rangle = \left\langle \, \mathcal{T} \!\left\{ \varphi(x) \,  \varphi^\dagger(y) \right\} \, \right\rangle \\
&=& \theta(x^0-y^0) \, G^>(x,y) + \theta(y^0-x^0) \, G^<(x,y) \notag \\ 
G^{\bar t}(x,y) &\equiv& \left\langle \, \mathcal{P} \!\left\{ \varphi_-(x) \,  \varphi_-^\dagger(y) \right\} \, \right\rangle = \left\langle \, \mathcal{T}^\dagger \!\left\{ \varphi(x) \,  \varphi^\dagger(y) \right\} \, \right\rangle  \\
&=& \theta(y^0-x^0) \, G^>(x,y) + \theta(x^0-y^0) \, G^<(x,y) \; . \notag
\eea
\end{subequations}
These Green's functions are the free or full propagators for fields in the interaction- or Heisenberg-pictures, respectively.
In particular, we see from Eq.~\eqref{eq:porder} that a perturbative expansion of $G^t(x,y)$ will necessarily involve contracting (+) and (--) fields together; these additional propagators are essential.  The CTP propagators can be assembled into the matrix
\be
\widetilde{G}(x,y) \equiv \left( \ba{cc} G^t(x,y) & - G^<(x,y) \\ G^>(x,y) & - G^{\bar t}(x,y) \ea \right) \;.
\ee
In this form, the propagators satisfy the relations
\begin{subequations}
\label{eq:schwing}
\bea
\widetilde{G}(x,y) &=& \widetilde{G}^{(0)}(x,y) + \int d^4 w \int d^4 z \left(  \widetilde{G}^{(0)}(x,z) \, \widetilde{\,\Pi}(z,w) \;  \widetilde{G}(w,y) \right) \\
\widetilde{G}(x,y) &=& \widetilde{G}^{(0)}(x,y) + \int d^4 w \int d^4 z \left(  \widetilde{G}(x,z) \, \widetilde{\,\Pi}(z,w) \; \widetilde{G}^{(0)}(w,y) \right) \;. 
\eea
\end{subequations}
These equations are the CTP version of the Schwinger-Dyson equations, where now both the propagator and the self-energy $\widetilde{\,\Pi}$ (defined by $\Lint$) are $2\times 2$ matrices.
The free propagator $\widetilde{G}^{(0)}(x,y)$ satisfies
\be
\left( \partial_x^2 + m_\varphi^2 \right) \, \widetilde{G}^{(0)}(x,y) = \left( \partial_y^2 + m_\varphi^2 \right) \, \widetilde{G}^{(0)}(x,y) = -i \, \delta^4(x-y) \, \widetilde{I} \;,
\ee
where $\widetilde{I}$ denotes the $2 \times 2$ identity matrix in CTP propagator space.

Taking the Wigner transform Eq.~\eqref{eq:wignermania} of the Green's functions in \eq{eq:schwing}, and then taking the sum and difference of the two equations, we obtain the constraint and kinetic equations for the CTP propagators,
\begin{subequations}
\begin{align}
\label{CE}
\left(\frac{\partial_X^2}{4} - k^2 + m^2\right)\widetilde G(k;X) &= -i\begin{pmatrix} 1 & 0 \\ 0 & 1 \end{pmatrix}  \\
&\quad - \frac{i}{2} e^{-i\Diamond}\left( \{ \widetilde\Pi(k;X)\} \{\widetilde G(k;X)\} + \{ \widetilde G(k;X)\} \{\widetilde \Pi(k;X)\}\right) \nonumber \\
\label{KE}
2k\cdot\partial_X \widetilde G(k;X) &= e^{-i\Diamond}\left( \{ \widetilde\Pi(k;X)\} \{\widetilde G(k;X)\} - \{ \widetilde G(k;X)\} \{\widetilde \Pi(k;X)\}\right)\,,
\end{align}
\end{subequations}
where the $\Diamond$ differential operator was defined in \eq{diamond}. 

The various CTP Green's functions contain information both on the spectrum of excitations of the $\varphi$ fields in the Lagrangian \eq{1flLag} and on the distribution of states in the thermal bath.  We can isolate Green's functions depending only on the spectrum by forming the \emph{retarded} and \emph{advanced} propagators,
\begin{subequations}
\label{GRA}
\begin{align}
G^R(x,y)&= (G^t - G^<)(x,y) = \theta(x^0-y^0)(G^> - G^<)(x,y) \\
G^A(x,y) &= (G^t - G^>)(x,y) = \theta(y^0 - x^0)(G^< - G^>)(x,y)\,,
\end{align}
\end{subequations}
whose poles in momentum space give the spectrum of $\varphi$ excitations. 
We also use the linear combinations,
\begin{subequations}
\label{Gh}
\begin{align}
G^h(x,y) &= \frac{1}{2}(G^t - G^{\bar t})(x,y) = \frac{1}{2}\epsilon(x^0-y^0)(G^> - G^<)(x,y) \\
\rho(x,y) &= (G^> - G^<)(x,y) = (G^R - G^A)(x,y)\,.
\end{align}
\end{subequations}
The latter, $\rho(x,y)$, is also called the \emph{spectral function}. Note in addition the relations
\begin{equation}
\label{GhGRGA}
G^{R,A}(x,y) = G^h(x,y) \pm \frac{1}{2}(G^> - G^<) (x,y)\,.
\end{equation} 
We will also define analogous components of the CTP self-energy $\widetilde \Pi$. In particular, if we define $2i\Gamma = \Pi^> - \Pi^<$, we obtain
\begin{equation}
\Pi^{R,A} = \Pi^h \pm i\Gamma\,.
\end{equation}
The $\Pi^h$ component of the self-energy induces a correction to the mass in the dispersion relations for $\varphi$ excitations, while $\Gamma$ induces a nonzero width.

\section{One Flavor Boltzmann Equation}
\label{appx:1flavor}

In this appendix we show how the CTP equations of motion above may be used to derive the Boltzmann equations for the simple case of a single scalar field with a constant mass, following the logic of Ref.~\cite{Calzetta:1986cq}. We consider the one-flavor version of the interaction \eq{eq:lint} between $\varphi$ and a scalar field $A$,
\begin{equation}
\label{1flLint}
\mathcal{L}_{\text{int}} =  -\frac{y}{2} \varphi^*\varphi A^2\,.
\end{equation}
We will work from the Schwinger-Dyson equations \eq{eq:schwing} and the constraint and kinetic equations \eqs{CE}{KE}, solving them  order-by-order in an expansion in $y$, and therefore, $\eint$. We will derive equations for the spectrum of excitations through the constraint equation and equations for the retarded and advanced propagators, and for the distribution functions from the kinetic equations for $G^\gtrless$.  At $\mathcal{O}(y)$, we will obtain a correction to the spectrum of $\varphi$ excitations, and at $\mathcal{O}(y^2)$, we will be able to derive Boltzmann equations for the distribution functions from the quantum kinetic equation with a collision term driving the distributions to equilibrium.

\subsection{Tree-level: Generic Form}

At tree level, the self-energies $\widetilde\Pi = 0$, so the constraint and kinetic equations \eqs{CE}{KE} are just
\begin{align}
\left(\frac{\partial_X^2}{4} - k^2 + m^2\right) \widetilde G(k;X) & =  -i \Id \label{CE0} \\
k\cdot \partial_X \widetilde G(k;X) &=0\label{KE0} \,.
\end{align}
We assume that the density matrix giving the initial state induces no spatial variation in $\vect{X}$ at later times (see examples in \cite{Calzetta:1986cq}). In addition, the tree-level kinetic equation \eq{KE0} implies that interactions do not induce further $X$ dependence $\partial_X\widetilde G$ until at least order $\eint$, so  in the constraint equation \eq{CE0} we can drop $\partial_X^2\widetilde G$ as it is $\mathcal{O}(\eint^2)$. 
These equations of motion and translational invariance imply that the most general solution for the Green's functions for $\varphi$ at tree level must be of the form \cite{Calzetta:1986cq} (cf. \eq{treesolution})
\begin{subequations}
\label{treeform}
\begin{align}
\label{treeform>}
G^{0>}(k;X)  &= 2\pi\delta(k^2 - m^2) [ \theta(k^0)(1+ f(k)) + \theta(-k^0) \bar f(-k)] \\
\label{treeform<}
G^{0<}(k;X)  &= 2\pi\delta(k^2 - m^2) [ \theta(k^0)f(k) + \theta(-k^0) (1+\bar f(-k))] 
\end{align}
\end{subequations}
for $G^\gtrless$, and 
\begin{equation}
G^{R,A}(k) = \frac{i}{k^2 - m^2 \pm i\epsilon k^0}\,,
\end{equation}
for the retarded and advanced propagators. Note that $G^\gtrless$ are proportional to the spectral function $\rho = G^R - G^A$. The functions $f,\bar f$ are required to appear in both \eqs{treeform>}{treeform<} since $\rho = G^> - G^<$ also. 
The time- and anti-time-ordered propagators can be found by using \eq{GRA}.  

The functions $f,\bar f$ give the distribution of particle and antiparticle states in the plasma. Since the right-hand side of the kinetic equation is zero, there is no relaxation or dissipation of $f,\bar f$, which remain static.

\subsection{First order: Shifted Spectrum}

At first order in $y$, the self-energy comes from the tadpole diagram with the same topology as in Fig.~\ref{fig:feyn}a,
and is given by
\begin{equation}
\widetilde\Pi_1(z,w) = -\frac{iy}{2}\delta^4(z-w)\begin{pmatrix} G_A^{t}(z,w) & 0 \\ 0 & G_A^{\bar t}(z,w) \end{pmatrix}\,,
\end{equation}
or in momentum space,
\begin{equation}
\label{Pi1}
\widetilde \Pi_1(k;X) = -\frac{iy}{2}\int\frac{d^4 p}{(2\pi)^4}G_A^{t}(p;X)\Id\,,
\end{equation}
which is diagonal since $\int d^4 p\, G_A^{t}(p) = \int d^4p\, G_A^{\bar t}(p)$, and in fact is independent of $k$. The propagator for $A$ is given by the form \eq{treeform} with a distribution function $f_A = \bar f_A$. We then insert this into the right-hand sides of the constraint and kinetic equations \eqs{CE}{KE}. Since we can use the tree-level expressions for all Green's functions on the right-hand sides, the $\Diamond$ operators give zero when acting on them.  Furthermore, the right-hand side of the kinetic equation vanishes identically since $\widetilde\Pi_1$ is diagonal (in CTP space) and commutes with $\widetilde G$. The only effect of $\widetilde \Pi_1$ is to add a correction to the mass term in the constraint equation:
\begin{align}
\left(k^2 - m^2 - i \Pi_1(X)\right) \widetilde G(k;X) & =  i \Id  \label{CE1} \\
k\cdot \partial_X \widetilde G(k;X) &= 0 \label{KE1} \,.
\end{align}
The vacuum part of the mass correction $\Pi_1$ is absorbed into renormalization coefficients, and the $f_A$ dependent part gives a medium-dependent contribution to the mass, as in Eq.~(\ref{eq:thmass}). The form of these equations is still the same as the tree-level equations, with the Green's functions translation invariant up to $\mathcal{O}(y)$, so the solution takes the same form as the tree-level solution, with a corrected mass:
\begin{subequations}
\label{1form}
\begin{align}
G^{1R}(k;X) &=  \frac{i}{k^2 - m^2[f_A] - i\epsilon k^0} \\
G^{1>}(k;X) &=  2\pi\delta(k^2 - m^2[f_A]) [ \theta(k^0)(1+ f(k)) + \theta(-k^0) \bar f(-k)]  \,.
\end{align}
\end{subequations}
Because of \eq{KE1}, there are still no collisions affecting $f(k;X)$ to this order.

\subsection{Second order: Collisions}

At second order in $y$, the self-energy receives contributions from the graphs with the same topology as in Fig.~\ref{fig:feyn}b. 

In position space,
\begin{equation}
\label{Pi2}
\begin{split}
\widetilde\Pi_2(z,w) = &-\frac{y^2}{2}
\begin{pmatrix} 
G_A^{t}(z,w)^2 G^{t}(z,w) & -G_A^{<}(z,w)^2 G^{>}(z,w) \\
G_A^{>}(z,w)^2 G^{>}(z,w)  & -G_A^{\bar t}(z,w)^2 G^{\bar t}(z,w) 
\end{pmatrix}
\,.
\end{split}
\end{equation}
$\Pi_2$ now generates a nonzero contribution to the right-hand side not only of the constraint equation \eq{CE} but also the kinetic equation \eq{KE}. We again drop the diamond operators and the $\partial_X^2$ in the constraint equation, because $\partial_X$ acting on $\widetilde G$ is already at least order $y^2$.

We can also rearrange the CTP components of the constraint equation into a more convenient equation for the retarded Green's function $G_R = G^t - G^>$ (equivalently for the advanced Green's function $G_A = G^t-G^<$). We then obtain for the constraint and kinetic equations,
\begin{align}
\bigl[k^2 - m^2 - i \Pi_1(X) &- i\Pi_{2R}(k;X)\bigr] G_R(k;X)  =  i  \label{CE2}  \\
2k\cdot \partial_X \widetilde G(k;X) &= \left[ \Pi_2^>(k;X)G^{<}(k;X) - \Pi_2^<(k;X)G^{>}(k;X)\right] \begin{pmatrix} 1 & -1 \\ 1 & -1 \end{pmatrix} \label{KE2} \,.
\end{align}
The poles of the retarded propagator given by \eq{CE2} give the medium-dependent masses and widths of quasiparticle excitations.

Note that the kinetic equation \eq{KE2} implies for $G_R$,
\begin{equation}
k\cdot\partial_X G_R(k;X) = 0\,,
\end{equation}
so the $k\cdot \partial_X$ derivative of $G_R$ (and $G_A$, and therefore the spectral function $\rho = G_R - G_A$) is at least third order in $y$.  This will allow us to neglect $k\cdot\partial_X$ acting on the spectral function in solving the kinetic equation.

Since the form of the $\mathcal{O}(y^2)$ equations of motion \eqs{CE2}{KE2} is no longer the same as the tree-level form, we must be more careful in justifying the form \eq{1form} for the Green's functions. The solution to \eq{CE2} for the retarded propagator,
\begin{equation}
G_R(k;X) = \frac{i}{k^2 - m^2 - i\Pi_1(X) - i\Pi_{2R}(k;X)} \,,
\end{equation}
and similarly for the advanced, gives the spectral function $\rho(k;X) = (G_R - G_A)(k;X)$. The spectral function also satisfies $\rho = G^> - G^<$. Thus, the most general form for $G^{\gtrless}$ is
\begin{subequations}
\label{cform}
\begin{align}
G^>(k;X) &= \rho(k;X)[1 + F(k;X)] + c(k;X) \\
G^<(k;X) &= \rho(k;X)F(k;X) + c(k;X)\,.
\end{align}
\end{subequations}
Since the Green's functions satisfied the form \eq{1form} at $\mathcal{O}(y)$, the function $c(k;X)$ must be explicitly at least of order $y^2$. It must also satisfy the constraint equation. Since $c$  already has an explicit $y^2$ it must satisfy the tree-level constraint equation. Then it is proportional to $\delta(k^2 - m^2)$, and can be absorbed into the $\rho(k;X)$ part of \eq{cform}. Redefining $F$, we bring $G^\gtrless$ into similar form as \eq{1form} in terms of $f$,
\begin{equation}
\label{2form}
G^>(k;X) =  \rho(k;X) [\theta(k^0)(1+ f(k;X)) - \theta(-k^0)\bar f(-k;X) ] \,.
\end{equation}

Now we evaluate the kinetic equation \eq{KE2} using the expression \eq{Pi2} for $\Pi_2$ and the form \eq{1form} for the Green's functions. Integrating over positive $k^0$, we find the Boltzmann equation for the particle distribution function,
\begin{equation}
\label{Boltzmann}
2k \cdot\partial_X f(k;X) = \int_0^\infty \frac{dk^0}{2\pi}\bigl[\mathcal{C}_{\text{ann}}(k;X) + \mathcal{C}_{\text{scatt}}(k;X) \bigr]\,,
\end{equation}
where the annihilation and scattering collision terms are the one-flavor analogs of  \eqs{eq:anncol}{eq:scatcol},
\begin{equation}
\label{1flann}
\begin{split}
\int_0^\infty \frac{dk^0}{2\pi}\mathcal{C}_{\text{ann}}(k;X) &= -\frac{y^2}{4\omega_{\vect{k}}}\int\frac{d^3 k'}{(2\pi)^3 2\omega_{\vect{k}'}} \frac{d^3 p}{(2\pi)^3 2\varepsilon_{\vect{p}}}\frac{d^3 p'}{(2\pi)^3 2\varepsilon_{\vect{p}'}} (2\pi)^4\delta^4(k +k' - p -p') \\
&\times\left\{ f(k) \bar f(k')[1 + f_A(p)] [1 + f_A(p')] - [1+f(k)] [1+\bar f(k')] f_A(p) f_A(p')\right\} 
 \,,
\end{split}
\end{equation}
and
\begin{equation}
\label{1flscatt}
\begin{split}
\int_0^\infty \frac{dk^0}{2\pi}\mathcal{C}_{\text{scatt}}(k;X) &= -\frac{y^2}{2\omega_{\vect{k}}}\int\frac{d^3 k'}{(2\pi)^3 2\omega_{\vect{k}'}} \frac{d^3 p}{(2\pi)^3 2\varepsilon_{\vect{p}}}\frac{d^3 p'}{(2\pi)^3 2\varepsilon_{\vect{p}'}} (2\pi)^4\delta^4(k - k' + p -p') \\
&\times\left\{ f(k) [1+ f(k')] f_A(p) [1 + f_A(p')] - [1+f(k)]f(k') [1+f_A(p)] f_A(p')\right\} 
 \, .
\end{split}
\end{equation}
We have performed the frequency integrals in the collision term using the free spectral function; corrections to this approximation are $\mathcal{O}(y^3)$.

Note that  \eq{Boltzmann} tells us that collisions create $\mathcal{O}(y^2)$ spacetime variations in the distribution functions. So if we were to continue to higher orders in perturbation theory, we could not keep neglecting the $\partial_X^2$ term in the constraint equation \eq{CE} and the $\Diamond$ operators in the constraint and kinetic equations \eqs{CE}{KE} as we have done so far.

\subsection{Equilibrium Solution}

To find the equilibrium solution for the distribution functions for $\varphi$ and $A$, we require that the collision terms in \eqs{1flann}{1flscatt} vanish.
It is easy to show that, due to the property
\begin{equation}
1 + n_B(k_0 - \mu) = e^{(k_0 - \mu)/T} \, n_B(k^0 - \mu)\,,
\end{equation}
and the momentum-conserving delta functions in \eq{Boltzmann}, the forms
\begin{equation}
\label{equilibrium}
\begin{split}
f(k;X) = n_B(k_0 - \mu(X))\,, &\quad \bar f(k;X) = n_B(k^0 + \mu(X)) \\
f_{A}(k;X) &= n_B(k_0)\,,
\end{split}
\end{equation}
are such a solution, as long as the $X$ variation of the chemical potentials is smaller than $\mathcal{O}(y^2)$.
Therefore, interactions at $\mathcal{O}(y^2)$ (namely, binary collisions or annihilation due to $y^2$ interactions) leave the distributions \eq{equilibrium} unchanged, as long as $\mu(X)$ satisfies the continuity 
equation $k \cdot \partial_X \mu = 0$.

\section{Two-Flavor Spectrum of Excitations to $\mathcal{O}(\epsilon)$}
\label{appx:constraint}

In this appendix we include further details on the two-flavor CTP Green's functions and solve for the spectrum of excitations to $\mathcal{O}(\epsilon)$ in the toy model introduced in \sec{sec:prelim}.

The spectrum of excitations can be found from the equations for the Wigner-transformed retarded and advanced propagators,
\begin{equation}
\label{constraintRA}
\begin{split}
\biggl(\frac{\partial_{t}^2}{4} - k^2\biggr)G^{R,A}(k;t) + \frac{1}{2}e^{-i\Diamond}\bigl\{m^2(t),G^{R,A}(k;t)\bigr\} -ie^{-i\Diamond}\bigl\{k^0\Sigma(t),G^{R,A}(k;t)\bigr\} \\
= - i - i e^{-i\Diamond}\bigl\{\Pi^{R,A},G^{R,A}\bigr\}
\end{split}
\end{equation}
and
\begin{equation}
\label{kineticRA}
\begin{split}
2k^0\partial_t G^{R,A}(k;t) &+ ie^{-i\Diamond}\bigl[m^2(t),G^{R,A}(k;t)\bigr] +\frac{1}{2}e^{-i\Diamond}\bigl[k^0\Sigma(t),G^{R,A}(k;t)\bigr]= e^{-i\Diamond}\bigl[\Pi^R, G^R\bigr] \,.
\end{split}
\end{equation}
After solving for $G^{R,A}$ and constructing the spectral function $\rho = G^R - G^A$, we can then use $\rho = G^>- G^<$ and the kinetic equations  \eq{kinetic} for $G^\gtrless$ to determine $G^\gtrless$.

We begin with the constraint equations at $\mathcal{O}(\epsilon^0)$: 
\begin{equation}
\label{2flCE0}
k^2\widetilde G - \frac{1}{2}\Bigl\{m^2,\widetilde G\Bigr\} = i \, \Id\,,
\end{equation}
proportional to the identity in both field space and CTP space. We gave the general form of the solution for $G^\gtrless_{ij}$ in \eq{treesolution}. The solution for the diagonal components of the retarded and advanced propagators are
\begin{equation}
G^{R,A}_{11,22} = \frac{i}{k^2 - m_{1,2}^2 \pm i\epsilon k^0}\,.
\end{equation}
Since all CTP components of the off-diagonal Green's function $\widetilde G_{12}$ obey the same equation \eq{2flCE0}, and since $G^{R,A} = G^t - G^{\lessgtr}$, this implies that the $\mathcal{O}(\epsilon^0)$ off-diagonal retarded and advanced propagators $G_{12}^{R,A} = 0$.

Next, keeping terms up to $\mathcal{O}(\epsilon)$ in \eq{constraintRA}, we obtain the equation for retarded and advanced propagators,
\begin{equation}
k^2 G^{R,A} -\frac{1}{2}\bigl\{ m^2, G^{R,A}\bigr\} +  ik^0\bigl\{\Sigma(t),G^{R,A}\bigr\} = i + i\bigl\{\Pi^{R,A},G^{R,A}\bigr\}\,.
\end{equation}
Writing the diagonal and off-diagonal components (in field space) separately, we find
\begin{subequations}
\begin{gather}
\label{constraintdiag1}
(k^2 - m_1^2 + 2ik^0\Sigma_{11} - 2i\Pi_{11}^R)G^R_{11} = i+i(\Pi^R_{12} - k^0\Sigma_{12})G_{21}^R +  (\Pi^R_{21} -k^0\Sigma_{21})G_{12}^R  \\
\label{constraintoff1}
\biggl(k^2 - \frac{1}{2}(m_1^2+m_2^2) + ik^0(\Sigma_{11}+\Sigma_{22}) - i(\Pi_{11}^R+\Pi_{22}^R)\biggr)G^R_{12} = i(\Pi_{12}^R - k^0\Sigma_{12})(G_{11}^R + G_{22}^R)\,,
\end{gather}
\end{subequations}
and similarly for $G_{22}^R$ and $G_{21}^R$ (and $G^A$).  

In the equations for the diagonal components, the off-diagonal components $G_{12,21}^R$ act as sources for $G_{11,22}$. As they multiply factors $\Pi,\Sigma$ explicitly of $\mathcal{O}(\epsilon)$, we can use the $\mathcal{O}(\epsilon^0)$ solutions for $G_{12,21}^R$ in these terms, which we noted above are zero. Therefore, the $\mathcal{O}(\epsilon)$ solutions for $G_{11,22}^{R,A}$ are simply
\begin{equation}
\label{GRdiag1}
G_{11,22}^{R,A}(k;t) = \frac{i}{k^2 - m_{1,2}^2(t) + 2ik^0\Sigma_{11,22}(t) - 2i\Pi_{11,22}^{R,A}(k;t)}\,.
\end{equation}
These solutions give us the shifted spectra of the excitations that had tree-level masses $m_{1,2}$ due to interactions $\Sigma$ with the wall  and $\Pi$ with other particles in the bath. They induce a corrected mass and width for the excitations. From these solutions, first, we construct the diagonal spectral functions $\rho_{11,22} = G^R_{11,22} - G^A_{11,22}$, and, second, we can use the spectral functions to construct the Green's functions $G^\gtrless_{11,22}$:
\begin{subequations}
\label{Gdiagspectral}
\bea
G_{11,22}^>(k;t) &= \rho_{11,22}(k;t) [\theta(k^0)(1+ f_{11,22}(k;t)) - \theta(-k^0) \bar f_{11,22}(-k;t)]  \\
G_{11,22}^{<}(k;t)  &= \rho_{11,22}(k;t) [\theta(k^0) f_{11,22}(k;t) - \theta(-k^0)(1+ \bar f_{11,22}(-k;t))]\,.
\eea
\end{subequations}
These Green's functions have the same form as the tree-level solutions \eq{treesolution}, with modified dispersion relations given by the spectral function.
These forms satisfy the constraint equations \eq{constraint}  for $G^\gtrless_{11,22}$ at $\mathcal{O}(\epsilon)$, while the  kinetic equations \eq{kinetic} at $\mathcal{O}(\epsilon)$ give nontrivial evolution of the distribution functions $f,\bar f$, as we derived in the main text. In this paper, since we were interested in the evolution of $f,\bar f$ only at $\mathcal{O}(\epsilon)$, we kept the tree-level spectral functions given in \eq{treesolution} when solving the kinetic equation.

For the off-diagonal components, substituting the solutions \eq{GRdiag1} for $G_{11,22}^{R,A}$ into the right-hand side of \eq{constraintoff1} implies that, to $\mathcal{O}(\epsilon)$, 
\begin{equation}
\label{GRoff1}
G_{12}^{R,A}(k;t) =  -\frac{2[\Pi_{12}^{R,A}(k;t) - k^0\Sigma_{12}(t)]}{(k^2 - m_1^2(t))(k^2-m_2^2(t))}\,.
\end{equation}
This solution induces an $\mathcal{O}(\epsilon)$ contribution to the spectral function $\rho_{12} = G^R_{12} - G^A_{12}$, which would add a modification to the dispersion relations appearing in the tree-level solutions \eq{treesolution} for $G_{12}^\gtrless$, so that these also obey the constraint equations \eq{constraint} at $\mathcal{O}(\epsilon)$. However, to solve for the $\mathcal{O}(\epsilon)$ evolution of $f_{12},\bar f_{12}$ with the kinetic equation, it suffices to keep the tree-level dispersion relations in \eq{treesolution}.

\end{document}